\documentclass[iop,referee,twocolumns]{emulateapj}
\usepackage{aas_macros}
\include{jdefs}

\usepackage{graphicx,xcolor,amssymb,amsmath,amsfonts,hyperref}

\newcommand{\Fermi}{\textit{Fermi}}
\newcommand{\ergs}{\,\text{erg}\,\text{s}^{-1}}
\newcommand{\kpc}{\,\text{kpc}}
\newcommand{\G}{\,\text{G}}
\newcommand{\mJy}{\,\text{mJy}}
\newcommand{\muJy}{\,\mu\text{Jy}}

\newcommand{\GeV}{\,\text{GeV}}
\newcommand{\GHz}{\,\text{GHz}}
\newcommand{\ms}{\,\text{ms}}
\newcommand{\h}{\,\text{h}}
\newcommand{\cm}{\,\text{cm}}
\newcommand{\pc}{\,\text{pc}}
\newcommand{\MeV}{\,\text{MeV}}
\newcommand{\K}{\,\text{K}}
\newcommand{\fex}{\textit{e.g.}}
\newcommand{\ie}{\textit{i.e.}}

\shorttitle{Radio searches for bulge MSPs from Fermi diffuse observations}
\shortauthors{F.~Calore et al.}

\begin{document}

\title{Radio detection prospects for a bulge population of millisecond pulsars
as suggested by Fermi LAT observations of the inner Galaxy}

\author{F.~Calore$^{1,\dagger}$, M.~Di Mauro$^{2}$, F.~Donato$^{3,4}$, J.W.T.~Hessels$^{5,6}$, C.~Weniger$^{1,\ddagger}$}
\affil{$^1$GRAPPA Institute, University of Amsterdam, Science Park 904, 1090 GL
  Amsterdam, Netherlands \\
  $^2$ Department of Physics and SLAC National Accelerator Laboratory, Stanford University, Stanford, CA 94305, USA \\
$^3$Physics Department, Torino University, via Giuria 1, 10125 Torino, Italy\\
$^4$Istituto Nazionale di Fisica Nucleare, Sezione di Torino, via Giuria 1, 10125 Torino, Italy\\
$^5$ ASTRON, the Netherlands Institute for Radio Astronomy, Postbus 2, 7990 AA, Dwingeloo, The Netherlands\\
$^6$ Anton Pannekoek Institute for Astronomy, University of Amsterdam, Science Park 904, 1098 XH Amsterdam, The Netherlands}

\email{$^\dagger$ email: f.calore@uva.nl}
\email{$^\ddagger$ email: c.weniger@uva.nl}

\begin{abstract}
The dense stellar environment of the Galactic center has been proposed to host a large population of as-yet undetected millisecond pulsars (MSPs).  Recently, this hypothesis has found support in an analysis of gamma rays detected with the Large Area Telescope onboard the \Fermi\ satellite, which revealed an excess of diffuse GeV photons in the inner 15 deg about the Galactic center. The excess can be interpreted as the collective emission of thousands of MSPs in the Galactic bulge, with a spherical distribution strongly peaked towards the Galactic center.  In order to fully establish the MSP interpretation, it is essential to find corroborating evidence in multi-wavelength searches, most notably through the detection of radio pulsations from individual bulge MSPs.  Based on globular cluster observations and gamma-ray emission from the inner Galaxy, we investigate the prospects for detecting MSPs in the Galactic bulge.  While previous pulsar surveys failed to identify this population, we demonstrate that upcoming large-area surveys of this region should lead to the detection of dozens of bulge MSPs.  Additionally, we show that deep targeted searches of unassociated \Fermi\ sources should be able to detect the first few MSPs in the bulge.  The prospects for these deep searches are enhanced by a tentative gamma-ray/radio correlation that we infer from high-latitude gamma-ray MSPs.   Such detections would constitute the first clear discoveries of field MSPs in the Galactic bulge, with far-reaching implications for gamma-ray observations, the formation history of the central Milky Way and strategy optimization for future deep radio pulsar surveys.
\end{abstract}
  

\maketitle

\section{Introduction}
\label{sec:intro}

Millisecond pulsars (MSPs) are rapidly spinning neutron stars that produce
observable pulsations (mostly in radio, but often also in gamma-rays, and occasionally in X-rays), 
have short spin periods and low surface magnetic fields (compared to other pulsars) that are
loosely in the range $P\leq30\ms$ and $B \leq10^{9}\G$.  MSPs are believed to
originate from pulsars in binary systems, in which the companion star transfers
material to the pulsar, reducing its magnetic field and increasing its angular
momentum.  During the accretion phase, and for low-mass companions, the system
can often be seen as a low-mass X-ray binary.  Afterwards, an MSP (for that
reason also called \emph{recycled} pulsar) is left behind and can emit observable
pulsations for
about $10^{10}$ years~\citep{1991PhR...203....1B}.

MSPs have a multi-wavelength emission spectrum, including both pulsed and un-pulsed types of emission,
 from radio frequencies up to TeV
gamma rays.  MSPs emit soft X-rays through the polar caps ($kT \leq 1$
keV,~\cite{2003A&A...398..639Z}).  They can also shine in GeV gamma rays
through curvature radiation as predicted by outer gap
models~\citep{2003A&A...398..639Z}.  We refer to the recent review by
\cite{Grenier:2015pya} for further details and references.  Strong pulsar
winds, accelerating relativistic electrons interacting with the surrounding
medium, might be responsible for non-pulsed X-ray emission through synchrotron
radiation~\citep{2000ApJ...539L..45C,2004ApJ...617..480C} and for TeV photons
through inverse Compton scattering~\cite{1997MNRAS.291..162A}.  The detailed
timing of the multi-wavelength emission provides useful information to study
emission models \citep[\fex][]{Kalapotharakos:2013sma}.

About 370 MSPs are currently known at radio frequencies: 237 of them are field
MSPs in the Galactic
disk,\footnote{\scriptsize\url{http://astro.phys.wvu.edu/GalacticMSPs/GalacticMSPs.txt}}
and 133 (with $P\leq30\ms$) are associated with 28 different globular
clusters.\footnote{\scriptsize\url{http://www.naic.edu/~pfreire/GCpsr.html}}
Historically, the first $\sim35$ field MSPs were found in the 1980s and 1990s
in large area radio surveys, mainly based on the Parkes southern sky survey and
the Arecibo survey at 430 MHz. Subsequently, various large area surveys using again
the Parkes telescope, Arecibo, and since 2002 also the Green Bank Telescope
(GBT), lead to the discovery of around 200 MSPs \citep[for a recent review
see][]{Stovall:2013gca}.  Additionally, $\sim70$ MSPs
were discovered in radio follow-ups of \Fermi~unassociated
sources~\citep{Ray:2012ue}, and at least one MSP was first detected by
observing gamma-ray pulsations~\citep{TheFermi-LAT:2013ssa}.  All MSPs in
globular clusters were instead found in deep targeted searches.  

\medskip

The presence of gamma-ray and radio MSPs in the Galactic disk and in globular
clusters is now well established~\citep{collaboration:2010bb,
TheFermi-LAT:2013ssa}.  Additionally, it has been long proposed that 
the Galactic center might harbor an MSP
population with a much larger number density than the Galactic
 disk. One traditional
argument~\citep{2015ApJ...805..172M} supporting this hypothesis is that the
high stellar density at the Galactic center is substantially different from the
disk.  In such a highly dense stellar environment the likelihood for the
formation of binary systems is enhanced.  This results in a higher probability
to produce MSPs, as it happens in the dense environment of globular
clusters~\citep{1982Natur.300..728A,1987IAUS..125..187V,2000ApJ...535..975C}.
On the other hand, these MSPs might be the fossils of tidally disrupted globular
clusters that fell in towards the Galactic center because of dynamical
friction.  They would release all their stellar content and contribute to the
nuclear stellar cluster and the Galactic bulge~\citep{1975ApJ...196..407T,
2014MNRAS.444.3738A,2014ApJ...785...71G,Brandt:2015ula}.

A population of $\sim6000$ MSPs at the Galactic center was first proposed
by~\cite{Wang:2005ti} in order to explain various multi-wavelength observations
at the same time: The large number of unidentified {\it Chandra} X-ray 
sources~\citep{2003ApJ...589..225M}, the EGRET GeV diffuse gamma-ray emission
in the inner $1.5^{\circ}$~\citep{1998A&A...335..161M}, and the TeV diffuse
emission as measured by HESS~\citep{2004A&A...425L..13A} (see
also~\cite{2013MNRAS.435L..14B} for interpretations of the TeV emission).

\medskip

Lately, \cite{Abazajian:2010zy} proposed a population of MSPs associated with
the bulge of the Galaxy as explanation for the extended excess emission of GeV
gamma-ray photons that has been found in observations of the inner Galaxy with
the \Fermi\ Large Area Telescope (LAT)~\citep{Goodenough:2009gk,
Vitale:2009hr}, dubbed the \Fermi\ GeV excess.  By now, numerous follow-up
studies by several independent groups~\citep{Hooper:2010mq, Hooper:2011ti,
Abazajian:2012pn, Gordon:2013vta, Macias:2013vya, Abazajian:2014fta,
Daylan:2014rsa, Zhou:2014lva, Calore:2014xka}, and lately also the LAT
collaboration~\citep{TheFermi-LAT:2015kwa}, have confirmed the existence of
this excess emission, which emerged above predictions from conventional
Galactic diffuse emission models.  

It is worth emphasizing that the word `excess' is here somewhat misleading and
potentially confusing.  In fact, \emph{none} of the Galactic diffuse emission
models that were used in the above analyses actually included any realistic
model for the gamma-ray emission of the Galactic bulge or center.  Significant
emission from the Galactic bulge hence necessarily shows up as `excess' above
the model predictions.  Since it is common in the literature, we will continue
to refer to this emission as \Fermi\ GeV excess, but note that a much more
appropriate and descriptive term would be `Galactic bulge emission'.

The \Fermi\ GeV excess shows specific spectral and spatial features (we follow
here the results from \cite{Calore:2014xka} and note that
\cite{TheFermi-LAT:2015kwa} come to similar results where the analyses
overlap).  The best fit to the energy spectrum is given by a broken power-law
($dN/dE\propto E^{-\alpha}$) with spectral indices
$\alpha(E<E_\text{b})=1.4^{+0.2}_{-0.3}$ and $\alpha(E>E_\text{b})=2.6\pm0.1$,
and break energy of $E_\text{b}=2.1\pm0.2\GeV$.  
However, also power-laws with an exponential cutoff fit the data well
when taking into
account the large systematic uncertainties related to the subtraction of
Galactic diffuse foregrounds.
\footnote{$dN/dE\propto E^{-\alpha}\exp[-E/E_\text{cut}]$, with cutoff energy of
$E_\text{cut}=2.5^{+1.1}_{-0.8}\GeV$ and a spectral index of
$\alpha=0.9^{+0.4}_{-0.5}$.}
This is in good agreement with
the stacked spectrum of gamma-ray MSPs as determined by \cite{McCann:2014dea}
(namely $E_\text{cut}=3.6\pm0.2\GeV$ and $\alpha=1.46\pm0.05$; see
\cite{Cholis:2014noa} for similar results).  Although the \Fermi\ GeV excess is
most clearly visible in the inner $5\deg$ of the Galactic center, indications
for an excess with a characteristic peak at around 2--3 GeV can be found up to
$15\deg$ above and below the Galactic plane~\citep{Daylan:2014rsa,
Calore:2014xka}.  The morphology of the excess is compatible with a spherical
symmetric volume emissivity that is strongly peaked towards the Galactic
center, and which follows a radial power-law of $d\mathcal{E}/dV\propto
r^{-\Gamma}$, with $\Gamma=2.56\pm 0.20$ in the inner $\sim15\deg$. 

The energy spectrum of the \Fermi\ GeV excess is indeed well in agreement with
\Fermi\ observations of Galactic field MSPs~\citep{Calore:2014nla}.  The combined
emission from thousands of MSPs, too dim to be resolved by the telescope as
individual objects, might produce the diffuse excess emission provided that the
density of sources steeply rises towards the Galactic
center~\citep{Abazajian:2010zy, Abazajian:2014fta,
Gordon:2013vta,Yuan:2014rca,Petrovic:2014xra}.  Such an extended, spherically
symmetric, spatial distribution could be generated as the debris from tidally
disrupted globular clusters~\citep{Brandt:2015ula}.  Also, secondary gamma-ray
emission can be produced from positron-electron pairs emitted by MSPs and
up-scattering low-energy ambient photons up to $\sim$ 100 GeV.  Such
emission could contribute to possible high-energy tails of the \Fermi\ GeV
excess~\citep{Petrovic:2014xra, Yuan:2014yda}.

Various other mechanisms have been proposed to account for or contribute to the
\Fermi\ GeV excess, and hence the gamma-ray emission from the Galactic bulge.
Interestingly, the properties of the observed emission are compatible with a
signal from the self-annihilation of dark matter particles in the dark matter
halo of the Galaxy, see \fex\ \cite{Calore:2014xka} and references therein.
Other astrophysical scenarios that were discussed are leptonic outbursts of the
supermassive black hole during an active past of the Galactic
center~\citep{Carlson:2014cwa, Petrovic:2014uda, Cholis:2015dea} and star
formation activity in the central molecular zone \citep{Gaggero:2014xla,
Carlson:2015ona}.
However, a generic feature of models that explain the excess with
inverse Compton emission of energetic leptons is that the excess spectrum
should vary with distance from the Galactic center, which is not observed in
the analysis of \cite{Calore:2014xka}. Also, the observed excess morphology can only be
accounted for with multiple finely tuned injection events (see \cite{Cholis:2015dea} for
details).

\medskip

Recently, \cite{Bartels:2015aea} and \cite{Lee:2015fea} found an enhanced
clustering of gamma-ray photons from the inner Galaxy, and showed that the most
likely cause is contributions from a population of sources just below the
detection threshold of \Fermi.  Furthermore, \cite{Bartels:2015aea} showed that
the inferred surface density and cutoff luminosity of the sub-threshold sources
is compatible with the expectations from a bulge population of MSPs that can
potentially account for 100\% of the emission associated with the \Fermi\ GeV
excess.  Significant contributions to the observed photon clustering from a
thick-disk population of MSPs, extragalactic or other Galactic sources were
ruled out, and un-modelled substructure in the gas emission seemed a rather
unlikely cause.  These results, together with the hard X-ray emission seen by
{\it NuSTAR}~\citep{2015Natur.520..646P}, make the case for a population of
MSPs at the Galactic center even stronger, and motivate additional
multi-wavelength observation strategies to probe the MSP interpretation of the
\Fermi\ GeV excess.

Lastly, it is worth mentioning that the stacked spectral energy distribution of
gamma-ray observed \emph{young} pulsars, $P\geq30$ ms and $B \geq10^{9}\G$, is also in agreement
with the spectral properties of the \Fermi\ GeV excess.  \cite{O'Leary:2015gfa}
argued that a population of young pulsars arising from star formation in the
inner Galaxy and the kinematical evolution in the Galactic potential can
account for most of the extended excess emission.  However, this scenario
does not account for the steep observed rise of the \Fermi\ GeV excess towards
the inner dozens of pc of the Galactic center~\citep[see,
\fex,][]{Daylan:2014rsa}, and it seems to lead to an oblate rather than a
spherical source distribution in the bulge.  In the present work, we will hence
assume that MSPs dominate the \Fermi\ GeV excess.
We note, however, the radio
  pulsation searches we investigate would also be at least equally
  sensitive to young pulsars, in addition to MSPs.

\medskip

Despite considerable efforts, MSP searches in the Galactic center region
have so far been unsuccessful up to now.  The main obstacles are the large
scatter-broadening of the pulsed signal along the line-of-sight towards the
inner Galaxy as well as the large distance to the sources.  This prevents the
detection of the pulsed radio emission in many cases~\citep{Stovall:2013gca},
because MSPs are in general weak radio sources (with flux densities in the
range $\muJy$ to $\mJy$).  The only MSPs observed in the inner 3 kpc 
($\sim$ 20 degrees at a distance of 8.5 kpc away)
are MSPs associated with the globular clusters M62, NGC 6440 and NGC 6522, and were
found in dedicated deep observations of these targets.  

Finding the bulge source population, at mid Galactic latitudes, with multi-wavelength
observations is certainly challenging. However, this possibility has never been
systematically explored.   Previous large radio surveys were shown to be
insensitive to MSPs at the Galactic center~\citep{2015ApJ...805..172M}.
Moreover, those same surveys were focused on the very inner few degrees about
the Galactic center, while, supported by the diffuse gamma-ray emission, we
expect the bulge MSP population to extend to latitudes of about $\pm 15^{\circ}$.

\bigskip

In this paper, we analyze the prospects for the detection of a bulge MSP
population (as suggested
  by the Fermi GeV excess) via searches for radio pulsations.  One of the most detailed descriptions of the \Fermi\ GeV excess at
$|b|>2^\circ$ latitudes was presented by \cite{Calore:2014xka}, and we will
base our modeling on these results.  We discuss various radio survey strategies
that could unveil the bulge MSP population with existing and future
instruments.  To this end, we will use observations of globular clusters as
well as high-latitude gamma-ray MSPs and unassociated \Fermi\ sources to
calibrate our predictions.  

\medskip

The paper is organized as follows: In Sec.~\ref{sec:model}, we describe the
modeling we adopt for the bulge MSP population, as motivated by the observation
of the GeV excess, and its radio luminosity function. In Sec.~\ref{sec:sens},
we estimate the sensitivity of current and future radio instruments to MSP
detection.  We present our results for large area radio surveys in
Sec.~\ref{sec:surveys}. In Sec.~\ref{sec:targeted}, we study the possibility to
detect the bulge sources in deep targeted observations, by exploiting an
observed loose correlation between gamma-ray and radio fluxes.  We discuss
various additional aspects and caveats of our results in
Sec.~\ref{sec:discussions}, where we also briefly comment on the possibility to
use X-rays to probe the bulge MSP population.  We conclude in
Sec.~\ref{sec:conclusions}.  

In the Appendix we furthermore investigate the MSP-candidates identified
by~\cite{Bartels:2015aea} as significant wavelet peaks in gamma-ray data from
the inner Galaxy.  In particular, we look for a possible correlation of
wavelets peaks with foreground sources, \ie~MSPs or young pulsars along the
line-of-sight but closer to us than bulge MSPs.  Finally, we provide a
multi-wavelength analysis of the 13 MSP candidates from~\cite{Bartels:2015aea}.

\section{Modeling the bulge MSP population}
\label{sec:model}
We start by constructing a phenomenological model for the radio emission
properties of the bulge MSP population as a whole.  The aim is to obtain a
reliable estimate for the surface density of \emph{radio-bright} MSPs in the
Galactic bulge.  To this end, we define as \emph{radio-bright} any MSP that has
a period-averaged flux density of at least $10\muJy$ at 1.4 GHz.  This is rather low compared to values
that are conventionally used in the literature, but will turn out to be
appropriate for the discussion in this work and is motivated by the
sensitivities of currently available radio telescopes.

We assume that bulge MSPs are responsible for the dominant part of the \Fermi\
GeV excess (hence the dominant part of the Galactic bulge emission), and we
will below adopt a spatial distribution that is consistent with \Fermi-LAT
observations.  We adopt here a phenomenological approach to the problem: We do
not pretend to model fully the dynamics and evolution of the Galactic bulge, but
we assume the spatial distribution required to explain the \Fermi~GeV
excess data.  Once the spatial distribution is fixed, however, estimating the
number of radio-bright MSPs in the bulge from diffuse gamma-ray observations is
rather challenging at first sight.  One would expect that it requires accurate
information about both the gamma-ray and radio luminosity functions and a
detailed understanding of beaming effects.  However, the discussion greatly
simplifies for the specific goals of this paper, as we shall see next.

\medskip

In most of the current paper we are interested in the \emph{combined} gamma-ray
emission of many bulge MSPs (averaged over regions of, say, $1\deg^2$).
This is what we can actually most readily determine with \Fermi-LAT
observations, in contrast to the much harder to detect gamma-ray emission of
individual bulge sources.  Details of the gamma-ray luminosity function, and
the potential correlation of gamma-ray with radio emission on a
source-by-source basis, are not directly relevant when studying the average
emission properties of MSPs in the Galactic bulge.  They will only become
relevant when discussing targeted observations in Sec.~\ref{sec:targeted}
below.

For our predictions, we need for a given random sample of $N_\text{tot}$ MSPs
at the distance of the Galactic bulge: 
\begin{itemize}
\item[(A)] An estimate for the number of
radio-bright MSPs in that population, $N_\text{rb}$.
\item[(B)] An estimate for
their combined gamma-ray emission, $L_\gamma$.  
\end{itemize}
Since our predictions for the
number of radio-bright MSPs in the bulge will only depend on the ratio
$N_\text{rb}/L_\gamma$, the total number $N_\text{tot}$ will drop out.

\medskip

The predictions in this paper rely on two critical assumptions.

1. We will assume that both, the population of bulge MSPs and of MSPs bound in
globular clusters, have similar gamma-ray and radio emission properties.  This is justified by the fact
  that -- while the formation of MSPs in globular clusters versus the
  field may in some cases follow different paths -- the fundamental
  physical processes creating the observed radio pulsations should in
  all cases be the same.  At the same time, globular cluster and field
  MSPs don't obviously have different age or luminosity
  distributions~\citep{2010MNRAS.409..259K}.  Thus, we can use the gamma-ray emission from
globular clusters as well as the radio observations of MSPs in globular
clusters as a proxy for the population of bulge MSPs.

2. We assume that \emph{all} of the gamma-ray emission from globular clusters
comes from MSPs.  If only a fraction $f_\text{MSP}$ of the gamma-ray emission
came from MSPs, this would simply \emph{increase} the number of radio-bright
MSPs in the bulge by a factor of $\propto f_\text{MSP}^{-1}$ with respect to
our predictions below. Therefore, this is a conservative assumption.

\subsection{Lessons from MSPs in globular clusters}

\begin{table*}
  \centering
  \begin{tabular}{lrrrrrrr}
    \hline\hline
    Globular cluster & $\ell$ [deg] & $b$ [deg] & $d$ [kpc] & $L_\gamma \rm\ [10^{34}\ergs]$
       & $N_\text{obs}$ & $N_{\rm rad}$ \\[1pt]
    \hline
    Ter 5    & $3.8$ & $1.7$ & 5.5 & $26.5\pm9.0$ & 25 & $82\pm16$ \\
    47 Tuc   & $305.9$ & $-44.9$ & 4.0 & $5.1\pm1.1$ & 14 & $37\pm10$ \\
    M 28 & $7.8$ & $-5.6$ & 5.7 & $6.4\pm2.0$ & 9 &$63\pm21$ \\
    NGC 6440 & $7.7$ & $3.8$ & 8.5 & $35.4\pm8.0$ & 6 & $48\pm21$ \\
    NGC 6752 & $336.5$ & $-25.6$ & 4.4 & $1.3\pm0.7$ & 5 & $21\pm10$ \\
    M 5 & $3.9$ & $46.8$ & 7.8 & $2.4\pm0.5$ & 5 & $13\pm6$ \\\hline
    Stacked &  &  &  & $77.1\pm 12.3$ & 64 & $264\pm 37$ \\    
    \hline\hline
  \end{tabular}
  \caption{List of globular clusters that we use as a proxy for the population of
    bulge MSPs. We show name, Galactic longitude and latitude, distance,
    gamma-ray luminosity \citep{TheFermi-LAT:2015hja,Zhou:2015cta}, the number
    of observed radio MSPs relevant to this work ($N_\text{obs}$), and the
    estimated \emph{total} number of radio MSPs ($N_\text{rad}$), based on our
    reference radio luminosity function.  Furthermore, in the last row, we show
    the \emph{stacked} gamma-ray luminosity ($L_\gamma^\text{stacked}$) and the
    estimated number of radio MSPs ($N_\text{rad}^\text{stacked}$). If not
    otherwise stated, parameters are taken from \cite{Bagchi:2011hs}, Model 3.
    Note that $N_\text{obs}$ refers to the number of observed MSPs with quoted
    flux density that were used in \cite{Bagchi:2011hs} to infer the radio
    luminosity function.}
  \label{tab:GCs}
\end{table*}

\begin{table}
  \centering
  \begin{tabular}{lrr}
    \hline\hline Luminosity function ($\mu$, $\sigma$)
    &  $N_{\rm rad}^\text{stacked}$ & $N_{\rm rb}^\text{stacked}(d\simeq8.5\kpc)$ \\[1pt]
    \hline
    Model 1 ($-1.1$, $0.9$)  & $514\pm 71$ & $74 \pm 10 $ \\
    Model 2 ($-0.61$, $0.65$)  &$339\pm 49$  & $80 \pm 12 $\\
    \textit{Model 3} ($-0.52$, $0.68$)  &$264\pm 37$  & $76 \pm 11$ \\
    \hline\hline
  \end{tabular}
  \caption{Estimated total number of radio MSPs ($N_\text{rad}^\text{stacked}$)
    and of radio-bright MSPs ($N_\text{rb}^\text{stacked}$) in the stacked
    globular clusters from Tab.~\ref{tab:GCs}, as inferred from the observed
    MSPs using three different luminosity functions \citep[][their models
    1--3]{Bagchi:2011hs}.  The reference luminosity function used in most of
    this paper is \emph{Model 3}.  We assume the MSPs are at a distance of
    $8.5\kpc$ (\ie~at the Galactic center) in order to determine whether they
    are radio-bright.  We find that, while the estimated \emph{total} number of
    radio MSPs in the stacked globular clusters depends on the rather uncertain
    low-luminosity tail of the radio luminosity function, the estimated number
    of MSPs that we would qualify as radio-bright remains consistent within the
    error bars, for all the three models.} 
  \label{tab:RadioBright}
\end{table}

To estimate the number 
of radio-bright sources expected from a population of MSPs located at the GC (A), 
we will use the radio luminosity function of detected globular clusters~\citep{Bagchi:2011hs}
and we rescale it to a distance of 8.5 kpc.
We will assume their combined gamma-ray luminosity (B) by stacking the measured \Fermi\ gamma-ray 
fluxes of the globular clusters in our sample. 
We will use the ratio between the stacked gamma-ray emission from globular
clusters and the expected number of radio-bright MSPs (at 8.5 kpc)
as a proxy for the relationship between the mean gamma-ray luminosity
and the mean number of radio-bright MSPs
in the Galactic bulge (see details below). In this way, we will be able to
get a robust estimate for the number of radio-bright MSPs in the Galactic bulge.

In Tab.~\ref{tab:GCs}, we list the globular clusters that we take into account
in the present discussion.  This is the subset of objects considered in
\cite{Bagchi:2011hs} for which gamma-ray measurements exist.  The number of
detected radio MSPs in the globular clusters in Tab.~\ref{tab:GCs} is
relatively large, ranging from 5 sources in NGC 6752 and M 5 to 25 sources in
Terzan 5~\citep[note that actually 33 MSPs, with $P < 30$ ms, are known in
Terzan 5, but only 25 were taken into account in the study of][]
{Bagchi:2011hs}.  We note that Terzan 5 and NGC 6440 are the most luminous
gamma-ray emitters, and we discuss their role for our results below.

\bigskip

The total number of radio MSPs, $N_\text{rad}$, in each globular cluster can be
estimated by a fit of a given radio luminosity function (with free
normalization but fixed shape) to the globular cluster MSPs that are
individually detected in radio.  The radio luminosity function of globular
cluster MSPs was studied in great detail by~\cite{Bagchi:2011hs}, using Monte
Carlo techniques that account for the finite observation depths.\footnote{Note
that although the study formally takes into account all pulsars in globular
clusters, the sample that they use is completely dominated by MSPs.}  They
found that the cumulative radio luminosity function of MSPs in globular
clusters is similar to the luminosity function of young and recycled pulsars in
the disk as derived by \cite{2006ApJ...643..332F}.
  
We will here adopt the best-fit model from \cite{Bagchi:2011hs} (their `Model
3') as a reference for the radio luminosity function. In
Sec.~\ref{sec:discussions}, we will comment on how our results depend on that
choice.  The luminosity function follows parametrically a log-normal
distribution,
\begin{equation} 
  f(L_\nu) = \frac{\log_{10} e}{L_\nu} \frac{1}{\sqrt{2\pi\sigma^2} }
  \exp\left[ \frac{-(\log_{10} L_\nu-\mu)^2}{2\sigma^2}\right]\;,
  \label{eqn:lognorm}
\end{equation}
with mean $\mu=-0.52$ and variance $\sigma=0.68$, and $L_\nu$ refers to the
`pseudo-luminosity' at $\nu=1.4\GHz$ $(\rm mJy\,kpc^2)$.  The pseudo-luminosity
is related to the measured flux density $S_\nu$ of a source by $L_\nu=S_\nu
d^2$, where $d$ denotes the distance to the source.  It is used
  because the beaming angle of the radio emission is unknown.

\begin{figure}
  \begin{center}
    \includegraphics[width=\linewidth]{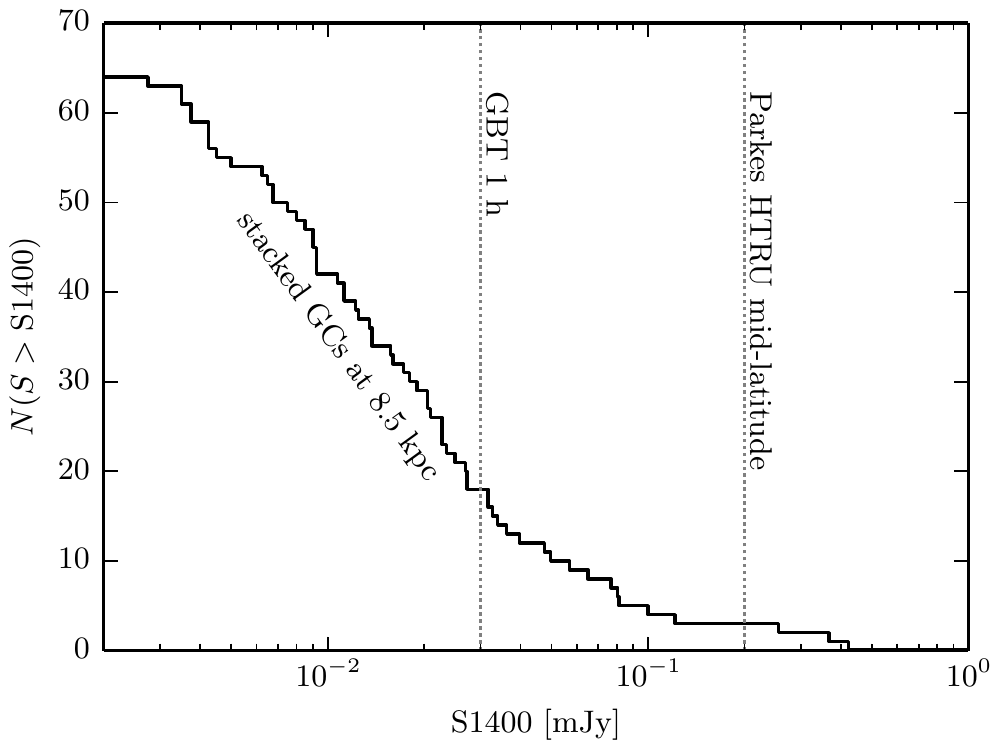}
  \end{center}
  \caption{Complementary cumulative distribution of flux densities at 1.4 GHz,
    $\rm S1400$, of the 64 pulsars in the globular clusters that are listed in
    Tab.~\ref{tab:GCs}, rescaled to a distance of $8.5\kpc$.  We show for
    comparison the limiting flux density, $\sim 0.2$ mJy, of the Parkes High Time Resolution
    Universe (HTRU) mid-latitude survey~\citep{2010MNRAS.409..619K} as well as
    the reference GBT survey, 0.03 mJy (discussed in Sec.~\ref{sec:sens}).  The
    plot illustrates that a survey that is significantly deeper than that with
    Parkes would start probing the radio luminosity function in a regime that
    is well supported by data.  Predictions for \emph{radio-bright} bulge MSPs
    ($\rm S1400\geq10\muJy$) are built upon 43 measured globular cluster MSPs.}
  \label{fig:fluxHist}
\end{figure}

Note that the above radio luminosity function has a high-luminosity tail that
predicts sources brighter than the brightest MSPs detected so far in globular
clusters (where the distance is relatively well known).  In order not
to unrealistically bias our prediction towards excessively bright sources, we
truncate the radio luminosity function to a maximum pseudo-luminosity of
$30\mJy \kpc^2$. Such a pseudo-luminosity corresponds to the maximum flux
density, 0.4 mJy, observed in stacked globular clusters rescaled to a distance
of 8.5 kpc, cf.~Fig.~\ref{fig:fluxHist}.\footnote{We also point out that there are only a 
handful of MSPs in the ATNF catalog that have pseudo-luminosity at 1.4GHz 
higher than $30\mJy \kpc^2$. These have pseudo-luminosities of about
$50 - 60 \mJy \kpc^2$,  with one exceptional source at B1820-30A at 100 $\mJy \kpc^2$.}

\medskip

Based on the radio luminosity function in Eq.~\eqref{eqn:lognorm}, the number
of radio MSPs in each globular cluster was inferred by~\cite{Bagchi:2011hs}.
The results, together with $1\sigma$ error bars from the fits, are listed in
Tab.~\ref{tab:GCs}.  In this table, we also show the total number of radio MSPs
in all considered globular clusters combined.  It is
$N_\text{rad}^\text{stacked} = 264\pm37$ (with errors summed in quadrature).
We note that the \emph{total} number of MSPs in the globular clusters is
definitively larger, since not all MSPs are expected to have a radio beam
pointing towards the Earth (although the beams
  are arguably wide in the case of MSPs); this,
however, is not relevant for our discussion. 

It is reassuring that, for a bulge population of MSPs, measuring flux densities
below 0.1 mJy (at 1.4 GHz) is enough to start probing the parts of the radio
luminosity function that are directly supported by observations (rather than by
an extrapolation beyond the brightest observed MSP).  To illustrate this point,
we rescale the flux densities of MSPs observed in the globular clusters from
Tab.~\ref{tab:GCs} to the distance of the Galactic center, for which we here
adopt $8.5\kpc$~\citep[consistent with][]{Gillessen:2008qv}.  We show the
resulting complementary cumulative distribution function of these flux
densities in Fig.~\ref{fig:fluxHist}.  In this figure, we also indicate for
comparison the maximum sensitivity of our reference Parkes and GBT observations
from Tab.~\ref{tab:telescopes}, which we will discuss in detail below.  

Lastly, in Tab.~\ref{tab:RadioBright}, we indicate the number of
\emph{radio-bright} MSPs in the stacked globular clusters, assuming that they
are at a distance of $8.5~\kpc$.\footnote{ We note that the number of
radio-bright sources in Fig.~\ref{fig:fluxHist}, which is based on various
flux-limited samples that were used in the analysis by \cite{Bagchi:2011hs}, is
as expected somewhat smaller than the corresponding values quoted in
Tab.~\ref{tab:RadioBright} that were obtained from the inferred luminosity
functions.} To this end, we use our above reference luminosity function
normalized to the number of radio pulsars as indicated in Tab.~\ref{tab:GCs},
but we also show results for the two other luminosity functions
from~\cite{Bagchi:2011hs} which reasonably bracket the uncertainties implied by
the observed MSPs (see their Fig.~3).  We find that, although the \emph{total}
number of radio MSPs (which is just obtained by integrating the appropriately
normalized radio luminosity function to the lowest luminosities) is uncertain
by at least a factor of a few, the number of \emph{radio-bright} MSPs is much
better constrained, since it has direct observational support.  Indeed, this is
also apparent from Fig.~\ref{fig:fluxHist} above.

\bigskip

The total gamma-ray luminosity from all considered globular clusters combined
is $L_\gamma^\text{stacked} = (7.71\pm1.23)\times10^{35}\ergs$, where the error
refers to \Fermi\ flux measurement errors that are added in quadrature.  The
stacked luminosity is dominated by Terzan 5 and NGC 6440, and we refer to
Sec.~\ref{sec:discussions} for further discussions about the effect of
individual globular clusters on our results.  Following
\cite{TheFermi-LAT:2013ssa}, we define gamma-ray luminosity as $L_\gamma=4\pi
d^2 G_{100}$, where $G_{100}$ is referring to the energy flux measured by
\Fermi-LAT above $100\MeV$.

Gamma-ray luminosity functions have in general very non-Gaussian tails, and one
might worry that the sample variance of the combined gamma-ray emission of the
six globular clusters is excessively large.  We estimate the sample variance of
this summed gamma-ray luminosity in a simple toy scenario.  To this end, and
\emph{only} for the purpose estimating the variance, we assume that the summed
gamma-ray emission of the globular clusters is caused by about 250 MSPs that
are randomly drawn from a power-law gamma-ray luminosity function with hard
lower and upper cutoffs at $10^{32}\ergs$ and $10^{35}\ergs$, respectively.
The upper cutoff is selected to be compatible with the brightest observed MSPs,
the lower cutoff is adjusted such that 250 sources yield the combined total
luminosity.  The index of the luminosity function is fixed to $-1.5$ \citep[see
discussions in][]{Strong:2006hf, Venter:2014zea, Petrovic:2014xra,
Cholis:2014noa}.  We find a mean total luminosity of $7.9\times10^{35}\ergs$,
comparable to the above value for $L_\gamma^\text{stacked}$, and the standard
deviation of the total luminosity over many samples is $1.5\times10^{35}\ergs$.
This implies that $L_\gamma^\text{stacked}$ can be considered as a reasonable
estimate for the population averaged gamma-ray luminosity, with a sample
variance uncertainty of about $20\%$.  Indeed, this is larger than the $6\%$
that would be expected from shot noise alone for a population with an average
number of 250 sources.  We will adopt the $20\%$ here as estimate for the
sample variance,  but we stress that the precise value depends on the not
well-constrained details of the gamma-ray luminosity function at high
luminosities.

\bigskip

We now calculate the ratio between the overall gamma-ray emission from globular
clusters and the number of radio-bright MSPs (assuming $8.5\kpc$ distance),
taking into account uncertainties in the number of total radio MSPs, \Fermi\
flux measurements and sample variance.  We will subsequently assume that this
ratio provides the relationship between the mean gamma-ray luminosity $\langle
L_\gamma^\text{bulge} \rangle$ and the mean number of radio-bright MSPs
$\langle N_\text{rb}^\text{bulge} \rangle$ in the Galactic bulge.  It is given
by
\begin{equation}
  \mathcal{R}_\text{rb}^\gamma  \equiv \frac{\langle{L_\gamma^\text{bulge}}\rangle}
  {\langle N_\text{rb}^\text{bulge}\rangle }
  \simeq \frac{L_\gamma^\text{stacked}} {N_\text{rb}^\text{stacked}}
  = (1.0\pm0.3)\times10^{34}\ergs\;.
  \label{eqn:R}
\end{equation}
We emphasize that the value of $\mathcal{R}_\text{rb}^\gamma$ does \emph{not}
provide a robust estimate for the average gamma-ray luminosity of radio-bright
MSPs, since not every gamma-ray emitting MSP must be bright in radio or vice
versa.  But it provides a reasonable relation between the overall gamma-ray
luminosity of a large population of MSPs and the number of radio-bright sources
in that same population at Galactic center distances.

The errors that we quote for $\mathcal{R}_\text{rb}^\gamma$ do not directly
take account the effect of varying the radio luminosity function.  However, as
we discussed above, and showed in Tab.~\ref{tab:RadioBright}, the systematic
uncertainties related to the adopted luminosity function are smaller than the
statistical error from fitting the luminosity function to the globular cluster
observations.  Given this, and the various other uncertainties that enter the
estimate in \eqref{eqn:R}, these variations can be neglected.

\bigskip

As we will see, the spin period is critical for the detectability of MSPs.  The
analysis of the spin period distribution of field MSPs by
\cite{Lorimer:2015iga} finds a modified log-normal distribution.  The mean is
$P_\text{mean}\simeq 5.3 \ms$ and hence in good agreement with the mean of the
observed periods of MSPs in globular clusters
($P_\text{mean}\simeq5.7\ms$)~\citep{2010MNRAS.409..259K}.  We will use here
the results from \cite{Lorimer:2015iga} as reference.

\subsection{Predicted radio-bright MSPs in the Galactic bulge}

Following the results of the gamma-ray analysis by~\cite{Calore:2014xka}, we
assume that the density of field MSPs in the Galactic bulge follows an inverse
power-law as function of the Galacto-centric distance $r$, with an index of
$\Gamma=2.56$.  For definiteness, we adopt a hard cutoff at $r=3\rm\, kpc$,
which is not critical for our results.  We fix the normalization of the
combined (and population averaged) gamma-ray intensity of this bulge population
in the pivot direction $(\ell,b)=(0^\circ,\pm5^\circ)$.  In this direction, and
for a reference energy of $E_\gamma=2\rm\, GeV$, the differential intensity of
the proposed bulge MSP population is given by
$\Phi=(8.5\pm0.7)\times10^{-7}\rm\, GeV^{-1}cm^{-2}s^{-1}sr^{-1}$
\citep{Calore:2014xka}.  We remark that the quoted gamma-ray intensity is not
the \emph{total} intensity of the excess emission (which is to some degree
  ill-defined, given the large uncertainties in the Galactic diffuse
foregrounds), but the fraction that can be reasonably attributed to MSP-like
spectra after accounting for foreground subtraction systematics~\citep[for
details see][]{Calore:2014xka}.

We assume that the energy spectrum of the combined gamma-ray emission of bulge
MSPs follows the stacked MSP spectrum inferred by \cite{McCann:2014dea} from 39
nearby sources.  As mentioned in the introduction, this spectrum is in good
agreement with the spectrum of the \Fermi\ GeV excess as derived by
\cite{Calore:2014xka}.  The above differential intensity at 2 GeV corresponds
then to an energy intensity (above 100 MeV) of $(5.5\pm 0.5)\times
10^{-12}\rm\,erg\, cm^{-2}\,s^{-1}\,deg^{-2}$.  Using the ratio
$\mathcal{R}_\text{rb}^\gamma$ as estimated in the previous subsection, this
implies a surface density of radio-bright bulge MSPs at 5 deg above and below
the Galactic center of around $(4.7\pm1.5)\deg^{-2}$.

\medskip

With the above assumptions, we find a total gamma-ray luminosity of the MSP
bulge population of
\begin{equation}
  L_\gamma^{\rm bulge} =  (2.7\pm0.2) \times10^{37}\ergs\;. 
\end{equation}
We note that variations of the spatial index $\Gamma$ by $\pm 0.2$, which is
the $1\sigma$ range found in \cite{Calore:2014xka}, would affect the total
gamma-ray luminosity by up to $40\%$.  However, we do not propagate this
additional uncertainty through the analysis, because most of our conclusions
will depend on the emission around the above-mentioned pivot directions, which
makes them relatively independent on the exact value of $\Gamma$.

Using the ratio $\mathcal{R}_\text{rb}^\gamma$ as estimated in the previous
subsection, we obtain \emph{an estimate for the number of radio-bright MSPs in
the Galactic bulge,}
\begin{equation}
  N_{\rm rb}^\text{bulge} = (2.7\pm0.9)\times10^{3}\;.
\end{equation}
As discussed above in context of Tab.~\ref{tab:RadioBright}, the number of
radio-bright sources is relatively weakly dependent on the adopted radio
luminosity function.  However, when simulating sources in the Galactic bulge,
we actually need the number of all radio MSPs.  We will in the remaining part
of the paper adopt `Model 3', for which we find a total number of radio MSPs of
$N_{\rm rad}^\text{bulge} = (9.2\pm3.1)\times10^3$.
About 1/3 of the radio MSPs
are thus radio-bright, \ie~$\geq 10 \, \mu$Jy.

\subsection{Comparison with the MSP thick-disk population}

\begin{figure}
  \centering
    \includegraphics[width=0.85\columnwidth]{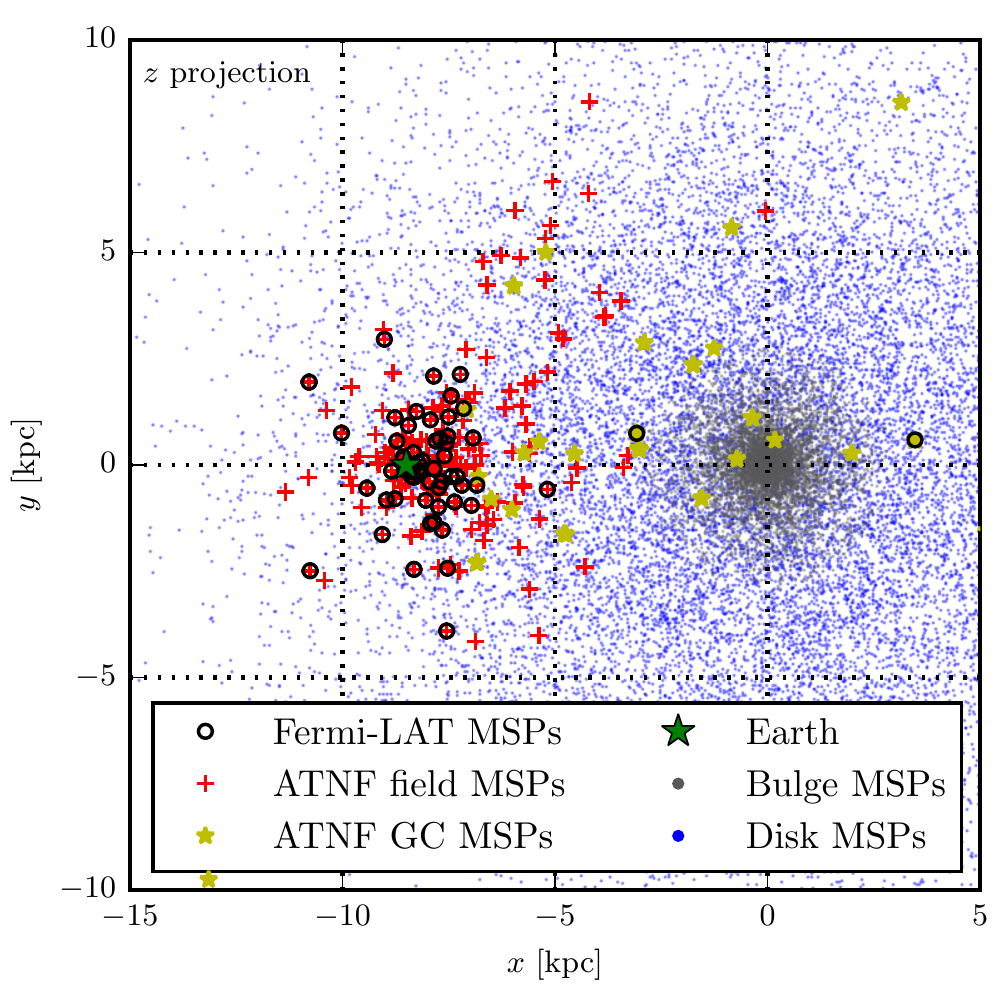}
    \includegraphics[width=0.85\columnwidth]{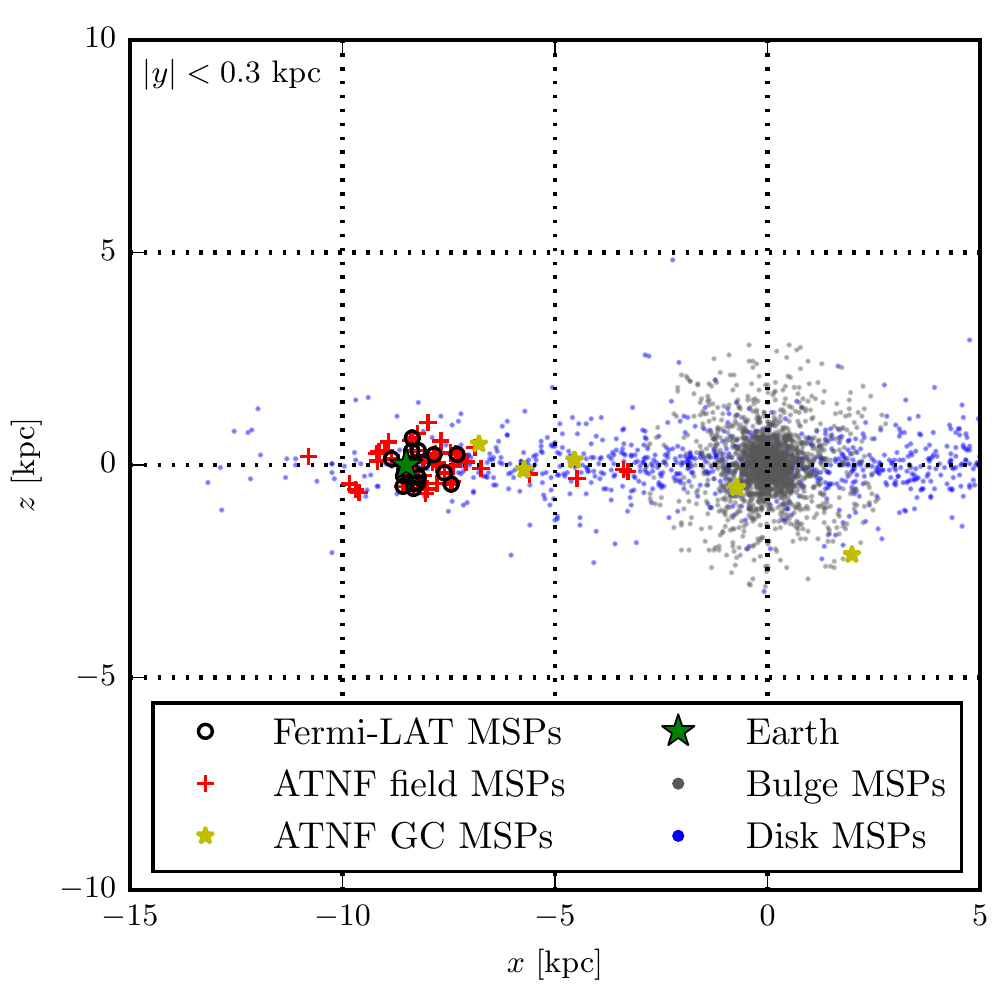}
    \caption{Predicted spatial distribution of MSPs in the bulge (\emph{grey
      dots}) and the disk (\emph{blue dots}), modeled based on gamma-ray and
      radio data as we describe in the text.  For comparison, we also show the
      position of measured radio pulsars with $P<30\ms$ from the Australia
      Telescope National Facility (ATNF) catalog, both sources in the field (\emph{red
      crosses}) and MSPs in globular clusters (\emph{yellow stars}).  We also
      show gamma-ray detected field MSPs (\emph{black circles}). Distance
      estimates for these sources are based on the NE2001
      model~\citep{Cordes:2002wz}, except for globular clusters were distances
      are better known and taken from the ATNF.
      We show projections both in the x--y
      (\emph{upper panel}) and the x--z plane (\emph{lower panel}), and mark
      the position of the Earth (in our convention at z=y=0 and x=-8.5 kpc).
      In the lower panel, we only show a thin slice with $|y|<0.3\kpc$ in order
      to better visualize the increased source densities in the inner Galaxy.}
    \label{fig:population}
\end{figure}

\begin{figure}
  \centering
    \includegraphics[width=0.90\columnwidth]{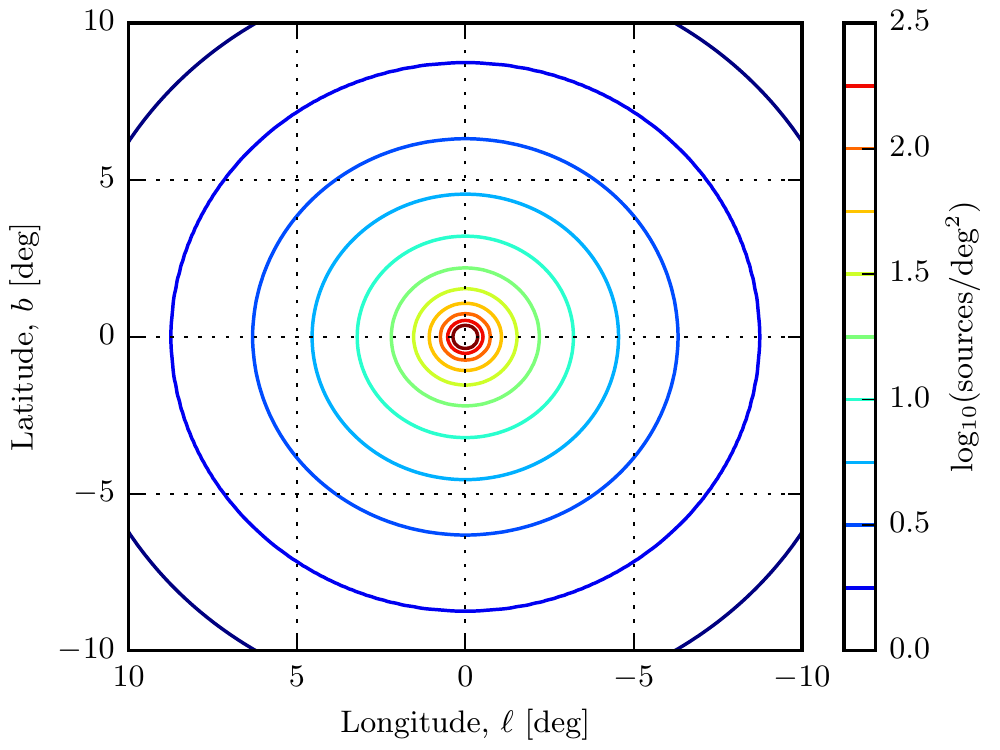}
    \caption{Surface density of radio-bright (\ie~$\geq 10 \, \mu$Jy) bulge 
    MSPs towards the inner
    Galaxy, per $\deg^2$. Beyond an angular distance of $5^\circ$ from the
    Galactic center, the density drops well below $\sim5\deg^{-2}$.}
    \label{fig:density}
\end{figure}

We illustrate the putative bulge population of radio MSPs in
Fig.~\ref{fig:population}.  There, we show the distribution of bulge radio MSPs
in Galacto-centric Cartesian coordinates, both in x--z and x--y projection, and
compare it with the actually observed MSPs and with a thick-disk MSP
population~\citep{2010JCAP...01..005F}.  We assume that the population of
thick-disk MSPs has a cylindrical symmetry with an exponential distribution,
and with a scale radius of 5 kpc~\citep{2010JCAP...01..005F} and a scale height
of 0.5 kpc~\citep{Calore:2014oga,Lorimer:2015iga}.  Following
\cite{2013MNRAS.434.1387L}, we attribute 20000 radio MSPs to the disk.  We note
that in this way we will somewhat over-predict the number of pulsars detectable
with the Parkes HTRU (as discussed below in Sec.~\ref{sec:discussions}).  This
is, however, not critical for our results, since having a smaller number of
thick-disk sources would make the bulge component even more pronounced.

Analogously to the bulge MSP population, the radio luminosity function of disk
MSPs is modeled according to our reference radio luminosity function.  From
Fig.~\ref{fig:population} it is very clear that the observed spatial
distribution of \emph{known} MSPs is almost exclusively driven by selection
effects that limit the maximum distance to which they can be found, and should
obviously not be used as a proxy for the real distribution of MSPs in the
Galaxy.  

Lastly, the implied \emph{surface density} of radio-bright bulge MSPs is shown
in Fig.~\ref{fig:density}.  At $(\ell, b)=(0^\circ,\pm5^\circ)$ it is
consistent with our above simple estimate (although we now take into account
the varying distance to the bulge sources that can be slightly closer or
further away than 8.5 kpc depending on their position).  Otherwise, it ranges
from $>300$ sources $\deg^{-2}$ around the Galactic center to just a hand full
of sources $\deg^{-2}$ a few degrees away from the Galactic center.

\section{Sensitivity of radio telescopes}
\label{sec:sens}

Here, we summarize briefly how we estimate the sensitivity of radio pulsation
searches. 

\subsection{Radiometer equation}

\begin{figure}
  \centering
  \includegraphics[width=\columnwidth]{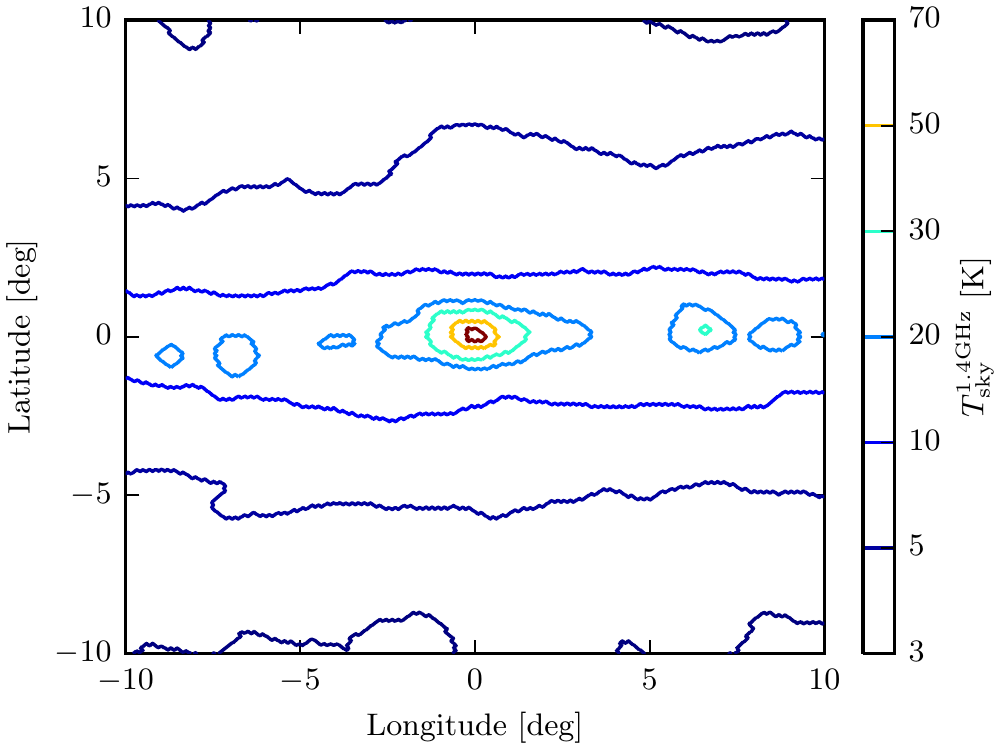}
  \caption{Sky temperature contours at 1.4 GHz, $T_\text{sky}^{1.4\GHz} (K)$,
    as derived from the Haslam 408 MHz radio maps~\citep{1982A&AS...47....1H}.
    The strong emission in the Galactic disk and Galactic center increases the
    background noise for MSP searches in these regions by a factor of a few.
    Note that point sources are not removed and affect our results close to the
    Galactic center.}
  \label{fig:Tsky}
\end{figure}

\begin{figure}
  \centering
  \includegraphics[width=0.85\columnwidth]{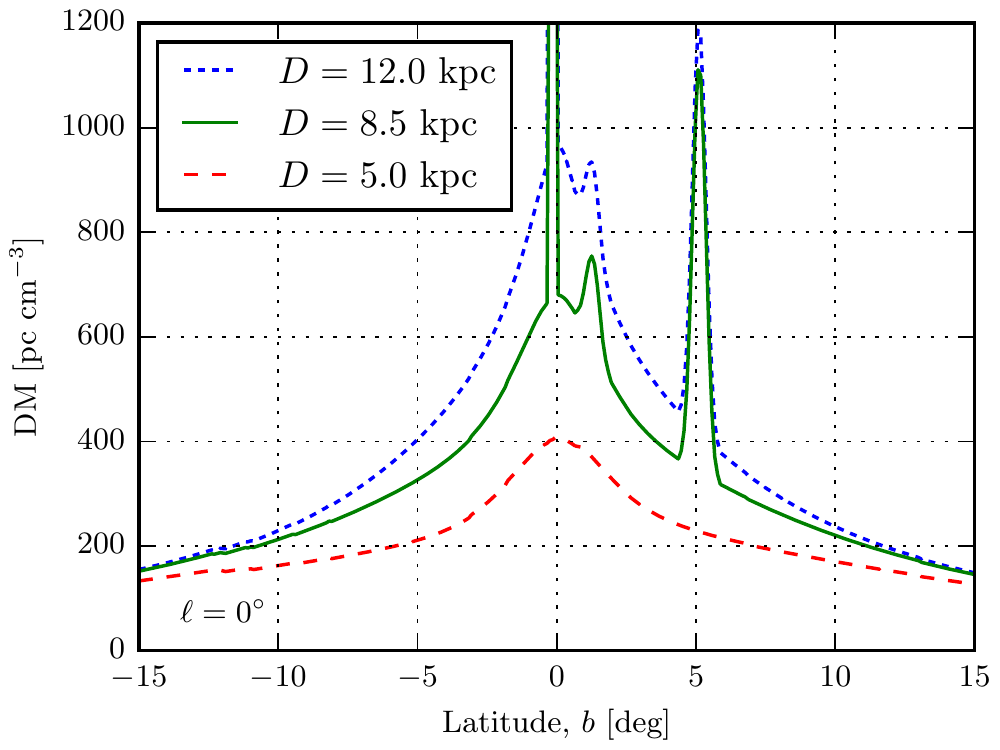}
  \caption{Latitude dependence of dispersion measure, DM, as derived from the
    NE2001 model \citep{Cordes:2002wz}, at zero Galactic longitude,
    $\ell=0^\circ$, for different line-of-sight distances between $D=5\kpc$ and
    $D=12.0\kpc$.  At longitudes in the range $\ell = [-10^\circ, 10^\circ]$
    the values typically differ by $<10\%$. The spikes in the otherwise smooth 
    curves correspond to discrete ``clumps" of enhanced free electron density that are 
    included in the NE2001 model (see Tables 5 -- 7 in~\cite{Cordes:2002wz}).}
  \label{fig:DMprofile}
\end{figure}

From the radiometer equation~\citep[see \fex][]{1984bens.work..234D}, the RMS
uncertainty of the flux density (in mJy) is given by
\begin{equation} 
  S_{\nu, \rm rms} = \frac{T_{\rm sys}}{G \, \sqrt{t_{\rm obs} \,  \Delta
    \nu \, n_p}} \left(\frac{W_{\rm obs}}{P-W_{\rm obs}}
\right)^{1/2} \,, 
\label{eq:Snurms}
\end{equation}
where $T_{\rm sys} = T_{\rm sky} + T_{\rm rx}$ is the system temperature (K)
given by the sum of sky and receiver temperatures, $G$ is the telescope gain
(K/Jy), $n_p$ is the number of polarizations, $\Delta \nu$ is the frequency
bandwidth (MHz), and $t_{\rm obs}$ is the integration time (s).  The sky
temperature is a function of Galactic longitude and latitude. For any given
line-of-sight we compute the corresponding sky temperature from the Haslam 408
MHz all-sky radio maps~\citep{1982A&AS...47....1H}, assuming a power-law
rescaling to the frequency of interest with index
$-2.6$~\citep{1987MNRAS.225..307L}.  In Fig.~\ref{fig:Tsky}, we show the
contours of constant $T_{\rm sky}$ for a $20^{\circ} \times 20^{\circ}$ region
around the Galactic center at 1.4 GHz.
As for the gain, the sensitivity calculations here assume an effective estimate
that accounts for the fact that the gain decreases by a factor of
two towards the FWHM edge of the telescope beam.  This effect should be taken
into account when planning actual surveys.

A reliable, blind pulsar detection requires a signal flux density 
$S_{\nu} \geq 10 \times
S_{\nu, \rm rms}$. In order to detect the pulsations, the observed (or
effective) pulse width, $W_{\rm obs}$ (ms), should be small with respect to the
source period, $P$ (ms). 
The observed pulse width can be estimated as ({\it e.g.}~\cite{Hessels:2007pq}):
\begin{equation}
  W_{\rm obs}\!=\! \sqrt{(w_{\rm int} P)^2 + \tau^2_{\rm DM} + \tau^2_{\rm scatt} +  \tau^2_{\rm samp}
    + \tau_{\Delta \rm DM}^2} \, , 
\label{eq:wobs}
\end{equation}
where $w_{\rm int} \sim 0.1$ is the intrinsic fractional pulse width typical
for MSPs, $\tau_{\rm DM}$ is the dispersive smearing across an individual
frequency channel that depends on the dispersion measure (DM) of the source,
$\tau_{\rm scatt}$ is the temporal smearing due to multi-path propagation from
scattering in a non-uniform and ionized interstellar medium, $\tau_{\rm samp}$
corresponds to the data sampling interval, and $\tau_{\Delta\rm DM}$ is the
smearing due to finite DM step size in the search.  
We note that typically intra-channel smearing, $\tau_\text{DM}$, can be
mostly ignored, as long as one assumes that the data is taken with a high-enough
frequency resolution. Here, we model the intra-channel smearing as 
$\tau_\text{DM}$ is related to the DM,  $\tau_\text{DM} =  8.3 \times 10^6  \,
\rm DM \, \Delta\nu_{\rm chan} / \nu^3$, where $\Delta\nu_{\rm chan}$
is the channel bandwidth, \ie~the total bandwidth divided by the number 
of channels~\citep{Hessels:2007pq}.
Throughout, we also neglect
$\tau_{\Delta\text{DM}}$, since sufficiently small DM step sizes can make this
contribution small as well. The only limitation comes then from the computing
resources that are available for the problem (besides of course temporal smearing).

The dispersion measure, DM, which enters in the definition of both $\tau_{\rm
DM}$ and $\tau_{\rm scatt}$, for any given line-of-sight and distance of the
source is computed using the Cordes-Lazio model for free-electron density in
the Galaxy, NE2001~\citep{Cordes:2002wz}.\footnote{
\scriptsize\url{http://www.nrl.navy.mil/rsd/RORF/ne2001/}}  In
Fig.~\ref{fig:DMprofile}, we show the latitude profile of the DM, as derived
from~\cite{Cordes:2002wz}, for $\ell = 0^{\circ}$ and for different distances
of the source from the Galactic center.  The scattering time is modeled
according to~\cite{2004ApJ...605..759B}.  We adopt a log-normal distribution
with mean $\mu = \log_{10}\tau_{\rm scatt}$, and a variance $\sigma = 0.8$ is
assumed to account for the large uncertainty affecting $\tau_{\rm scatt}$.
Indeed, while DM just depends on the column density of free electrons, the
amount of scattering depends on how these electrons are distributed along the
line-of-sight.  Note that, typically, temporal scattering has
  the effect of smearing out the radio pulsations of almost all MSPs
  within a degree of the Galactic disk to the point of undetectability -- for the assumed observing
  frequency of 1.4 GHz. Unlike dispersive broadening, 
  it is not possible to correct the measurement for scattering broadening, which is thus
  a fundamental limit for detection.

We note that, since most MSPs are found in
  binary systems, the effect of Doppler smearing due to orbital motion
  also has a significant impact on the ability to blindly detect new
  pulsars.  This is particularly true for the shortest (a few hour)
  orbital periods and most massive companions~\citep{2001PhDT.......123R}.

\begin{figure}
\centering
  \includegraphics[width=0.85\columnwidth]{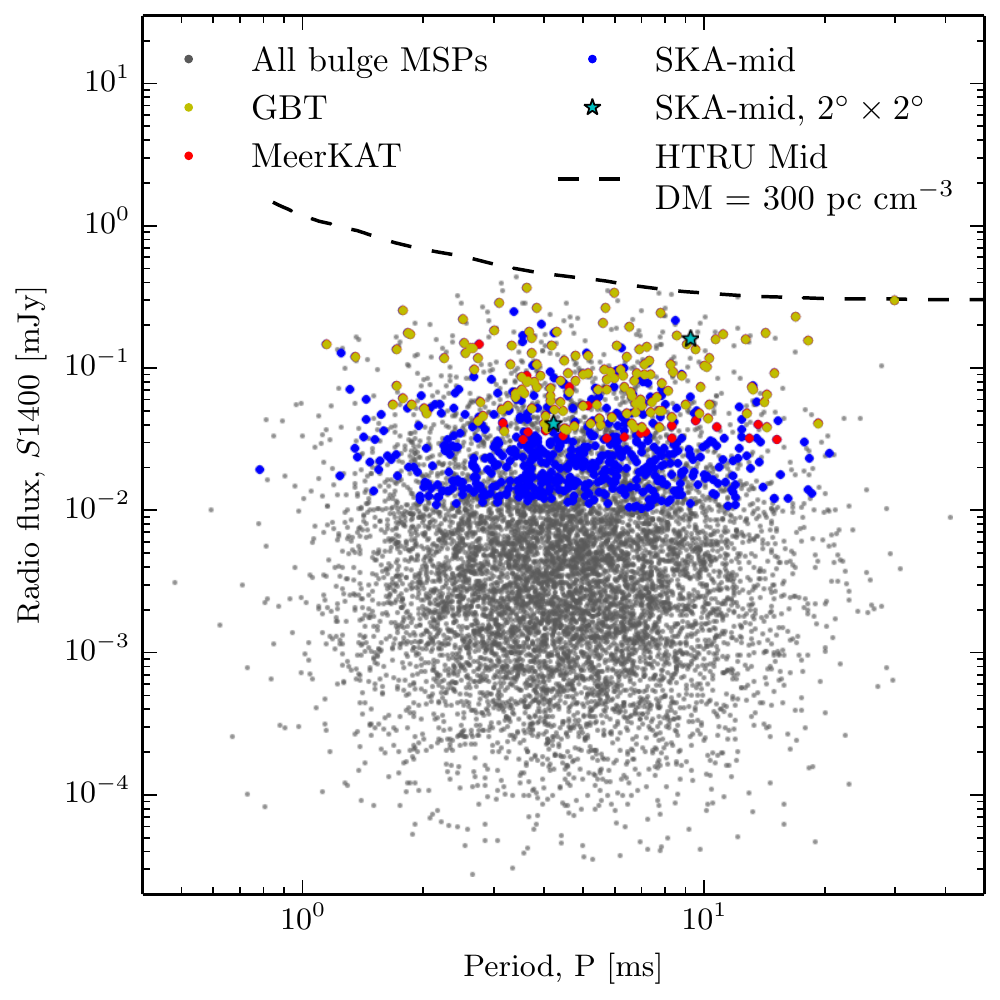}
  \includegraphics[width=0.85\columnwidth]{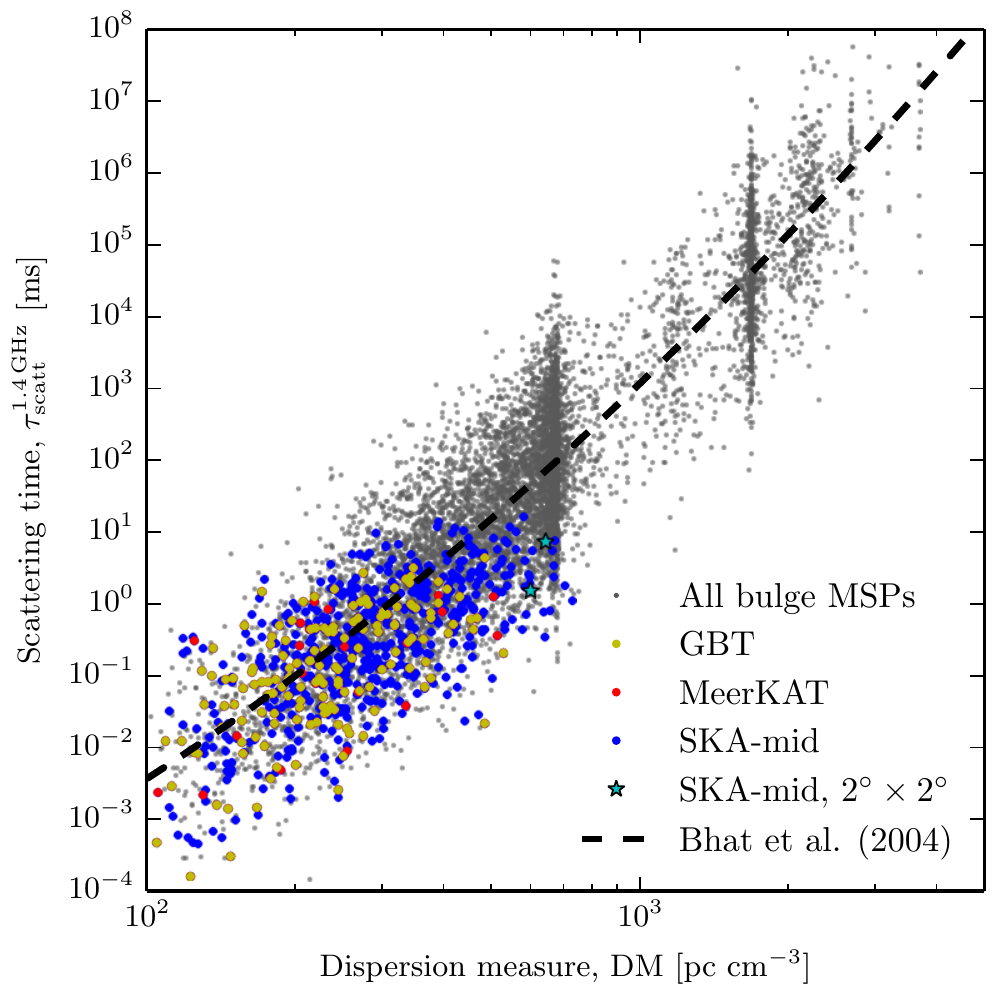}
  \caption{We show the simulated bulge population of MSPs, modeled from
    gamma-ray observations as described in the text, both in the period
    vs.~flux density plane (\emph{top panel}), and in the dispersion measure
    vs.~scattering time plane (\emph{bottom panel}).  \emph{Grey dots} denote
    the entire MSP bulge population. 
    The \emph{colored dots} show which of these sources
    would be detectable with the various observational scenarios that are
    described in Tab.~\ref{tab:telescopes}. Namely, \emph{yellow points} correspond
    to sources that will be detectable by GBT, MeerKAT and SKA-mid, 
    \emph{red points} to sources detectable by MeerKAT and SKA-mid, and \emph{blue points}
    to sources detectable only by SKA-mid.
    The \emph{dashed black line} in the
    \emph{upper panel} corresponds to the minimum flux sensitivity of the Parkes HTRU
    mid-latitude survey at a reference value of $\text{DM} = 300\pc \cm^{-3}$,
    and rescaled for the 10\% duty cycle we adopt in the present work. In the
    \emph{bottom panel}, we show also the average relation from
    \cite{2004ApJ...605..759B} as \emph{dashed black line}. The visible
    structures correspond to specific sky regions with very large DM, see
    Fig.~\ref{fig:DMprofile}.}
 \label{fig:compare}
\end{figure}

\bigskip

\begin{table*}
  \centering
  \begin{tabular}{lrrrr}
    \hline\hline
    Parameters & HTRU (mid) & GBT & MeerKAT & SKA-mid \\\hline
    $\nu$ [GHz] & 1.35  & 1.4 & 1.4 & 1.67 \\
    $\Delta\nu$ [MHz] & 340 & 600 & 1000 & 770\\
    $t_{\rm samp}$ [$\mu$s] & 64 & 41 & 41 & 41 \\
    $\Delta\nu_{\rm chan}$ [kHz] & 332 & 293 & 488 & 376\\
    $T_{\rm rx}$ [K] &  23 & 23 & 25  & 25  \\
    $G$ [K/Jy] & 0.74  & 2.0 & 2.9  & 15  \\
    Max. Base. Used [km] & -- & -- & 1.0 & 0.95 \\
    Eff. $G$ sub-array [K/Jy] & 0.74  & 2.0 & 2.0  & 8.5 \\
    Ele. $\theta_{\rm FWHM}$ [arcmin] &  14 & 8.6 & 65 & 49 \\
    Ele. FoV [deg$^2$] & 0.042 & 0.016 & 0.92 & 0.52 \\
    Beam $\theta_{\rm FWHM}$ [arcmin] &  14 & 8.6 & 0.88 & 0.77 \\
    Beam FoV [deg$^2$] & 0.042 & 0.016 & 0.00017 & 0.00013 \\
    \# Beams & 13  & 1  & 3000  & 3000   \\
    Eff. FoV [deg$^2$] & 0.55 & 0.016 & 0.51 & 0.39 \\
    \hline
    $T_{\rm point}$ [min] & 9 & 20 & 20 & 20 \\
    $T_{108\deg^2}$ [h] & 29 & 2250 & 71 & 92 \\
    \# Bulge(Foreground) MSPs & 1(6) & 34(37) & 40(41) & 207(112) \\
    \hline\hline
  \end{tabular}
  \caption{Relevant instrumental and observational parameters for existing
    (Parkes HTRU, GBT) and future (MeerKAT, SKA-mid) telescopes that we
    consider in this work.  Where possible, values are taken from Table~1 of
    the SKA Baseline Design report.  We quote the survey central observing
    frequency, $\nu$; effective bandwidth, $\Delta \nu$; sampling time, $t_{\rm
    samp}$; channel bandwidth, $\Delta\nu_{\rm chan}$; receiver temperature,
    $T_{\rm rx}$; gain of the whole array $G$; maximum baseline used (where
    applicable), `Max. Base.  Used'; the effective gain of the sub-array that
    can be used for wide-field pulsar surveys, `Eff. $G$ sub-array'; the
    beam-width of the elements in the array, `Ele. $\theta_{\rm FWHM}$'; the
    field of view of the array elements, `Ele. FoV'; the beam-width of the
    synthesized beam, Beam $\theta_{\rm FWHM}$; the field of view of the
    synthesized beam, `Beam FoV'; the number of beams recorded per pointing,
    `\# Beams'; and the effective field of view per pointing, `Eff. FoV'.  Next
    we give the integration time per pointing, the time required to cover $108
    \deg^2$ of sky, and the total expected yield of bulge and foreground MSPs
    from a region of that size.  The target region is here defined as
    ($|\ell|<5^\circ$ and $3^\circ<|b|<7^\circ$) plus ($|\ell|<3^\circ$
    and $1^\circ<|b|<3^\circ$) plus ($|\ell|,|b|<1^\circ$).}
    \label{tab:telescopes}
\end{table*}

\subsection{Instrumental parameters}

In the present work, we provide the predicted yields of bulge MSPs for three observational
scenarios based on the performances of currently operating and upcoming radio
telescopes: GBT, MeerKAT and SKA-mid.  As a
reference, and for comparison with past results, we choose to present results
for surveys at 1.4 GHz.  This turns out to be close to optimal in many cases,
and we discuss how our sensitivity predictions change at higher and lower
frequencies in Sec.~\ref{sec:discussions}.  In Tab.~\ref{tab:telescopes}, we
quote the parameters used for each instrument.  Parameters for the GBT are
based on the GUPPI back-end and taken from the Proposer's Guide for the
GBT.\footnote{\scriptsize\url{https://science.nrao.edu/facilities/gbt/proposing/GBTpg.pdf}}
Sensitivities for the future MeerKAT and SKA-mid are based on the SKA Phase 1 System
Baseline Design
report.\footnote{\scriptsize\url{http://www.skatelescope.org/wp-content/uploads/2012/07/SKA-TEL-SKO-DD-001-1_BaselineDesign1.pdf},
see Tab.~1.} We implement the performances of the MeerKAT and of the SKA-mid
(350--3050 MHz) Antenna Array configuration.  The quoted antenna gain in
Tab.~\ref{tab:telescopes} ($G=T_\text{sys}/\rm SEFD$) is derived from the
system-equivalent flux density (SEFD) assuming a receiver temperature of $25
\K$ (for the specific purpose of deriving the antenna gain from published results we here
neglect the sky temperature, however we do fully account for it when deriving the sensitivity predictions.)
For other parameters entering in Eq.~\ref{eq:wobs}, such as the number of channels 
and the sampling interval, we refer to the corresponding values quoted for each telescope 
in the references provided above. As for GBT, we use a sampling time 
of 41 $\mu$s and 2048 channels. 
We emphasize that our estimates for MeerKAT and SKA-mid are only of indicative
value, and should be updated once these telescopes are operational and accurate
telescope performance parameters are known.  Furthermore, the amount of data
that can be collected with these instruments in a short time is enormous, and
the likely bottleneck for pulsar searches will be the available computer
processing resources for exploring the full telescope field-of-view and
relevant astrophysical parameter space. Since not all data can be stored and
analyzed offline, our estimated observation times for MeerKAT and SKA-mid are
almost certainly too optimistic, probably by a factor of a few.  In the same
way, we assume that the entire arrays are used in the search.  However, when
doing the measurement only a limited baseline (and hence only a subset of the
full array) should be used in order to increase the size of the synthesized
beam which then decreases the computation time.

In Tab.~\ref{tab:telescopes} we also show the parameters for the HTRU survey
performed recently with the 13-beam Multibeam receiver on the Parkes radio
telescope at 1.4 GHz~\citep{2010MNRAS.409..619K}. This is the most recent and
relevant large area survey of the southern sky, performed at high latitudes
(from the Galactic plane up to $b=\pm 15^{\circ}$).  In what follows, we adopt
the HTRU mid-latitude survey as a reference to check the consistency of our
results with previous surveys.

\medskip

In Tab.~\ref{tab:telescopes} we
also quote other relevant parameters for the present analysis, as,
for example, the adopted per pointing observation dwell times, 
along with the corresponding total time needed to
cover a $108 \deg^2$ area of sky.  
We here assume that beams are non-overlapping.
These effects need to be taken into account when setting up an actual
observation strategy, and will increase the required observation time for a
given field by a factor of less than two.

\section{Results for large area searches}
\label{sec:surveys}

In this section, we will first discuss prospects for current and future radio
telescopes to detect bulge MSPs in large area surveys (meaning several square
degrees of sky), and then quantify the number of MSP detections that would be
required to unambiguously confirm the existence of a bulge population in
addition to the observed thick-disk population of MSPs.

\subsection{General reach of current and future radio surveys}

For each simulated MSP in the bulge, modeled according to Sec.~\ref{sec:model},
we compute the corresponding 10$\sigma$ detection sensitivity flux, following
Eqs.~\ref{eq:Snurms} and~\ref{eq:wobs} for the observation scenarios in
Tab.~\ref{tab:telescopes}.  In Fig.~\ref{fig:compare} (top panel), we show the
distribution of all bulge MSPs in the flux density (at 1.4 GHz) versus period
plane. As mentioned above, the adopted period distribution
\citep{Lorimer:2015iga} has a mean of 5.3 ms. 
We note that this value is slightly higher than what is typically
adopted as mean MSP period, $P \sim$ 3 ms. 

Assuming a lower mean spin period would somewhat reduce our estimates since finding
fast-spinners is harder due to scattering and Doppler smearing in binaries.
However, since the threshold sensitivities
in the top panel of Fig.~\ref{fig:compare} depend only mildly on the spin
period, we do not expect a large effect.

We simulate sources with period between 0.4 and 40 ms.
The
corresponding radio fluxes at 1.4 GHz span from about $10^{-5}$ mJy up to about
0.9 mJy (we note that the lower flux limit is a consequence of the adopted
luminosity function and observationally neither relevant nor well constrained).
However, not all the sources with high flux densities can be detected for our
three reference scenarios.  Colored dots show which of the sources would be
detected by our assumed measurements with GBT, MeerKAT and SKA-mid with
10$\sigma$ significance.  The GBT will be able to detect sources down to about
0.03 mJy and periods in the range $ 1 \, \rm ms \leq P \leq 40 \, \rm ms$.
MeerKAT and SKA-mid, instead, will probe radio fluxes as low as 0.03 mJy and
0.01 mJy respectively, in the full period range of the population above 0.8
ms. 
We also overlay the sensitivity of the currently most sensitive survey covering the 
relevant sky area,
the Parkes HTRU mid-latitude survey (assuming $\text{DM} = 300\pc \cm^{-3}$).
No source lies above this line, showing that such a survey is not quite yet 
sensitive to detect the bulge MSPs, however it is evident that it starts to
scratch the high-luminosity tail of this population.  On the other
hand, it is clear that there will be a progressive improvement in the number of
sources detectable by the three telescopes we consider.  Already with GBT the
gain in sensitivity would result in hundreds of sources being above threshold
with only 20 minutes integration time per sky position (although the total time to
survey a large enough region of the sky still remains very large, as we will
see below).
 
\smallskip

The bottom panel of Fig.~\ref{fig:compare} clarifies what is the distribution
of DM for the simulated bulge population and the corresponding scattering time,
$\tau_{\rm scatt}$.  Most of the sources have DM in the range 100--800 pc
cm$^{-3}$. The sharp, and dense, features at around 800 pc cm$^{-3}$ and 1800
pc cm$^{-3}$ correspond to regions very close to the Galactic center and are
due to discrete ``clumps" of enhanced free electron density that are included in 
the NE2001 model (see Tables 5 --7 in~\cite{Cordes:2002wz}; these are also visible in Fig.~\ref{fig:DMprofile}).
The scattering times follow as expected the trend of
the adopted reference model from \cite{2004ApJ...605..759B}, with a significant
scatter.  In general, scattering times larger than 5--10 ms
 prevent the sources to
be detected and the limiting factor in Eq.~\ref{eq:wobs} is indeed $\tau_{\rm
scatt}$. For scattering times smaller than 5--10 ms, instead, a source might be
detected or not depending on its spin period. The GBT and MeerKAT can detect most sources
with DM up to 550 pc $\rm cm^{-3}$, while none with DM $\sim$ 600--800 pc
cm$^{-3}$. On the other hand, SKA-mid  will be able to detect MSPs
that suffer from larger scattering, up to about 800 pc cm$^{-3}$.  In
particular, we can see that with SKA-mid we will be able to detect a few sources with
high DM ($\sim$ 600--800 pc cm$^{-3}$) and in the few inner degrees of the Galactic
center,  namely the inner $2^{\circ} \times 2^{\circ}$ degrees.
In general, SKA can probe more sources because of the higher sensitivity.
Since the luminosities are uncorrelated with spin period and other parameters, 
it can pick out the sources that have high DM but luckily have anomalously low scattering. 
Moreover, the central observing frequency of SKA (assumed here) is 1.67 GHz, which is slightly higher
than GBT and MeerKAT. Given the strong frequency dependence of the scattering time, it
reduces temporal scattering by a factor of around two.

\subsection{Optimal target regions}

\begin{figure}
  \centering
  \includegraphics[width=0.85\columnwidth]{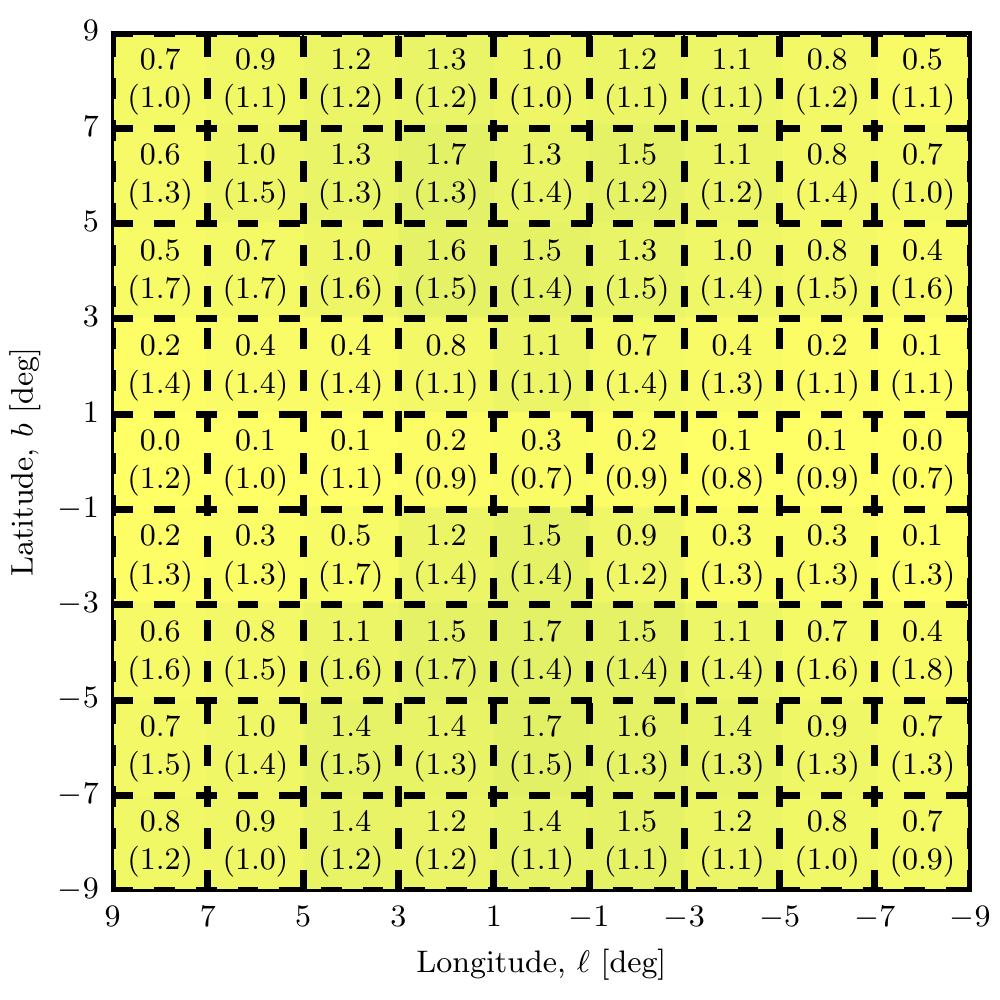}
  \includegraphics[width=0.85\columnwidth]{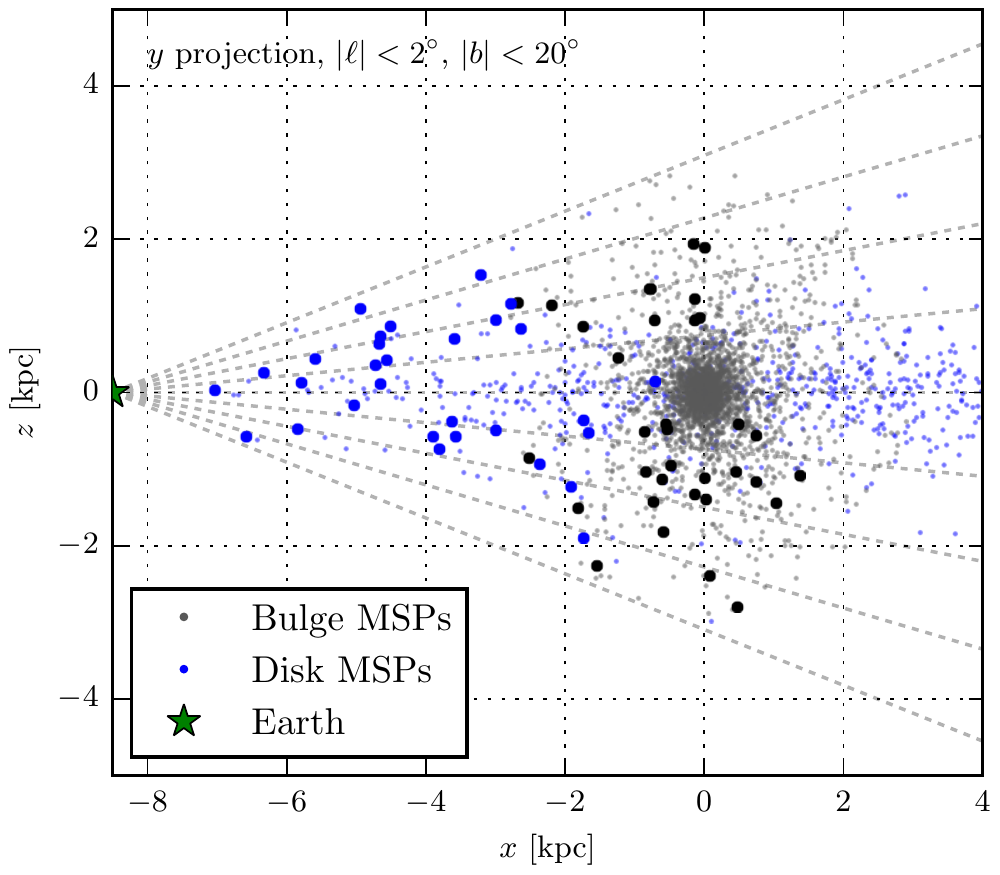}
  \caption{\emph{Top panel:} GBT detected sources from bulge (disk) population
    for 20 minutes integration time per pointing (250 h for each field of
    $2^\circ\times2^\circ$) at 1.4 GHz.  The number of sources detectable is
    also represented by the colored background.  \emph{Bottom panel:} x--z
    projection of simulated bulge (\emph{thin black dots}) and disk (\emph{thin
    blue dots}) MSPs.  \emph{Thick black dots} refer to bulge MSPs detectable
    towards the inner Galaxy, $|\ell| < 2^{\circ}$ and $|b| < 20^{\circ}$, with
    the GBT survey.  \emph{Thick blue dots} are instead the disk MSPs that would
    be detected by the survey in the same region of interest.}
  \label{fig:sensGBT}
\end{figure}

\begin{figure}
  \centering
  \includegraphics[width=0.85\columnwidth]{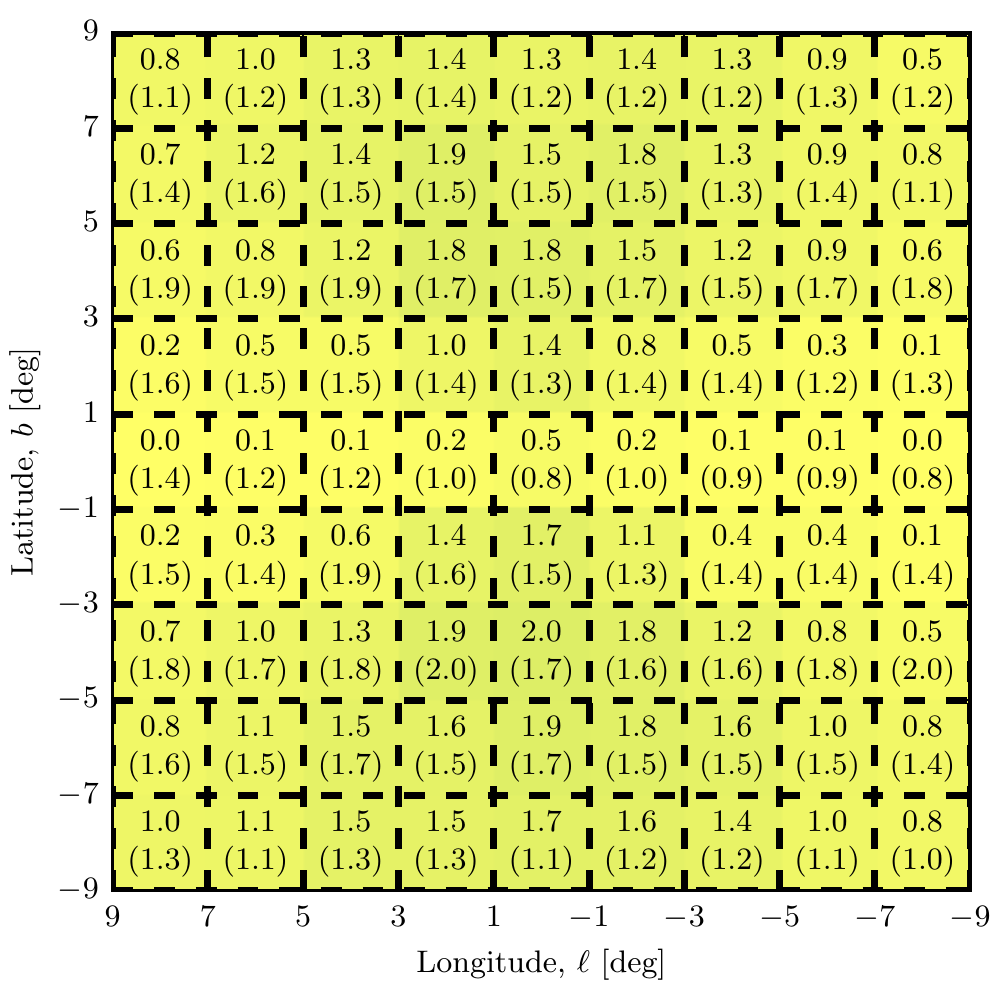}
  \includegraphics[width=0.85\columnwidth]{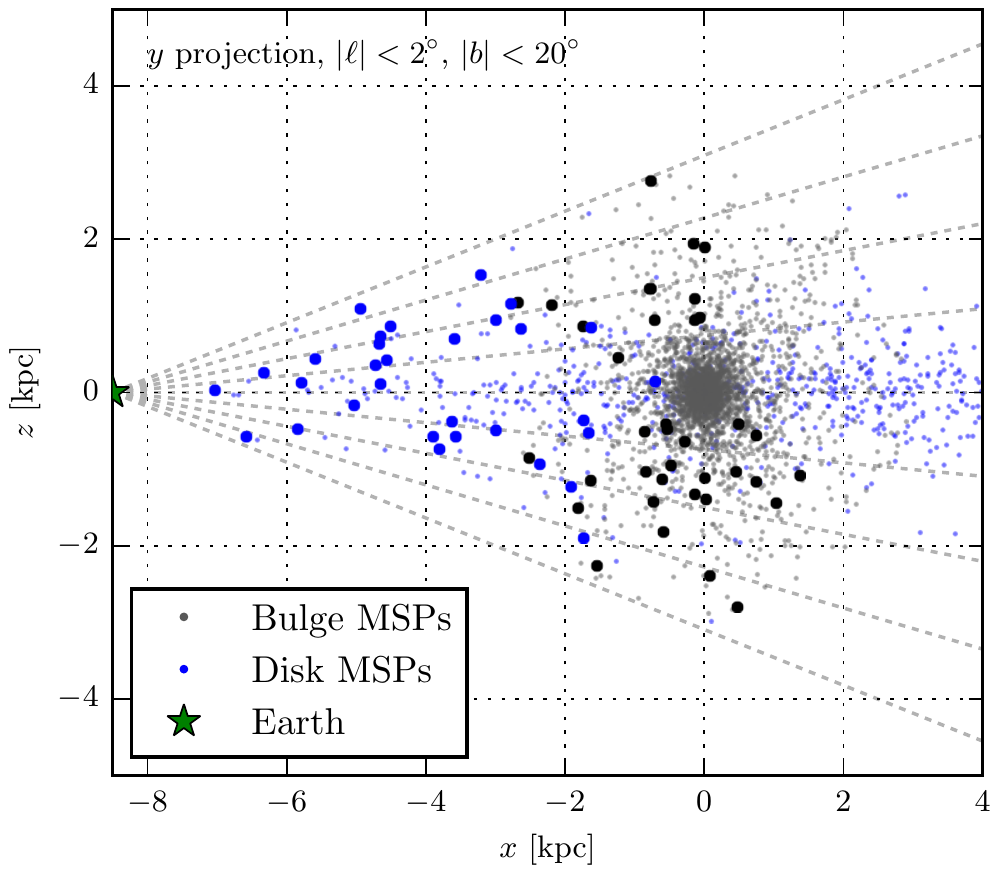}
  \caption{Same as Fig.~\ref{fig:sensGBT}, but for a MeerKAT-like survey with
  parameters as described in Tab.~\ref{tab:telescopes}.}
  \label{fig:sensMeerKAT}
\end{figure}

\begin{figure}
  \centering
  \includegraphics[width=0.85\columnwidth]{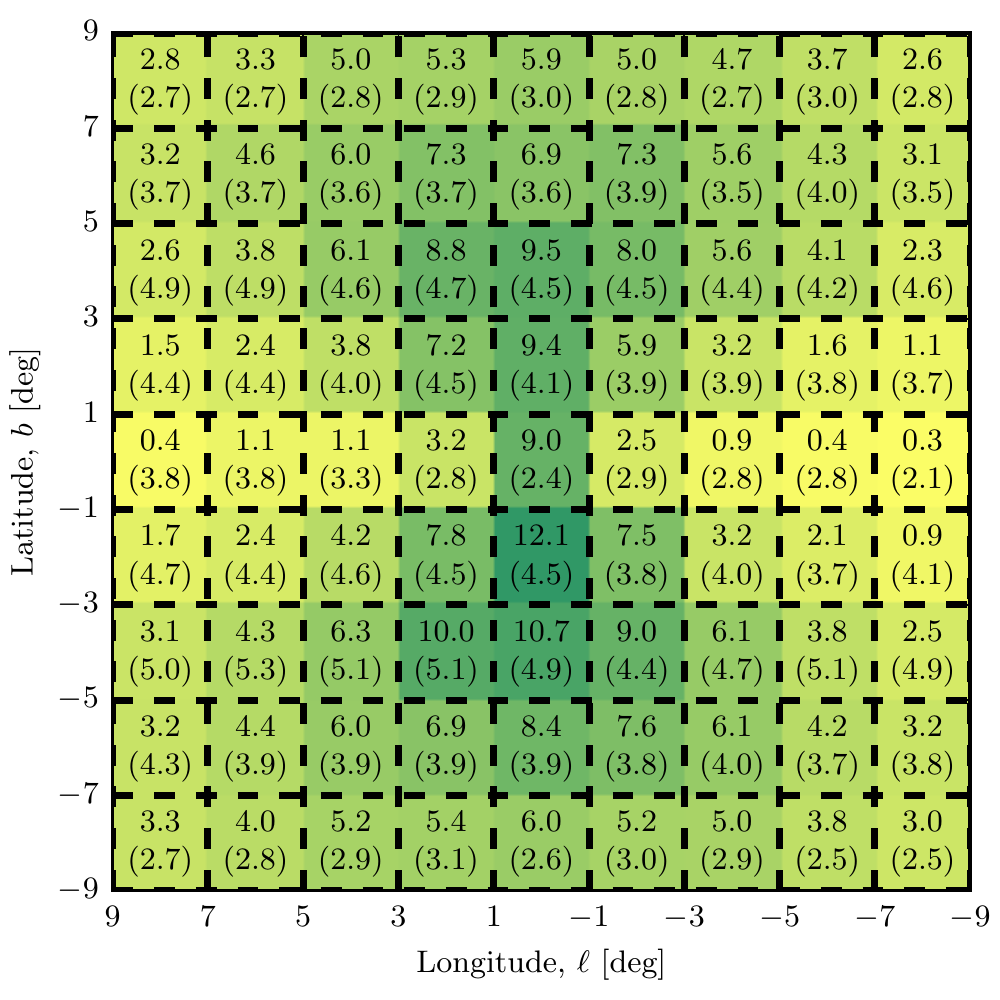}
  \includegraphics[width=0.85\columnwidth]{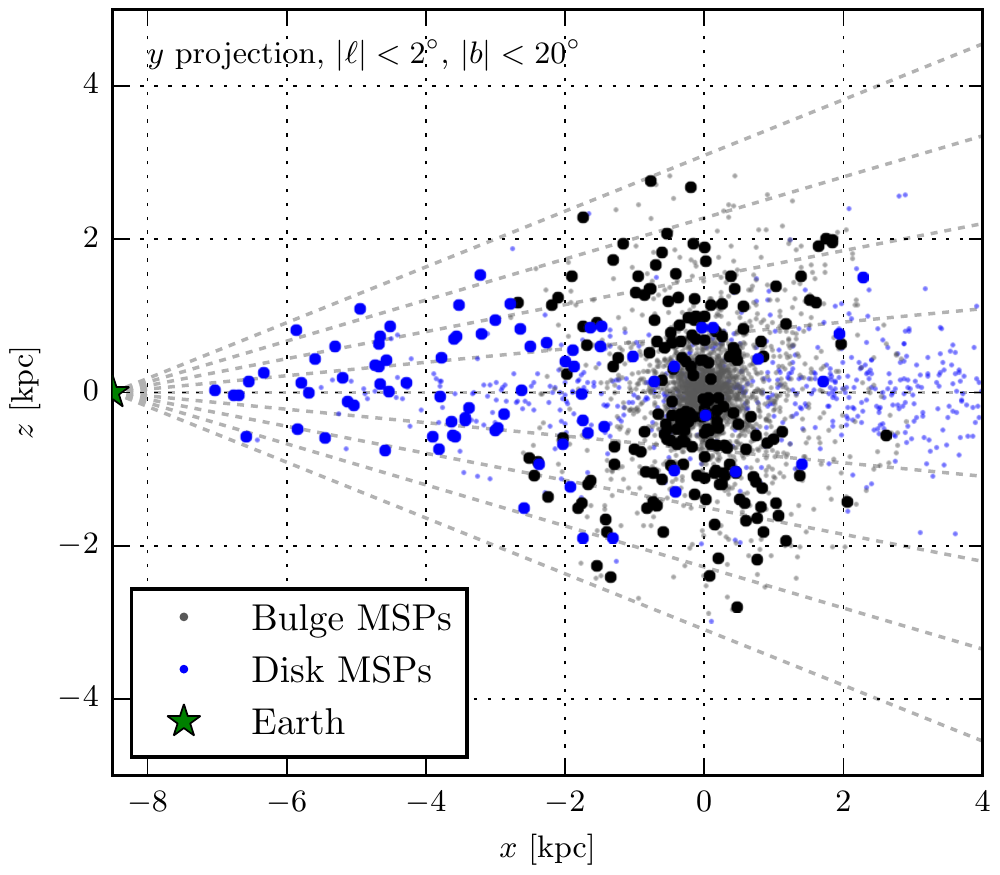}
  \caption{Same as Fig.~\ref{fig:sensGBT}, but for a SKA-mid-like survey with
  parameters as described in Tab.~\ref{tab:telescopes}.  Here, one can also
  nicely see a dearth of detectable MSPs in the shadow of the Galactic center
  as well as in front of the Galactic center. In both cases presumably due to  scattering and uncorrected dispersive smearing.}
  \label{fig:sensSKA}
\end{figure}

We now investigate what are the detection prospects for large-area surveys
performed with the three instrumental reference scenarios (namely with GBT, MeerKAT and 
SKA-mid configurations).  For each instrument
we show, in the top panels of Figs.~\ref{fig:sensGBT}--\ref{fig:sensSKA}, the
number of bulge MSPs that can be detected with 10$\sigma$ significance and the
corresponding number of detectable disk MSPs in parenthesis (as modeled in
Sec.~\ref{sec:model}). We analyze a region in the inner Galaxy defined by $|
\ell| < 9^{\circ}$ and $|b| < 9^{\circ}$, and we split it in squared subregions
of size $2^\circ\times2^\circ$.  Integration times per pointing and central
observing frequencies are as shown in Tab.~\ref{tab:telescopes}.

An alternative way to visualize the prospects for detection of the bulge
population above the disk population is to plot in the x--z plane the sources
detectable along the lines of sight towards the inner Galaxy.  Emphasizing
sources detectable from these directions helps in understanding (a) what is the
contamination from foreground disk sources and (b) how deep towards the
Galactic center we can probe the bulge population.  In the bottom panels of
Figs.~\ref{fig:sensGBT}--\ref{fig:sensSKA}, we show the spatial distribution
of the simulated bulge and disk MSPs in the x--z plane and we highlight the
sources  that can be detected in the region $|\ell| < 2^{\circ}$ and $|b| <
20^{\circ}$ (which corresponds to the inner Galaxy region analyzed
by~\cite{Calore:2014xka}). 

\medskip

In Figs.~\ref{fig:sensGBT}, \ref{fig:sensMeerKAT}, and~\ref{fig:sensSKA}
we show the number of detectable sources
with GBT,  MeerKAT and SKA-mid, respectively, for 20 min observation dwell time per pointing. 
For the GBT scenario the number of detectable bulge MSPs 
is always lower than 2 for each sky subregion and depending on the
subregion, the number of detectable disk MSPs is comparable. 
On the other hand, in the case of MeerKAT and even more 
for SKA-mid, there is an optimal search
region, which is a few degrees south of the Galactic center, at approximately
$|\ell|\leq1^\circ$ and $- 5^\circ\leq b \leq - 3^\circ$, where the number of
detectable bulge MSPs is the largest.  
While for MeerKAT the number of bulge MSPs in such an optimal spot is still comparable with 
the number of foreground
thick-disk MSPs, in the case of SKA-mid (for which the 
optimal target region slightly shifts towards lower latitudes, $|\ell|\leq1^\circ$
and $- 3^\circ\leq b \leq - 1^\circ$) the number of detectable bulge sources is
as high as 12 per $4\deg^2$ and the corresponding 
detectable disk MSPs are always about
half of the number of bulge MSPs detectable in the same subregion.

Typically,  the suppression of the number of detectable sources along the Galactic disk comes
from strong scattering effects discussed in Sec.~\ref{sec:sens}. We will
discuss the advantage (against scattering effects) of using higher frequency
surveys in Sec.~\ref{sec:discussions}.
While from the bottom panels of
Figs.~\ref{fig:sensGBT} and \ref{fig:sensMeerKAT}
it is evident that, for the GBT, the bulge MSPs that
lie truly at the Galactic center and along the Galactic disk remain hard to
identify for those two scenarios, the predictions improve with SKA-mid.
From the bottom panel of Fig.~\ref{fig:sensSKA}, indeed, we can  see how
the detectability of bulge MSPs from the very central region of the bulge is less
affected by pulse broadening and the contamination along
directions towards the inner Galaxy is lower.
Interestingly, SKA-mid will be able to probe sources residing in the innermost
degree, $|\ell|\leq1^\circ$ and $|b| \leq 1^\circ$ (those same sources are the
ones highlighted in Fig.~\ref{fig:compare}; note that Fig.~\ref{fig:sensSKA}
shows average values).  These sources happen to have a
very low scattering broadening, which is in our case possible even in the inner
Galaxy, since we adopt a large variance in the scattering time of individual
sources.  The bottom panel of Fig.~\ref{fig:sensSKA} clearly demonstrates the
detection power of SKA-mid.  While the number of detectable thick-disk MSPs
remains limited to a few objects (simply because the density of thick-disk
sources is relatively small), the number of bulge MSPs that can be observed is
very large.

\smallskip

For GBT, observations of sky areas as large as $4 \deg^2$ are mainly limited by the
small size of the telescope beam at high frequencies and to cover a $2^\circ
\times 2^\circ$ region of sky with the GBT at 1.4 GHz, a total observation time
of about 83 hours is required. This makes the survey of larger areas
unfeasible, and in any case it would lead to a maximum of 2 detections per
$4\deg^2$.
The much larger field-of-view of MeerKAT, with respect to the GBT beam size,
allows to survey the same $4 \deg^2$ area in a much shorter time, \ie~about 2.5
hours.  Analogously, for SKA-mid about 3.5 hours are required to survey the region.
This might enable $\sim$ 100-hour-long surveys that can scan
sky areas about 40 times larger than our $4\deg^2$ subregion and thus
probe $\sim$ 100 bulge MSPs (in the most promising sky regions). 

As
mentioned above, limiting factors like a reduced maximum baseline and limited
computation power will likely increase the required observation times by a
factor of two or more. 

\smallskip

To understand the interplay among area surveyed, total integration time and
predicted number of detectable bulge MSPs (and foreground thick-disk MSPs),
in Tab.~\ref{tab:telescopes} we quote the number of bulge and foreground thick-disk
MSPs that would be detectable by the GBT, MeerKAT and SKA-mid
for a large-area survey of 108 $\deg^2$ and 20 minutes of dwell time per pointing.
The chosen large-area survey is defined by the 27 $4\deg^2$ sky areas that 
have a large yield of detectable sources (larger than 6) for the SKA-mid scenario.
This region corresponds to ($|\ell|<5^\circ$ and $3^\circ<|b|<7^\circ$) 
plus ($|\ell|<3^\circ$ and $1^\circ<|b|<3^\circ$) plus ($|\ell|,|b|<1^\circ$).
It is evident that GBT and MeerKAT might lead to comparable numbers of detected 
MSPs from the bulge ($\sim 30-40$ sources). 
Analogously, for both observational scenarios the number of detectable thick-disk MSPs
is comparable with the bulge ones and thus this is not really a promising strategy, given 
the strong contamination from disk sources.
Moreover, the time needed for GBT to survey a 108 $\deg^2$ area is about 30 
times larger than the total time required for the same survey with MeeKAT. In this respect,
large-area surveys will not be feasible with the GBT but might be promising with
MeerKAT.
SKA-mid clearly improves those predictions: It allows a discrimination between 
bulge and thick-disk MSPs in a reasonable total integration time (92 hours).
A large-area survey with time per pointing of about 20 minutes can thus be
an optimal strategy for SKA-mid to identify bulge MSPs.

\bigskip

In conclusion, prospects for large-area surveys are extremely good for upcoming
radio telescopes, albeit they are less promising for current observations
through the GBT.  With GBT the main limitation is represented by the very large
integration time required to survey a small sky area, and the relatively low
number of detectable bulge and disk sources, which would make it harder to
disentangle the two populations.  On the other hand, with MeerKAT and later
with SKA-mid the smaller required total integration time, together with the
higher sensitivity, will allow to quickly probe large areas and detect a very
significant fraction of the MSP bulge population.

\subsection{Discrimination of bulge and thick-disk populations}

\begin{figure}
  \centering
    \includegraphics[width=0.85\columnwidth]{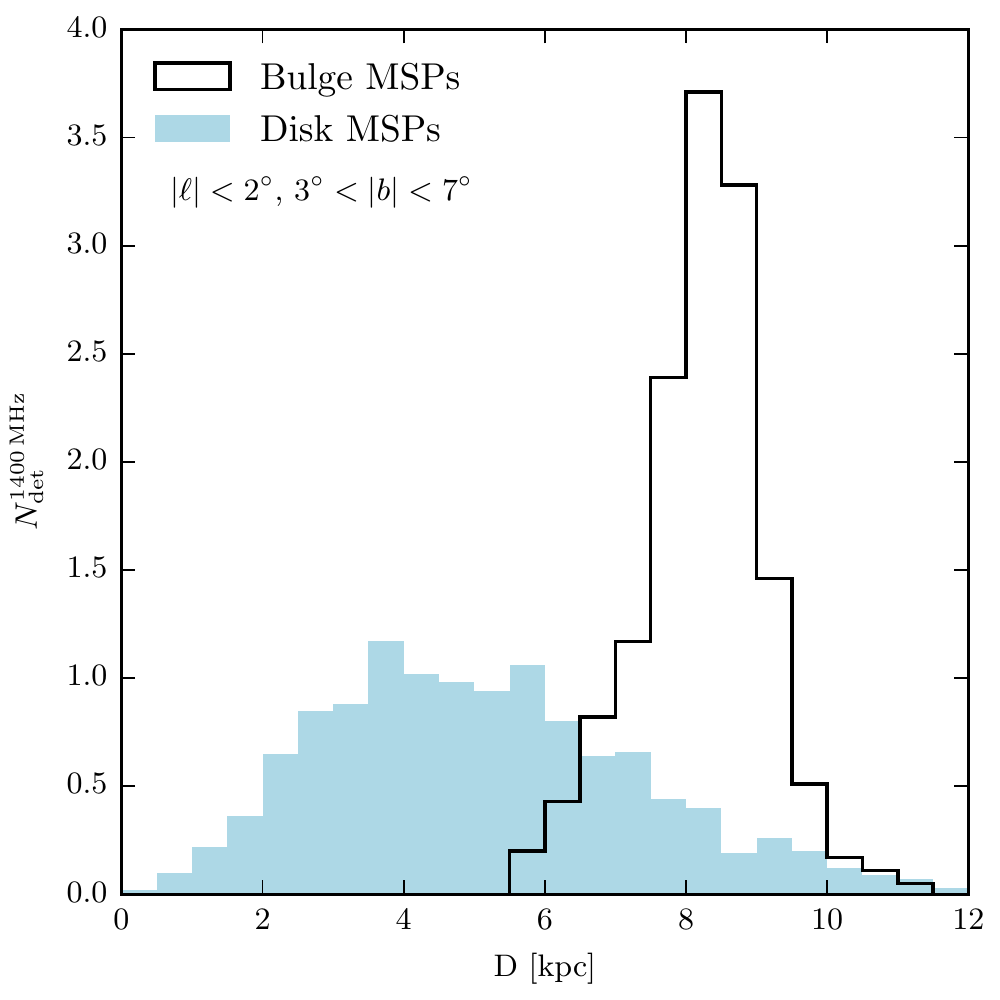}
    \caption{Histogram of distances of detected bulge (\emph{black}) and disk
      MSPs (\emph{blue}), assuming the MeerKAT reference survey in
      Tab.~\ref{tab:telescopes}.  Bulge and disk components can be clearly
      separated.  The bulge component should appear as a clear excess of
      sources with dispersion measures that indicate distances around $8.5\kpc$.}
    \label{fig:DMhist}
\end{figure}

In Fig.~\ref{fig:DMhist}, we show a histogram of the distances of all MSPs that
would be detected by our MeerKAT reference survey in eight $4\deg^2$ subregions
below and above the Galactic center, $|\ell|<2^\circ$ and
$3^\circ<|b|<7^\circ$. The adopted survey region is exemplary, and chosen
because it
provides a good MSP yield (see Fig.~\ref{fig:sensMeerKAT}) while at the same
time having a relatively low contamination with foreground sources.
Furthermore, we concentrate on MeerKAT to obtain conservative estimates.
The deeper observations with SKA would only increase the relative number of
bulge sources, and simplify a discrimination from foreground MSPs.
For the adopted survey and target region, the number of detected bulge sources
would be 14.3.  The number of detected disk sources in our reference scenario would
be 12.2.  Already visually it is clear that the distance distributions are very
different, with the thick-disk distribution peaking very broadly at $4\kpc$,
whereas the bulge population has a pronounced peak around $8.5\kpc$.  

In order to provide a first estimate for the \emph{minimum} number of bulge
MSPs that need to be detected in order to identify the bulge population with a
statistical significance of $99.7\%$~confidence level (CL) above the foreground of thick-disk
MSPs, we perform a simple statistical test as follows.  Let $\mu_i^\text{disk}$
and $\mu_i^\text{bulge}$ be respectively the expectation values for the disk
and bulge components as shown in Fig.~\ref{fig:DMhist} ($i$ refers to
individual distance bins).  We consider the ``Asimov data set'' \citep{Cowan11}
$c^A_i = \zeta(\mu_i^\text{bulge}+\mu_i^\text{disk})$, where $c^A_i$ denotes
the number of measured MSPs in a certain distance bin, and $\zeta$ is a
rescaling factor with respect to the number of sources shown in
Fig.~\ref{fig:DMhist}.  It accounts for the effect of surveying a smaller
region of the sky.  We calculate now the Poisson likelihood both for the null
hypothesis $\mu_i^\text{null} = \zeta\mu_i^\text{disk}$ and the alternative
hypothesis $\mu_i^\text{alt}= \zeta(\mu_i^\text{bulge}+\mu_i^\text{disk})$.  We
numerically solve for $\zeta$ by requiring that the minus-two log-likelihood
ratio $-2\ln\left(\mathcal{L}_\text{null}/\mathcal{L}_\text{alt}\right)$ equals
9.  The value that we find is $\zeta=0.24$, which corresponds to the detection
of 2.9 disk and 3.4 bulge sources.  Note that we implicitly assume here that
the normalization of the disk component can be constrained from other regions
of the sky (since we keep $\zeta$ fixed when calculating
$\mathcal{L}_\text{null}$).  Indeed, the main reason for the low number of only
3.4 required bulge detections is the low background from the disk at distances
around $\sim8.5\kpc$ distance.

We conclude that the detection of a handful of bulge sources is enough,
provided their distances can be estimated accurately enough, to start
discriminating the bulge and disk components in a statistically meaningful way.
The NE2001 model provides DM-based distance predictions, typically with 25\% 
fractional uncertainty.  This will be useful for associating MSP discoveries with a bulge population.  
Parallax distance measurements (or lower limits) using very-long-baseline radio interferometry
(VLBI) could also be used, but for the weakest sources the sensitivity of current VLBI 
arrays may be insufficient for detection.
However, we stress that a robust statistical statement should be ideally based
on a physical model for the bulge distribution (which might not necessarily
include sources in the inner kpc) and be marginalized appropriately over disk
and bulge profile uncertainties, the total number of disk and bulge sources,
and include uncertainties in the DM-based distance measure.  However, our above
estimates suggest that a robust detection of the bulge MSP component should be
possible once radio pulsation from the first couple of bulge sources has been
observed.

\section{Results for targeted searches}
\label{sec:targeted}

Deep searches for radio pulsations towards unassociated \Fermi~gamma-ray
sources have been extremely successful in discovering new
MSPs~\citep{Grenier:2015pya, TheFermi-LAT:2013ssa,Ray:2012ue}.  This is mostly
due to the fact that targeted searches allow deeper observations than
time-intensive large area surveys.  It is thus natural to assume that the same
strategy should also be useful for identifying the bulge population of MSPs.
Interesting targets in this case are unassociated \Fermi\ sources in the inner
Galaxy, but also potential sources that remained below the \Fermi\ source
detection threshold could be valuable targets.  Candidates for the latter were
recently identified as wavelet peaks in the analysis of \cite{Bartels:2015aea}
and as hotspots in the analysis of \cite{Lee:2015fea}.  We will from here on
refer to all of these potential sources as \emph{MSP candidates}, and discuss
the prospects for identifying their radio pulsation signal.

In contrast to the above discussion about large area surveys, the prospects for
radio targeted searches depend strongly on the details of gamma-ray and radio
beaming.  The reason is that the success of deep, targeted, follow-up radio
searches hinge on whether gamma-ray bright sources are also bright in radio.
Although even a strong gamma-ray/radio correlation would leave our above
discussion about prospects for large area surveys completely untouched, it
would be very beneficial for targeted searches.  

\smallskip

Obviously, not every MSP candidate found in \Fermi\ data will correspond to an
MSP.  The odds for this depend on the density of MSPs and other sources in the
inner Galaxy, the statistical significance of the MSP candidate, its spectrum
and its variability.  However, we will focus here on the radio detection
sensitivity and the effect of a possible gamma-ray/radio correlation.  To this
end, we will simply assume that all of our MSP candidates correspond in fact to
MSPs, and that their localization is known with much better accuracy than the
beam size of the GBT.

As an instructive example, we will here use the 13 unassociated
3FGL~\citep{TheFermi-LAT:2015hja} sources that were identified as MSP
candidates in \cite{Bartels:2015aea}, based on their spectrum and the absence
of variability.  We stress that \emph{this does not mean that these sources are
necessarily the best targets for follow-up searches}.  However, their gamma-ray
brightness, as well as their positions in the inner Galaxy, have typical values
that should be comparable in \emph{any} list of follow-up targets.  Studying
the radio sensitivity for targeted observations at the position of these
sources is hence indicative for targeted observations of any sources related to
the \Fermi\ GeV excess.

\subsection{On the gamma-ray radio correlation}

\begin{figure}
  \centering
    \includegraphics[width=0.90\columnwidth]{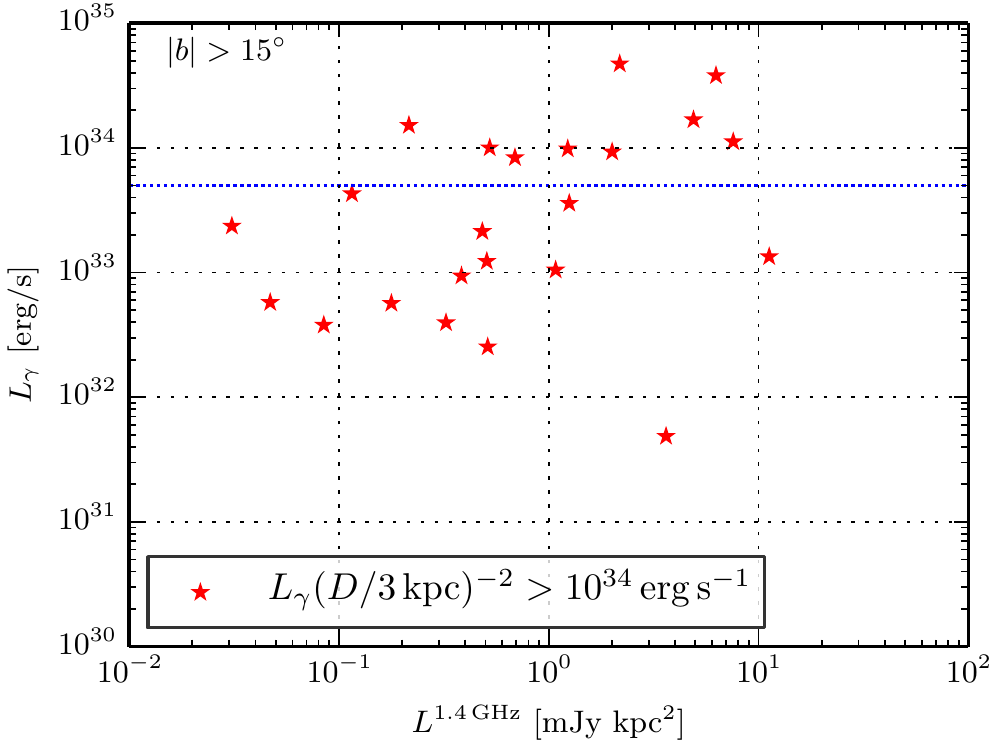}
    \caption{Gamma-ray luminosity vs.~radio pseudo luminosity at 1.4 GHz, for
      high-latitude ($|b|>15^\circ$) MSPs from \cite{TheFermi-LAT:2013ssa} that
      pass the flux threshold as defined in the figure.  We also show the
      gamma-ray luminosity threshold ($L_\gamma > 5\times 10^{33} \ergs$) that
      we use for selecting radio luminosities for luminous gamma-ray MSPs (see
      text for details).}
    \label{fig:LgammaLradio}
\end{figure}

\begin{figure}
  \centering
    \includegraphics[width=0.90\columnwidth]{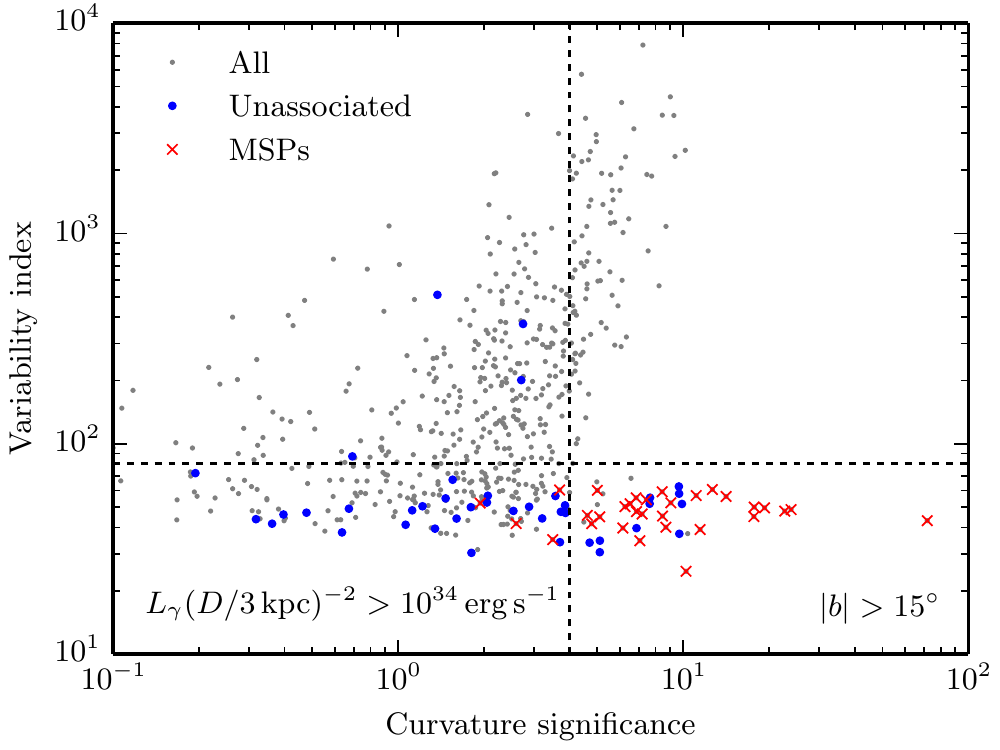}
    \caption{Curvature significance vs.~variability index, for \emph{all}
      high-latitude sources that pass the flux threshold as indicated in the
      text and in the figure.  We furthermore indicate unassociated sources and
      MSPs. The horizontal line separates variable from non-variable sources,
      the vertical line separates sources with a significantly curved spectrum
      from those whose spectra are power-law like. The full source list and
      definitions can be found in \cite{TheFermi-LAT:2015hja}.}
    \label{fig:pscs}
\end{figure}

As a very rough estimate, only bulge MSPs with a luminosity of at least
$L_\gamma \gtrsim 10^{34}\ergs$ will show up as MSP candidates in \Fermi-LAT
gamma-ray observations (potentially with very low significance).  The required
luminosities for detection are typically higher \citep[see][]{Petrovic:2014xra,
TheFermi-LAT:2013ssa, Bartels:2015aea}, but the exact value does not matter for
the following discussion.  We will show that for such gamma-ray bright MSPs,
also the radio emission is very well above the average, and exploit it when
predicting prospects for radio follow-up observations.

We emphasize that the adopted
estimate depends critically on possible selection effects.  In almost all
cases, \Fermi\ sources were identified as MSPs by the observation of radio
pulsation.  This will in general bias a relation that is just based on
radio-observed MSPs, since radio-quiet MSPs would be listed as unassociated
Fermi sources.  We will below conservatively take this effect into account by
assuming that all unassociated non-variable high-latitude sources are
radio-quiet MSPs.

Roughly 1/3 of the MSPs discovered in \Fermi\ targeted searches have been shown 
to be in eclipsing ``black widow" or ``redback" systems (Ray et al. 2012).  While eclipses can lead 
to MSPs being missed in a survey, we conservatively estimate that this is about a 15\% 
reduction in the potential yield of a wide-field survey -- assuming that 30\% of the sources are 
eclipsed 50\% of the time.
In the following discussion, we will study the gamma-ray and radio emission
properties of MSPs and unassociated sources, based on the sources listed in the
Second Pulsar Catalog, 2PC~\citep{TheFermi-LAT:2013ssa}, and in the
3FGL~\citep{TheFermi-LAT:2015hja}.  In order to select bright gamma-ray
sources, we adopt a \emph{flux} threshold that corresponds to $L_\gamma =
10^{34}\ergs$ at $3\kpc$ distance.  This trivially includes all luminous
(namely $L_\gamma > 10^{34}\ergs$) MSPs within $3\kpc$ distance from the Sun,
but also all unassociated sources that could be luminous MSPs in that volume.
As a \emph{spatial} cut, we adopt $|b|>15^\circ$, which practically removes all
young pulsars and other disk sources, and leaves only high-latitude sources
(predominantly active galactic nuclei).

\smallskip

In Fig.~\ref{fig:LgammaLradio}, we show the gamma-ray luminosity and the radio
pseudo-luminosity of high-latitude \Fermi\ MSPs from the 2PC
\citep{TheFermi-LAT:2013ssa}.  
In addition, we also include the MSPs PSR
J1816+4510, PSR J1311$-$3430, PSR J0610$-$2100, PSR J1903$-$7051 and PSR J1745+1017,
for which we take the gamma-ray fluxes from the 3FGL, and radio fluxes and
distance measures from \cite{Barr:2013qh, Camilo:2015caa, Pallanca:2012dc,
Ray:2012ms, Stovall:2014gua}. 
Almost all sources with $L_\gamma \geq 5\times
10^{33}\ergs$ have radio luminosities above around $0.5\rm\, mJy\, kpc^2$. This
is \emph{above} the median of our reference radio luminosity function
($0.3\mJy\,\rm kpc^2$).  Somewhat contrary to the conventional wisdom that
gamma-ray and radio luminosities are truly uncorrelated,
this does suggest a loose correlation between these
quantities.\footnote{ A simple estimate for the $p$-value for this happening by
chance can be obtained as $p\sim0.5^8\simeq0.004$, given that we have seven
sources, which corresponds to $2.8\sigma$.}  However, given the low number of
sources, little can be said about the nature of the correlation (\fex, whether
it is linear in log-log space, or whether it continues to lower luminosities).
We will for now take this observation at face value, and comment below in
Sec.~\ref{sec:discussions} how the results might change when any correlation is
neglected.

\smallskip

In order to estimate how many MSPs that are bright in gamma rays could have
remained undetected in radio, we show in Fig.~\ref{fig:pscs} high-latitude
MSPs, unassociated and other sources from the 3FGL, as a function of the
variability index and the curvature significance \citep[for definitions
see][]{TheFermi-LAT:2015hja}.  We only show sources that pass the flux
threshold that we discussed above.\footnote{Note that Fig.~\ref{fig:pscs} shows
31 MSPs, Fig.~\ref{fig:LgammaLradio} shows 23.  The 8 MSPs that are
missing in Fig.~\ref{fig:LgammaLradio} are either without radio detection (in
two cases) or the detected flux is not yet published.}
These parameters provide useful discriminators, and help to
separate pulsar-like sources from other sources at high latitudes, such as
active galactic nuclei.  One can clearly see that MSPs consistently have  a low
variability index (values below around 80 indicate non-variable sources), and
most of them feature a curved spectrum that leads to a large curvature
significance.  Many of the unassociated sources appear to be non-variable as
well, and a few of them feature high curvature significances.  On the other
hand, most of the remaining bright high-latitude sources are variable, since
the dominant fraction of the extragalactic sources is formed by (variable)
active galactic nuclei.

If we focus on the indicated region in Fig.~\ref{fig:pscs} with non-variable
sources and high curvature significance (lower-right corner), it is clear that
there is only little room for bright gamma-ray MSPs to `hide' as unassociated
sources.  The number of MSPs in that region could be at most a fraction
$\sim30\%$ larger with respect to what is already known.
These additional MSPs, which would not yet have shown up in radio searches,
could be potentially radio quiet, and weaken the above loose
gamma-ray/radio correlation.

\medskip

In order to model the radio luminosity of MSP candidates from \Fermi\
observations in a way that is motivated by actual radio observations, we adopt
the following simple strategy.  In 60\% of the cases, we will draw a random
radio luminosity from the nine MSPs in Fig.~\ref{fig:LgammaLradio} with a
gamma-ray luminosity $L_\gamma>5\times10^{33}\ergs$, since only such bright
sources would appear as MSP candidates associated with the bulge population. In
the other 40\% of the cases we will assume that radio luminosity is zero, to
account for fact that some or most of the unassociated sources could be
actually radio-dim MSPs, and for the fact that that some of the MSPs in
Fig.~\ref{fig:pscs} are either radio-quiet or have no published fluxes.  This
procedure is somewhat \emph{ad hoc}, but is completely data driven and should
give a reasonably accurate description of the detection prospects of MSP
candidates.  However, the uncertainties associated with this method are
certainly large, and likely affect the resulting detection probability by a
factor of roughly two (which we estimate from the typical
Poisson error associated with drawing from just nine sources).

\subsection{Detectability}

\begin{figure}
  \centering
    \includegraphics[width=0.90\columnwidth]{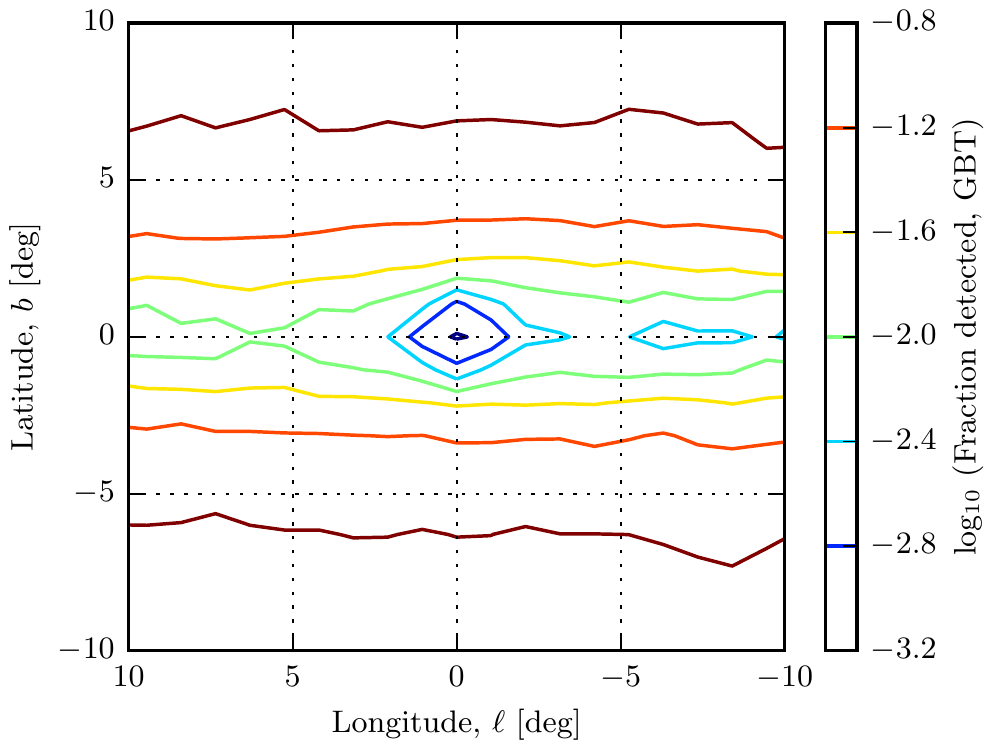}
    \caption{Fraction of gamma-ray bright bulge MSPs along the line-of-sight
    that can be detected with the GBT survey from Tab.~\ref{tab:telescopes}.
    See text for details of the empirically derived radio luminosity of the MSP
    population.}
    \label{fig:pGBT}
\end{figure}

In Fig.~\ref{fig:pGBT}, we show the detection probability of gamma-ray bright
bulge MSPs in different regions of the inner Galaxy, assuming that each source
is observed by the GBT as summarized in Tab.~\ref{tab:telescopes}.
We note that here we adopt integration time per pointing of 60 minutes for
all three observational scenarios (see below).  We adopt the
empirically derived radio luminosity function for gamma-ray bright MSPs as
discussed above, and calculate the probability that a bulge MSP along the
line-of-sight can be detected, weighted by the source density in the bulge and
the volume factor.

At high latitudes, the probability is nearly $10\%$, whereas close
to the Galactic disk it is well below $0.1\%$.  This already indicates that
follow-up observations of individual MSP candidates are rather challenging,
even if their position is known precisely.  This is true in particular close to
the Galactic disk. 

\begin{figure}
  \centering
    \includegraphics[width=0.90\columnwidth]{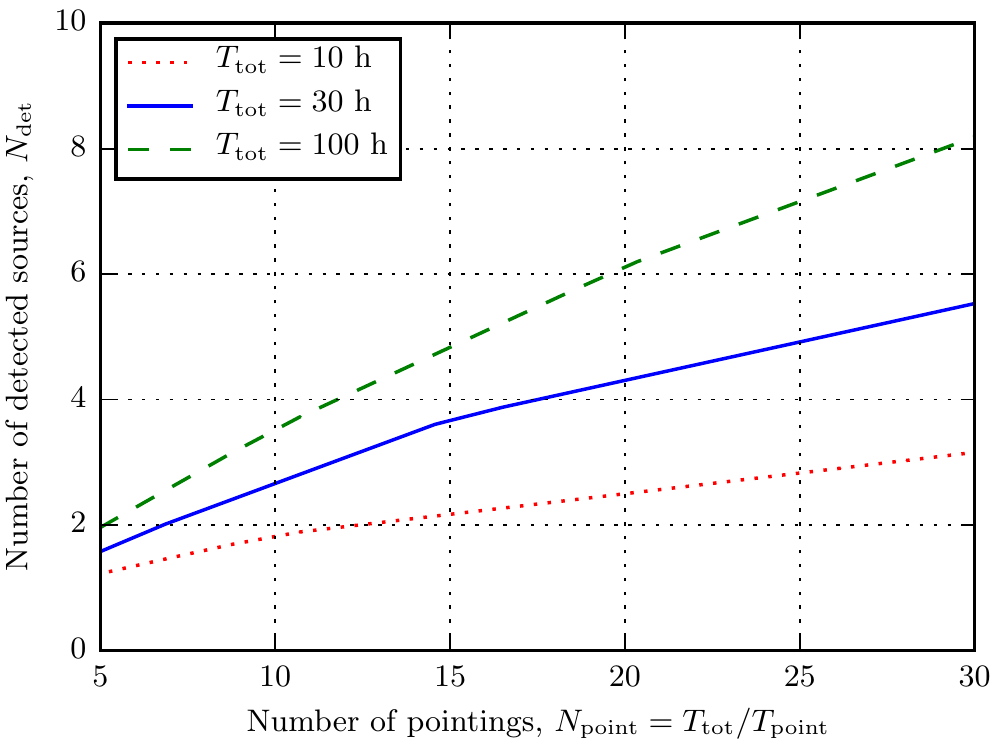} \\
    \caption{Number of detectable sources as a function of the number of
    targeted observations using GBT with total integration time of 10 hours
    (\emph{dotted red}), 30 hours (\emph{solid blue}) and 100 hours (\emph{dashed
  green}).}
    \label{fig:fractionDet}
\end{figure}

\begin{table}
  \centering
  \begin{tabular}{llrr}
    \hline\hline
    Instrument & $t_\text{obs}$& \multicolumn{2}{c} {Detection of MSP candidates}\\
               & total & Probability & Number (20 total) \\\hline
    GBT & $20\h$ & 18.4\%  & 3.7 \\
    MeerKAT & $20\h$ & 20.5\% & 4.1 \\
    SKA-mid & $20\h$  & 40.8\% & 8.2 \\
    \hline\hline
  \end{tabular}
  \caption{Projected number of detections for follow-up radio searches in 20
    MSP candidates, assuming that all of the MSP candidates are indeed
    gamma-ray luminous MSPs in the bulge region.  The radio luminosity of
    gamma-ray luminous MSPs is estimated from a flux limited sample of
    high-latitude MSPs and unassociated sources.  Although the results were
    obtained in an observation-driven approach, they are uncertain by at
    least a factor of two and of indicative value only.  Caveats are discussed
    in the text.
    }
  \label{tab:det13}
\end{table}

\smallskip

In order to get an estimate for the detection probability of a typical bulge
MSP candidate, we average the detection probability over the 13 reference 3FGL
sources from \cite{Bartels:2015aea}.  The resulting probabilities are
summarized in Tab.~\ref{tab:det13}, for the different observational scenarios
from Tab.~\ref{tab:telescopes}.  We find average probabilities of 18\% in the
case of GBT, which grow to 40\% in the case of SKA-mid.

\medskip

Our results indicate that, on a short timescale, radio follow-up observations
of MSP candidates with the GBT or similar instruments are the most promising
strategy to actually find the first MSPs from the bulge region.  The numbers in
Tab.~\ref{tab:det13} are very promising.  However, as mentioned above,
additional effects need to be taken into account that will further reduce the
detection probabilities.  Firstly, not every MSP candidate will correspond to
an MSP.  
This will reduce the number of possible detections by the likelihood for a
given MSP candidate to correspond to an MSP (probably by up to a factor of two, 
see~\cite{Bartels:2015aea}).
Secondly, source localization is critical.  The GBT
beam size of $0.14\deg$ FWHM is comparable to the localization accuracy that
can be reached with \Fermi\ at 68\% CL.  Hence, several pointings might be
necessary to fully cover the area in which the radio emission from an MSP
candidate could lie.  Both of the caveats need to be carefully taken into
consideration when planning actual observations.  
Furthermore, we note that targeted searches using long,
  60-min integration times have the additional issue that MSPs often
  reside in binary systems and Doppler smearing of the pulsed signal
  is difficult to correct in a blind search if the integration time is
  a significant fraction of the orbital period.
This
is further discussed in Sec.~\ref{sec:discussions}.

Finally, in Fig.~\ref{fig:fractionDet} we show the number of sources that will
be detectable with increasing GBT targeted observations for a fixed total
integration time.  In general, it is more promising to use a shorter dwell time
and allow more pointings. While with a total
integration time of 10 hours only a few sources, out of 30 pointings, can be
detected, a total integration time of 100 hours, distributed over 30 spots, in
the sky would enable the detection of about 8 sources.

\section{Discussion}
\label{sec:discussions}

The predicted radio emission of the MSP bulge population has to be consistent
with the results of existing pulsar radio surveys.  We will here concentrate on the
consistency with the Parkes HTRU mid-latitude survey, which covers latitudes in
the range $3.5^\circ < |b| < 15^\circ$, and hence regions of the sky that we
find to be the most promising for finding MSP bulge sources (at lower latitudes
scattering becomes increasingly important).  We find that, with the
configuration listed in Tab.~\ref{tab:telescopes}, the HTRU mid-latitude survey
should have detected around 7 MSPs from our reference bulge population and
luminosity function (`Model 3').  For the alternative luminosity functions
Model 1 (2) we find that 10 (4) bulge MSPs should have been seen.  

Interestingly, the HTRU mid-latitude survey has detected only one field MSP within
3 kpc of the Galactic center, J1755$-$3716 at 6.38 kpc distance~\citep{Ng:2014mca}
This source could be just on the edge of the bulge 
population.
This is on first sight slightly inconsistent with the number of bulge MSPs
that Parkes should have seen according to our above estimates.  For reasons
that we discuss next, we do not consider this discrepancy as severe, given that
the HTRU sensitivity is just scratching the brightest of the bulge MSP sources.
However, it is an indication that the bulge MSPs are in principle in reach of
current instruments.

There are a number of possible interpretations for the apparent non-observation
of a few bulge MSPs with Parkes HTRU.  The first possibility is that the bulge MSP
population has different properties than derived in this work, since it
\fex~does not fully account for the observed gamma-ray excess in the inner
Galaxy.  This is certainly a possibility, but the inconsistency between Parkes HTRU 
predicted and actual detected sources is not
strong enough to make definitive statements here (this would likely change if
future surveys do not find bulge MSPs either).  Another concern might be that
we overestimate the sensitivity of the Parkes HTRU.  This seems unlikely as our
faintest simulated sources detected with Parkes HTRU (mid-latitude) have fluxes
around $0.18 \mJy$, which is compatible with the faintest measured MSPs with
Parkes~\citep{2013MNRAS.434.1387L}.  However, given that estimates of detection
thresholds are very sensitive to a large number of parameters, we cannot
exclude this possibility.

It could be that the radio luminosity function of bulge MSPs is significantly
different from what is observed in globular clusters.  Given the possibly
different formation histories of MSPs in globular clusters and the bulge, this
cannot be excluded.  Lastly, it could be that a number of bulge sources were
already discovered by the Parkes HTRU, but the DM-based distance measure is
biased to lower values such that the MSPs appear closer and less luminous than
they actually are.

We emphasize that most of the above caveats related to the sensitivity of the
Parkes HTRU do not directly apply to the other reference surveys from
Tab.~\ref{tab:telescopes}.  Already observations with the GBT will probe
significantly fainter sources, which reduces the dependence on the details
of the radio luminosity function in the bright tail.  Indeed, we find that the
number of sources detectable by the GBT for Model (1, 2, 3) is (162, 127,151),
and hence varies by less than $15\%$ (see Tab.~\ref{tab:RadioBright})
from our reference result.  However, a possible bias of DM-based distance
measures cannot be excluded and would also affect results by the GBT and other
instruments.

\medskip

About three quarters of all field MSPs are bound in \emph{binary systems}, with
orbital periods ranging from 94 min to hundreds of days \citep{Stovall:2013gca,
Stovall:2014gua}.  Given the many free orbital parameters, the induced Doppler
shift in the observed pulse period can make an identification of the pulsation
extremely difficult because it smears out the periodic signal in the Fourier
domain.  Using acceleration search techniques
\citep[\fex][]{2001PhDT.......123R}, it is possible to compensate for orbital
motion; however, such techniques are only sensitive in cases where the
observing dwell time is less than about a tenth of the orbital period.  As
such, this imposes a practical limitation to the beneficial dwell time per sky
pointing.  

Although the observation time per pointing in our described targeted searches
are comparable to the smallest observed orbital period, which would cause
problems for our reference searches, most other observed orbital periods are
much larger, and we do not expect a very strong effect on our results.  As we
discussed above, orbits that are at least ten times longer than the dwell time
per survey pointing should be enough.

\medskip

Conventionally it is assumed that gamma-ray and radio luminosities are
uncorrelated.  However, we showed that high-latitude gamma-ray MSPs and
unassociated \Fermi\ sources suggest a loose gamma-ray/radio correlation.  We
used this relation when estimating the radio detection probabilities for bright
gamma-ray MSPs in the bulge.  If we would neglect this correlation, and assume
instead that a given MSP candidate source has a radio luminosity that is
randomly drawn from our reference luminosity function `Model 3', the detection
prospects in the case of, \fex, GBT in Tab.~\ref{tab:det13} would reduce from
$\sim 18\%$ to $<10\%$.  Hence, the presence or absence of a gamma-ray/radio
correlation has a significant impact on the prospects for radio follow-up
searches for MSP candidates.  In this context, we emphasize that if there are
only a few dozen MSP candidates, then searching each one for 1 hour or more
would still take much less time than blindly
  searching the dozens of square degrees of sky needed to potentially
  lead to the same number of MSP detections.

\medskip

From Fig.~\ref{fig:compare} it is clear that the main limitation to the
detection is scattering.  In principle, this can be mitigated by observing 
higher frequencies, since the scattering time roughly scales with $\nu^{-4.4}$.
However, the price for this lower scattering time is a reduced signal flux
because of the steep source spectrum.  We use $\alpha_\nu$ = 1.7 as spectral
index to rescale the flux density from one frequency to another, with flux
density $S_\nu \propto \nu^{-\alpha_\nu}$.  This is in agreement with the average value
found for MSPs~\citep{1998ApJ...501..270K,2000A&AS..147..195M} ($\alpha_\nu$ =
1.6--1.8), while \cite{Bates:2013ear} found $\alpha_\nu$ = 1.4 for slowly
rotating pulsars.  

\begin{figure*}
  \centering
  \includegraphics[width=0.33\linewidth]{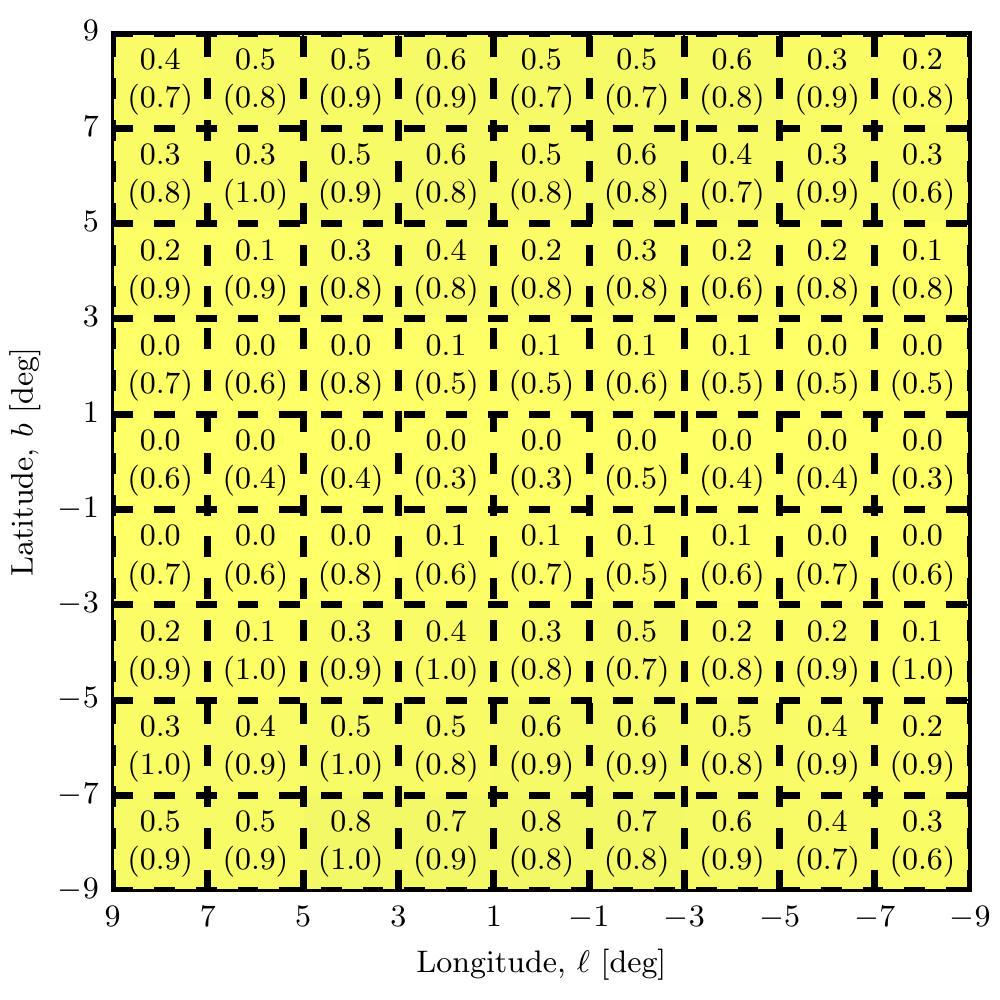}
  \includegraphics[width=0.33\linewidth]{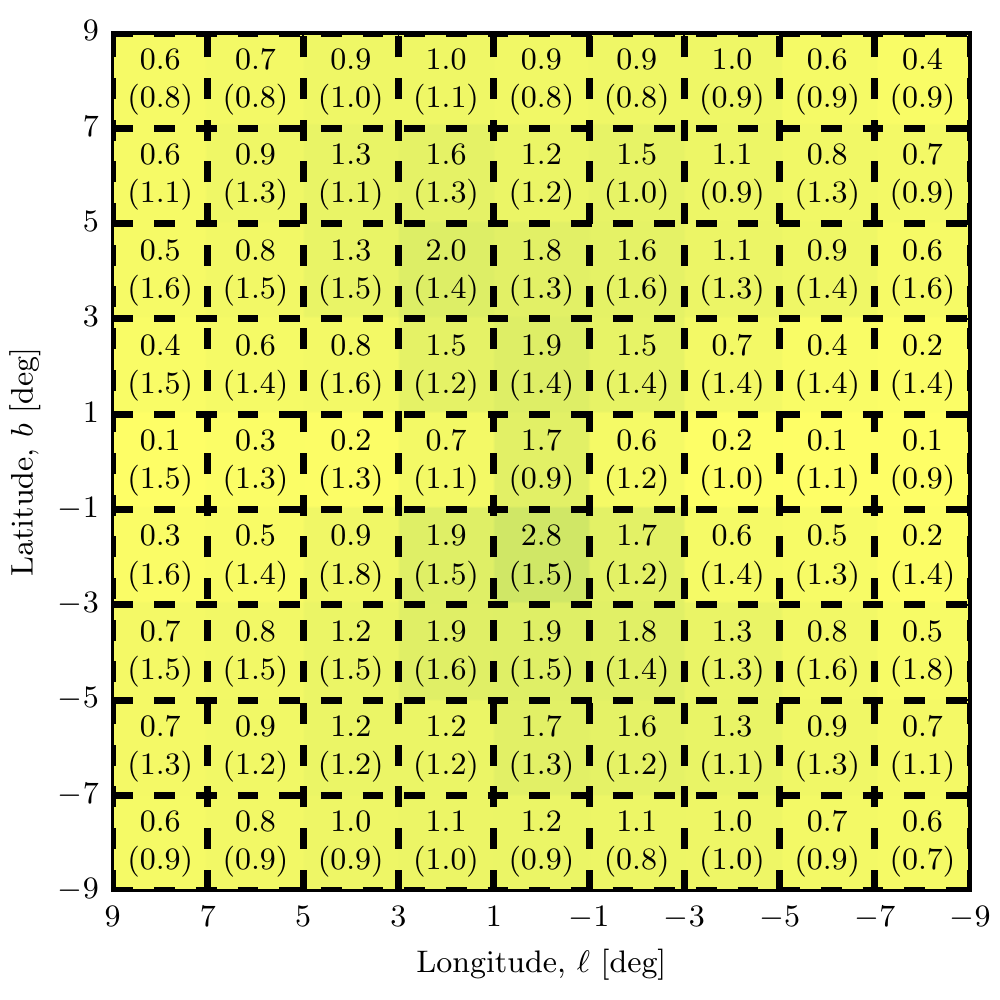}
  \includegraphics[width=0.33\linewidth]{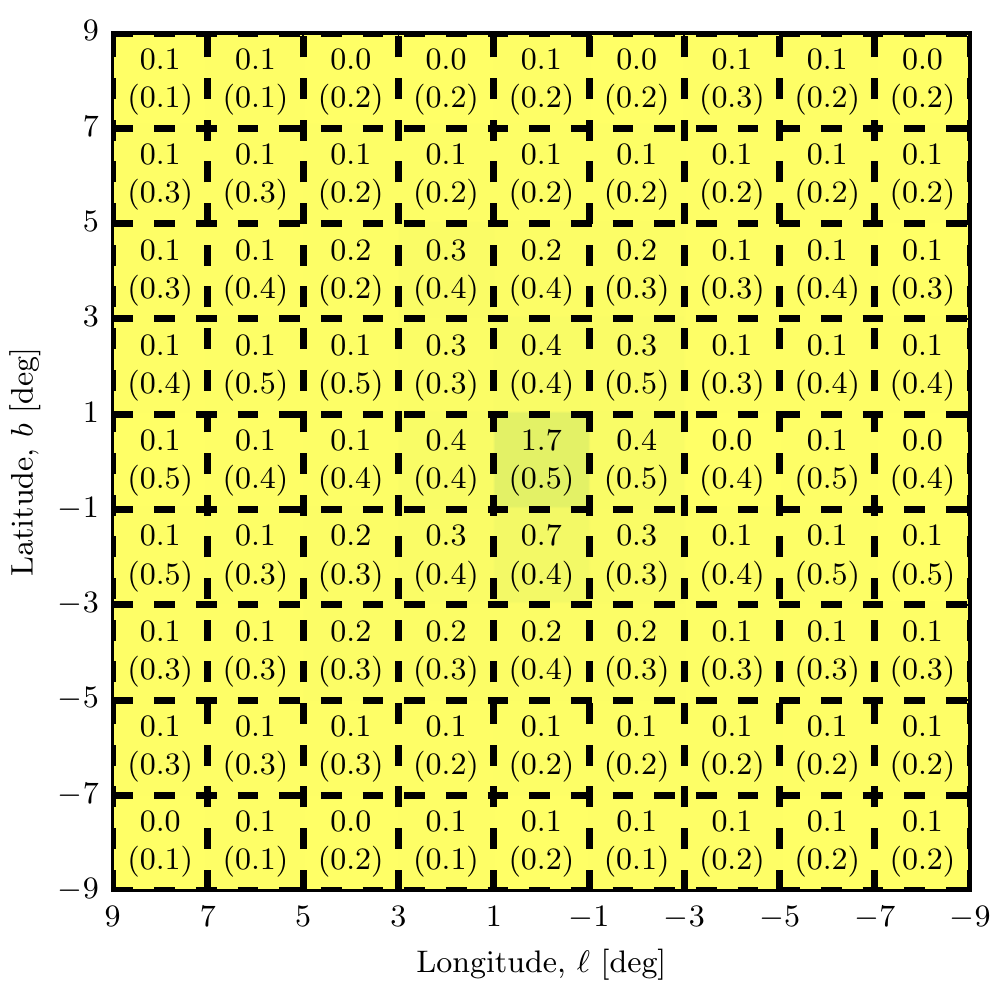}
  \caption{Same as Fig.~\ref{fig:sensGBT}, but for a survey performed with the
  GBT at 850 MHz (left panel), 2 GHz (central panel) or 5 GHz (right panel).}
  \label{fig:sensGBT_XXX}
\end{figure*}

In Fig.~\ref{fig:sensGBT_XXX} we show the detectability predictions for GBT
observations at 850 MHz, 2 GHz and 5 GHz respectively.  While at 850 MHz the
effect of scattering prevents the detection of sources in the inner region of
the Galaxy and, in particular, along the Galactic plane, 2 GHz turns out to be
probably the optimal frequency for large area surveys at mid- and
low-latitudes\footnote{We mention however that wide-area surveying at 
2 GHz is more challenging, because the beam is even smaller than at 1.4 GHz.}.  
Indeed, at 2 GHz, on the one hand, the relevance of scattering
is reduced with respect to 1.4 GHz (as seen by comparing the number of sources
detected in the Galactic plane and in the sub-region around the Galactic
center) and, on the other hand, the reduction of the signal flux is not as
relevant as at 5 GHz. At 5 GHz, indeed, the number of sources that can be
detected with the same observation time is much smaller than the number of
sources detectable at 2 GHz for all $2^{\circ} \times 2^{\circ}$ sub-regions.
The only exception is the region centered on the Galactic center, where the
effect of scattering is still relevant, in agreement with the latest works considering 
the detectabiliy of MSPs at the Galactic center~\citep{Macquart:2015jfa}.  However,
these central sources are only detectable if they lie in the low-scattering
tail of the scattering-time-DM relation.  This, and hence the detection
prospects in the inner 1 deg, are very uncertain. 
We note that past radio surveys of the GC region at high frequencies 
\citep{[see for example][]{2009ApJ...702L.177D,2006MNRAS.373L...6J}}
were intended to find pulsars at the GC, in the very inner degree or less,
with a very narrow field of view, and thus they were not sensitive to MSPs detection,
as explained in~\cite{Macquart:2015jfa}.

\medskip

As described above, we assume that \emph{all} of the gamma-ray emission from
the considered globular clusters comes from MSPs.  In the case of, \fex, NGC
6440, which contains a young pulsar that is very bright in radio, it could be
that the dominant part of the observed gamma-ray emission is actually due to
this young pulsar, or another source along the
line-of-sight~\citep{collaboration:2010bb}.  In that case, namely if we neglect
NGC 6440 with its very high gamma-ray luminosity in our analysis, our estimate
in Eq.~\eqref{eqn:R} would systematically \emph{decrease}.  This would then
\emph{increase} the number of predicted radio-bright MSPs in the bulge, in the
case at hand, by a factor of 1.5 and thus make our predictions more optimistic.

\medskip

Finally, we comment on another relevant wavelength for MSP studies, namely
X-rays.  The observation of MSPs in the X-ray band has been pursued by several
experiments in the past, and recently by the {\it Chandra} and {\it XMM-Newton}
observatories. Up to now,  62 MSPs (with period $P<20$ ms) have been detected
\citep{2015arXiv151107713P}.  MSPs are very faint X-ray sources with typical
luminosities ranging from $L_X \sim 10^{30}-10^{31} {\rm erg} \; {\rm s}^{-1}$.
For this reason, their detection in the X-ray band is challenging, and requires
very deep exposures.  A large fraction of the MSPs detected in X-rays belongs
to globular clusters \citep{2006ApJ...646.1104B}.  In general, no systematic
differences exist between MSPs in globular clusters and those in the field of
the Galaxy \citep{2006ApJ...646.1104B}.  MSPs around the Galactic center are
very difficult to probe via soft X-rays (0.5 -- 2 keV), since their faint emission would be
mostly absorbed by the intervening material.  The hard spectral component
could be seen by {\it NuSTAR}, which in turn suffers from poor angular
resolution and makes it difficult to determine whether the source is an MSP
\citep{2015Natur.520..646P}.  The need for very deep exposures combined with
the typical angular resolution of current X-ray observatories (i.e. 0.5 arcsec
for {\it Chandra} and 6 arcsec for {\it XMM-Newton}) makes the exploration of a
single $2^\circ\times2^\circ$ sky area (e.g., see Fig. \ref{fig:sensGBT})
very time consuming. The discovery of a bulge population by means of X-ray
campaigns seems therefore unfavored with respect to present day and next
generation radio telescopes.

\section{Conclusions}
\label{sec:conclusions}

It has been proposed that the extended excess of GeV photons that was found in
\Fermi-LAT data from the inner Galaxy is caused by the combined emission of a
large number of hitherto undetected MSPs in the Galactic bulge.  We presented
the first comprehensive study of the prospects for detecting radio pulsations
from this new MSP population.  Based on observations of globular clusters,
which we consider as \emph{versions in miniature} of the MSP bulge population,
we constructed a radio emission model for the bulge population as a whole.  We
found a loose correlation between the gamma-ray and radio emission of
individual sources in a flux-limited sample of high-latitude \Fermi\ MSPs and
unassociated sources.  We quantitatively showed how existing radio pulsar
surveys are not quite sensitive enough to detect a first sample of MSPs from the
bulge population.  Finally, we discussed in detail how future \emph{deep
targeted searches} as well as \emph{large area surveys} can detect the bulge
MSPs as a distinct population with high confidence in the upcoming years.  Our
main findings can be summarized as follows.

\medskip \noindent
(1) \Fermi-LAT data from the inner Galaxy suggests that around $\sim 3000$
radio-bright MSPs ($S_{1.4 \GHz} > 10\muJy$) are present as distinct population
in the Galactic bulge.

\smallskip
Our estimates are based on an extrapolation of the gamma-ray and radio emission
of six globular clusters.  The largest uncertainties come from the details of
diffuse gamma-ray emission from the inner $200\pc$ of the Galactic center, and
the actual spatial extent of the MSP bulge population beyond $1.5\kpc$.

\medskip \noindent
(2) The expected \emph{surface density} of radio-bright bulge MSPs a few
degrees above and below the Galactic center can be determined with good
accuracy.

\smallskip
For instance, at Galactic longitudes $\ell\simeq 0^\circ$ and latitudes
$|b|\simeq 5^\circ$ we predict a surface density of radio-bright bulge MSPs of
$(4.7\pm 1.5)\deg^{-2}$.  This quoted error takes into account uncertainties
related to the radio luminosity function, sampling variance of the relatively
small numbers of MSPs in globular clusters, the diffuse gamma-ray emission from
the inner Galaxy and the gamma-ray emission from globular clusters.  Closer to
the Galactic center the surface density becomes much higher (but so do the
challenges of finding millisecond radio pulsations).

\medskip \noindent
(3) We find that frequencies around $1.4\GHz$ are best for radio pulsation
searches for bulge MSPs at mid-latitudes.  The effects of scatter-broadening
at these frequencies are rather large in the Galactic plane.  Detection
prospects are hence best at intermediate Galactic latitudes, $2^\circ \lesssim
|b|\lesssim8^\circ$. 

\smallskip
Due to broadening from scattering, observations at lower frequencies (850 MHz) yield
in general a worse result, whereas observations at 5 GHz suffer from the
pulsar's intrinsically decreased flux.  Optimal frequencies are in the range 1.4--2.0~GHz.  At
intermediate latitudes, the most sensitive large area survey in the inner
Galaxy is the Parkes HTRU survey at $1.4\GHz$.  The brightest bulge MSPs with a
few hundred $\mu$Jy just scratch the sensitivity of this survey, which is
consistent with current results.

\medskip \noindent
(4) Deep targeted observations of \Fermi\ unassociated sources at mid-latitudes
with the GBT, and with integration times per pointing of around one hour, can
likely lead to the first discoveries of bulge MSPs.

\smallskip
We show that \Fermi\ observations of nearby MSPs and bright unassociated
sources at high Galactic latitudes suggest a loose but significant correlation
between the MSP gamma-ray and radio luminosities.  Taking this
  relation into account, we estimate that there is roughly an 18\%
  probability (with uncertainties of at least a factor of two) that a
  1-hour deep observation with GBT at 1.4 GHz could detect a bulge MSP
  that is seen in gamma rays.  The
success of such a targeted campaign will crucially depend on the careful
preparation of a list of promising targets.  

\medskip \noindent
(5) In the upcoming years, large area surveys using, \fex, MeerKAT and later
SKA, can cover hundred square degrees within a hundred hours of observation time,
and they should find dozens to hundreds of bulge MSPs, both in the inner few
degrees of the Galactic center and up $10^\circ$ Galactic latitude or more. 

\smallskip
Thanks to the much larger field-of-view and gain, the prospects for detecting a
large number of bulge MSPs with upcoming radio telescopes are excellent.  The
largest limitation of these searches will likely not directly come from the
instrumental capabilities, but from the enormous computing time required to
process all recorded data.

\medskip \noindent
(6) We showed that, for observations a few degrees off the Galactic plane, the
detection of\, $\gtrsim 4$ MSPs with a DM $\sim$ 300--400 pc cm$^{-3}$ at
latitudes around $|b|\sim5^\circ$ could already be enough to detect the bulge
component above the thick-disk MSP population with high statistical
significance.

\smallskip
The bulge MSP population would increase the number of MSPs that are detectable
at $7$--$10\kpc$ distances in the inner Galaxy by a large factor with respect
to the expectations from only a thick-disk population, and hence at
mid-latitudes easily identifiable as a distinct population.  However, due to the
large scatter broadening, even with SKA it will remain rather challenging to
detect bulge MSPs in the inner $1\deg$ of the Galactic center (although a few
sources might lie along lines-of-sights with reduced scattering).  It is hence
rather likely that in the foreseeable future the \Fermi\ observations of
diffuse gamma rays from the Galactic center will continue to provide the best
(though somewhat indirect) constraints on a possible MSP bulge population in
the inner $\sim200\pc$ of the Galactic center.

\medskip

\emph{In summary}, if the \Fermi\ GeV excess is indeed due to a population of
MSPs in the Galactic bulge, the first discovery of this bulge population could
be achieved with current technology in the next couple of years.  Such a
discovery would likely be based on targeted radio searches in \Fermi\
unassociated sources, or source candidates just below the 3FGL threshold.  It
is hence now most pressing to build a list of the most promising targets from
\Fermi\ gamma-ray data, with reliable probabilistic statements about possible
source types.  

In the more distant future, on the time scale of at least five years and more,
large area surveys with upcoming radio instruments should start to detect many
dozens or even hundreds of bulge MSPs.  The scientific implications of such
detections would be significant.  They would allow a systematic study of a
potentially very large sample of field MSPs in the bulge, of their gamma-ray
and radio emission properties, and of their formation history.  They would
clarify the origin of the long-debated \Fermi\ GeV excess, and allow to
disentangle emission from unresolved point sources from the truly diffuse
emission from the Galactic bulge, with possible contributions from the \Fermi\
bubbles, the activity of the supermassive black hole, or even a signal from
dark matter annihilation.  Lastly, they would open a completely new window
for the systematic study of the formation history of the Galactic bulge and
center and the objects that they contain.

\begin{acknowledgments}
  \paragraph{Acknowledgments.}
  We very warmly acknowledge discussions with Francesco Massaro about
  multi-wavelength associations of unassociated \Fermi\ sources.  We
  furthermore acknowledge useful discussions with Jonathan E.~Grindlay, Tim
  Linden, Scott Ransom, Marco Regis, Pasquale~D.~Serpico and Meng Su in
  different stages of the project.  J.W.T.H. acknowledges funding from an NWO
  Vidi fellowship and from the European Research Council under the European
  Union's Seventh Framework Programme (FP/2007-2013) / ERC Starting Grant
  agreement nr. 337062 (``DRAGNET").  F.C.~and C.W.~acknowledge funding from an
  NWO Vidi fellowship.
\end{acknowledgments}

\bibliography{paper}

\begin{thebibliography}{94}
\expandafter\ifx\csname natexlab\endcsname\relax\def\natexlab#1{#1}\fi

\bibitem[{Abazajian(2011)}]{Abazajian:2010zy}
Abazajian, K.~N. 2011, JCAP, 1103, 010

\bibitem[{Abazajian {et~al.}(2014)Abazajian, Canac, Horiuchi, \&
  Kaplinghat}]{Abazajian:2014fta}
Abazajian, K.~N., Canac, N., Horiuchi, S., \& Kaplinghat, M. 2014, Phys.Rev.,
  D90, 023526

\bibitem[{Abazajian \& Kaplinghat(2012)}]{Abazajian:2012pn}
Abazajian, K.~N., \& Kaplinghat, M. 2012, Phys.Rev., D86, 083511

\bibitem[{Abdo {et~al.}(2010)}]{collaboration:2010bb}
Abdo, A., {et~al.} 2010

\bibitem[{Abdo {et~al.}(2013)}]{TheFermi-LAT:2013ssa}
---. 2013, Astrophys.J.Suppl., 208, 17

\bibitem[{Acero {et~al.}(2015)}]{TheFermi-LAT:2015hja}
Acero, F., {et~al.} 2015, Astrophys.J.Suppl., 218, 23

\bibitem[{{Aharonian} {et~al.}(2004){Aharonian}, {Akhperjanian}, {Aye},
  {Bazer-Bachi}, {Beilicke}, {Benbow}, {Berge}, {Berghaus}, {Bernl{\"o}hr},
  {Bolz}, {Boisson}, {Borgmeier}, {Breitling}, {Brown}, {Bussons Gordo},
  {Chadwick}, {Chitnis}, {Chounet}, {Cornils}, {Costamante}, {Degrange},
  {Djannati-Ata{\"i}}, {O'C.~Drury}, {Ergin}, {Espigat}, {Feinstein}, {Fleury},
  {Fontaine}, {Funk}, {Gallant}, {Giebels}, {Gillessen}, {Goret}, {Guy},
  {Hadjichristidis}, {Hauser}, {Heinzelmann}, {Henri}, {Hermann}, {Hinton},
  {Hofmann}, {Holleran}, {Horns}, {de Jager}, {Jung}, {Kh{\'e}lifi}, {Komin},
  {Konopelko}, {Latham}, {Le Gallou}, {Lemoine}, {Lemi{\`e}re}, {Leroy},
  {Lohse}, {Marcowith}, {Masterson}, {McComb}, {de Naurois}, {Nolan},
  {Noutsos}, {Orford}, {Osborne}, {Ouchrif}, {Panter}, {Pelletier}, {Pita},
  {Pohl}, {P{\"u}hlhofer}, {Punch}, {Raubenheimer}, {Raue}, {Raux}, {Rayner},
  {Redondo}, {Reimer}, {Reimer}, {Ripken}, {Rivoal}, {Rob}, {Rolland},
  {Rowell}, {Sahakian}, {Saug{\'e}}, {Schlenker}, {Schlickeiser}, {Schuster},
  {Schwanke}, {Siewert}, {Sol}, {Steenkamp}, {Stegmann}, {Tavernet},
  {Th{\'e}oret}, {Tluczykont}, {van der Walt}, {Vasileiadis}, {Vincent},
  {Visser}, {V{\"o}lk}, \& {Wagner}}]{2004A&A...425L..13A}
{Aharonian}, F., {Akhperjanian}, A.~G., {Aye}, K.-M., {Bazer-Bachi}, A.~R.,
  {Beilicke}, M., {Benbow}, W., {Berge}, D., {Berghaus}, P., {Bernl{\"o}hr},
  K., {Bolz}, O., {Boisson}, C., {Borgmeier}, C., {Breitling}, F., {Brown},
  A.~M., {Bussons Gordo}, J., {Chadwick}, P.~M., {Chitnis}, V.~R., {Chounet},
  L.-M., {Cornils}, R., {Costamante}, L., {Degrange}, B., {Djannati-Ata{\"i}},
  A., {O'C.~Drury}, L., {Ergin}, T., {Espigat}, P., {Feinstein}, F., {Fleury},
  P., {Fontaine}, G., {Funk}, S., {Gallant}, Y., {Giebels}, B., {Gillessen},
  S., {Goret}, P., {Guy}, J., {Hadjichristidis}, C., {Hauser}, M.,
  {Heinzelmann}, G., {Henri}, G., {Hermann}, G., {Hinton}, J.~A., {Hofmann},
  W., {Holleran}, M., {Horns}, D., {de Jager}, O.~C., {Jung}, I.,
  {Kh{\'e}lifi}, B., {Komin}, N., {Konopelko}, A., {Latham}, I.~J., {Le
  Gallou}, R., {Lemoine}, M., {Lemi{\`e}re}, A., {Leroy}, N., {Lohse}, T.,
  {Marcowith}, A., {Masterson}, C., {McComb}, T.~J.~L., {de Naurois}, M.,
  {Nolan}, S.~J., {Noutsos}, A., {Orford}, K.~J., {Osborne}, J.~L., {Ouchrif},
  M., {Panter}, M., {Pelletier}, G., {Pita}, S., {Pohl}, M., {P{\"u}hlhofer},
  G., {Punch}, M., {Raubenheimer}, B.~C., {Raue}, M., {Raux}, J., {Rayner},
  S.~M., {Redondo}, I., {Reimer}, A., {Reimer}, O., {Ripken}, J., {Rivoal}, M.,
  {Rob}, L., {Rolland}, L., {Rowell}, G., {Sahakian}, V., {Saug{\'e}}, L.,
  {Schlenker}, S., {Schlickeiser}, R., {Schuster}, C., {Schwanke}, U.,
  {Siewert}, M., {Sol}, H., {Steenkamp}, R., {Stegmann}, C., {Tavernet}, J.-P.,
  {Th{\'e}oret}, C.~G., {Tluczykont}, M., {van der Walt}, D.~J., {Vasileiadis},
  G., {Vincent}, P., {Visser}, B., {V{\"o}lk}, H.~J., \& {Wagner}, S.~J. 2004,
  \aap, 425, L13

\bibitem[{{Aharonian} {et~al.}(1997){Aharonian}, {Atoyan}, \&
  {Kifune}}]{1997MNRAS.291..162A}
{Aharonian}, F.~A., {Atoyan}, A.~M., \& {Kifune}, T. 1997, \mnras, 291, 162

\bibitem[{Ajello {et~al.}(2015)}]{TheFermi-LAT:2015kwa}
Ajello, M., {et~al.} 2015

\bibitem[{{Alpar} {et~al.}(1982){Alpar}, {Cheng}, {Ruderman}, \&
  {Shaham}}]{1982Natur.300..728A}
{Alpar}, M.~A., {Cheng}, A.~F., {Ruderman}, M.~A., \& {Shaham}, J. 1982, \nat,
  300, 728

\bibitem[{{Arca-Sedda} \& {Capuzzo-Dolcetta}(2014)}]{2014MNRAS.444.3738A}
{Arca-Sedda}, M., \& {Capuzzo-Dolcetta}, R. 2014, \mnras, 444, 3738

\bibitem[{Bagchi {et~al.}(2011)Bagchi, Lorimer, \&
  Chennamangalam}]{Bagchi:2011hs}
Bagchi, M., Lorimer, D.~R., \& Chennamangalam, J. 2011,
  Mon.Not.Roy.Astron.Soc., 418, 477

\bibitem[{Barr {et~al.}(2013)}]{Barr:2013qh}
Barr, E.~D., {et~al.} 2013, Mon. Not. Roy. Astron. Soc., 429, 1633

\bibitem[{Bartels {et~al.}(2015)Bartels, Krishnamurthy, \&
  Weniger}]{Bartels:2015aea}
Bartels, R., Krishnamurthy, S., \& Weniger, C. 2015

\bibitem[{Bates {et~al.}(2013)Bates, Lorimer, \& Verbiest}]{Bates:2013ear}
Bates, S.~D., Lorimer, D.~R., \& Verbiest, J. P.~W. 2013, Mon. Not. Roy.
  Astron. Soc., 431, 1352

\bibitem[{{Bednarek} \& {Sobczak}(2013)}]{2013MNRAS.435L..14B}
{Bednarek}, W., \& {Sobczak}, T. 2013, \mnras, 435, L14

\bibitem[{{Bhat} {et~al.}(2004){Bhat}, {Cordes}, {Camilo}, {Nice}, \&
  {Lorimer}}]{2004ApJ...605..759B}
{Bhat}, N.~D.~R., {Cordes}, J.~M., {Camilo}, F., {Nice}, D.~J., \& {Lorimer},
  D.~R. 2004, \apj, 605, 759

\bibitem[{{Bhattacharya} \& {van den Heuvel}(1991)}]{1991PhR...203....1B}
{Bhattacharya}, D., \& {van den Heuvel}, E.~P.~J. 1991, \physrep, 203, 1

\bibitem[{{Bogdanov} {et~al.}(2006){Bogdanov}, {Grindlay}, {Heinke}, {Camilo},
  {Freire}, \& {Becker}}]{2006ApJ...646.1104B}
{Bogdanov}, S., {Grindlay}, J.~E., {Heinke}, C.~O., {Camilo}, F., {Freire},
  P.~C.~C., \& {Becker}, W. 2006, \apj, 646, 1104

\bibitem[{Brandt \& Kocsis(2015)}]{Brandt:2015ula}
Brandt, T.~D., \& Kocsis, B. 2015, Astrophys. J., 812, 15

\bibitem[{Calore {et~al.}(2015{\natexlab{a}})Calore, Cholis, McCabe, \&
  Weniger}]{Calore:2014nla}
Calore, F., Cholis, I., McCabe, C., \& Weniger, C. 2015{\natexlab{a}}, Phys.
  Rev., D91, 063003

\bibitem[{Calore {et~al.}(2015{\natexlab{b}})Calore, Cholis, \&
  Weniger}]{Calore:2014xka}
Calore, F., Cholis, I., \& Weniger, C. 2015{\natexlab{b}}, JCAP, 1503, 038

\bibitem[{Calore {et~al.}(2014)Calore, Di~Mauro, Donato, \&
  Donato}]{Calore:2014oga}
Calore, F., Di~Mauro, M., Donato, F., \& Donato, F. 2014, Astrophys.J., 796, 1

\bibitem[{{Camilo} {et~al.}(2000){Camilo}, {Lorimer}, {Freire}, {Lyne}, \&
  {Manchester}}]{2000ApJ...535..975C}
{Camilo}, F., {Lorimer}, D.~R., {Freire}, P., {Lyne}, A.~G., \& {Manchester},
  R.~N. 2000, \apj, 535, 975

\bibitem[{Camilo {et~al.}(2015)}]{Camilo:2015caa}
Camilo, F., {et~al.} 2015, Astrophys. J., 810, 85

\bibitem[{Campana {et~al.}(2008)Campana, Stella, \& Kennea}]{Campana:2008vf}
Campana, S., Stella, L., \& Kennea, J.~A. 2008, Astrophys. J., 684, L99

\bibitem[{Carlson {et~al.}(2015)Carlson, Linden, \& Profumo}]{Carlson:2015ona}
Carlson, E., Linden, T., \& Profumo, S. 2015

\bibitem[{Carlson \& Profumo(2014)}]{Carlson:2014cwa}
Carlson, E., \& Profumo, S. 2014, Phys. Rev., D90, 023015

\bibitem[{{Cheng} {et~al.}(2004){Cheng}, {Taam}, \&
  {Wang}}]{2004ApJ...617..480C}
{Cheng}, K.~S., {Taam}, R.~E., \& {Wang}, W. 2004, \apj, 617, 480

\bibitem[{{Chevalier}(2000)}]{2000ApJ...539L..45C}
{Chevalier}, R.~A. 2000, \apjl, 539, L45

\bibitem[{Cholis {et~al.}(2015)Cholis, Evoli, Calore, Linden, Weniger, \&
  Hooper}]{Cholis:2015dea}
Cholis, I., Evoli, C., Calore, F., Linden, T., Weniger, C., \& Hooper, D. 2015

\bibitem[{Cholis {et~al.}(2014)Cholis, Hooper, \& Linden}]{Cholis:2014noa}
Cholis, I., Hooper, D., \& Linden, T. 2014

\bibitem[{{Condon} {et~al.}(1998){Condon}, {Cotton}, {Greisen}, {Yin},
  {Perley}, {Taylor}, \& {Broderick}}]{1998AJ....115.1693C}
{Condon}, J.~J., {Cotton}, W.~D., {Greisen}, E.~W., {Yin}, Q.~F., {Perley},
  R.~A., {Taylor}, G.~B., \& {Broderick}, J.~J. 1998, \aj, 115, 1693

\bibitem[{Cordes \& Lazio(2002)}]{Cordes:2002wz}
Cordes, J.~M., \& Lazio, T. J.~W. 2002

\bibitem[{{Cowan} {et~al.}(2011){Cowan}, {Cranmer}, {Gross}, \&
  {Vitells}}]{Cowan11}
{Cowan}, G., {Cranmer}, K., {Gross}, E., \& {Vitells}, O. 2011, Eur. Phys. J.
  C, 71, 1554

\bibitem[{{Daylan} {et~al.}(2016){Daylan}, {Finkbeiner}, {Hooper}, {Linden},
  {Portillo}, {Rodd}, \& {Slatyer}}]{Daylan:2014rsa}
{Daylan}, T., {Finkbeiner}, D.~P., {Hooper}, D., {Linden}, T., {Portillo},
  S.~K.~N., {Rodd}, N.~L., \& {Slatyer}, T.~R. 2016, Physics of the Dark
  Universe, 12, 1

\bibitem[{{Deneva} {et~al.}(2009){Deneva}, {Cordes}, \&
  {Lazio}}]{2009ApJ...702L.177D}
{Deneva}, J.~S., {Cordes}, J.~M., \& {Lazio}, T.~J.~W. 2009, \apjl, 702, L177

\bibitem[{{Dewey} {et~al.}(1984){Dewey}, {Stokes}, {Segelstein}, {Taylor}, \&
  {Weisberg}}]{1984bens.work..234D}
{Dewey}, R., {Stokes}, G., {Segelstein}, D., {Taylor}, J., \& {Weisberg}, J.
  1984, in Birth and Evolution of Neutron Stars: Issues Raised by Millisecond
  Pulsars, ed. S.~P. {Reynolds} \& D.~R. {Stinebring}, 234

\bibitem[{{Faucher-Gigu{\`e}re} \& {Kaspi}(2006)}]{2006ApJ...643..332F}
{Faucher-Gigu{\`e}re}, C.-A., \& {Kaspi}, V.~M. 2006, \apj, 643, 332

\bibitem[{{Faucher-Gigu{\`e}re} \& {Loeb}(2010)}]{2010JCAP...01..005F}
{Faucher-Gigu{\`e}re}, C.-A., \& {Loeb}, A. 2010, \jcap, 1, 5

\bibitem[{Gaggero {et~al.}(2015)Gaggero, Urbano, Valli, \&
  Ullio}]{Gaggero:2014xla}
Gaggero, D., Urbano, A., Valli, M., \& Ullio, P. 2015, Phys. Rev., D91, 083012

\bibitem[{Gillessen {et~al.}(2009)}]{Gillessen:2008qv}
Gillessen, S., {et~al.} 2009, Astrophys. J., 692, 1075

\bibitem[{{Gnedin} {et~al.}(2014){Gnedin}, {Ostriker}, \&
  {Tremaine}}]{2014ApJ...785...71G}
{Gnedin}, O.~Y., {Ostriker}, J.~P., \& {Tremaine}, S. 2014, \apj, 785, 71

\bibitem[{Goodenough \& Hooper(2009)}]{Goodenough:2009gk}
Goodenough, L., \& Hooper, D. 2009

\bibitem[{Gordon \& Macias(2013)}]{Gordon:2013vta}
Gordon, C., \& Macias, O. 2013, Phys.Rev., D88, 083521

\bibitem[{{Grenier} \& {Harding}(2015)}]{Grenier:2015pya}
{Grenier}, I.~A., \& {Harding}, A.~K. 2015, Comptes Rendus Physique, 16, 641

\bibitem[{{Haslam} {et~al.}(1982){Haslam}, {Salter}, {Stoffel}, \&
  {Wilson}}]{1982A&AS...47....1H}
{Haslam}, C.~G.~T., {Salter}, C.~J., {Stoffel}, H., \& {Wilson}, W.~E. 1982,
  \aaps, 47, 1

\bibitem[{Hessels {et~al.}(2007)Hessels, Ransom, Stairs, Kaspi, \&
  Freire}]{Hessels:2007pq}
Hessels, J. W.~T., Ransom, S.~M., Stairs, I.~H., Kaspi, V.~M., \& Freire, P.
  C.~C. 2007, Astrophys. J., 670, 363

\bibitem[{Hooper \& Goodenough(2014)}]{Hooper:2010mq}
Hooper, D., \& Goodenough, L. 2014, Phys. Lett., B697, 412

\bibitem[{Hooper \& Linden(2011)}]{Hooper:2011ti}
Hooper, D., \& Linden, T. 2011, Phys. Rev., D84, 123005

\bibitem[{{Johnston} {et~al.}(2006){Johnston}, {Kramer}, {Lorimer}, {Lyne},
  {McLaughlin}, {Klein}, \& {Manchester}}]{2006MNRAS.373L...6J}
{Johnston}, S., {Kramer}, M., {Lorimer}, D.~R., {Lyne}, A.~G., {McLaughlin},
  M., {Klein}, B., \& {Manchester}, R.~N. 2006, \mnras, 373, L6

\bibitem[{Kalapotharakos {et~al.}(2014)Kalapotharakos, Harding, \&
  Kazanas}]{Kalapotharakos:2013sma}
Kalapotharakos, C., Harding, A.~K., \& Kazanas, D. 2014, Astrophys. J., 793, 97

\bibitem[{{Keith} {et~al.}(2010){Keith}, {Jameson}, {van Straten}, {Bailes},
  {Johnston}, {Kramer}, {Possenti}, {Bates}, {Bhat}, {Burgay}, {Burke-Spolaor},
  {D'Amico}, {Levin}, {McMahon}, {Milia}, \& {Stappers}}]{2010MNRAS.409..619K}
{Keith}, M.~J., {Jameson}, A., {van Straten}, W., {Bailes}, M., {Johnston}, S.,
  {Kramer}, M., {Possenti}, A., {Bates}, S.~D., {Bhat}, N.~D.~R., {Burgay}, M.,
  {Burke-Spolaor}, S., {D'Amico}, N., {Levin}, L., {McMahon}, P.~L., {Milia},
  S., \& {Stappers}, B.~W. 2010, \mnras, 409, 619

\bibitem[{{Konar}(2010)}]{2010MNRAS.409..259K}
{Konar}, S. 2010, \mnras, 409, 259

\bibitem[{{Kramer} {et~al.}(1998){Kramer}, {Xilouris}, {Lorimer}, {Doroshenko},
  {Jessner}, {Wielebinski}, {Wolszczan}, \& {Camilo}}]{1998ApJ...501..270K}
{Kramer}, M., {Xilouris}, K.~M., {Lorimer}, D.~R., {Doroshenko}, O., {Jessner},
  A., {Wielebinski}, R., {Wolszczan}, A., \& {Camilo}, F. 1998, \apj, 501, 270

\bibitem[{{Lawson} {et~al.}(1987){Lawson}, {Mayer}, {Osborne}, \&
  {Parkinson}}]{1987MNRAS.225..307L}
{Lawson}, K.~D., {Mayer}, C.~J., {Osborne}, J.~L., \& {Parkinson}, M.~L. 1987,
  \mnras, 225, 307

\bibitem[{{Lee} {et~al.}(2016){Lee}, {Lisanti}, {Safdi}, {Slatyer}, \&
  {Xue}}]{Lee:2015fea}
{Lee}, S.~K., {Lisanti}, M., {Safdi}, B.~R., {Slatyer}, T.~R., \& {Xue}, W.
  2016, Physical Review Letters, 116, 051103

\bibitem[{{Levin} {et~al.}(2013){Levin}, {Bailes}, {Barsdell}, {Bates}, {Bhat},
  {Burgay}, {Burke-Spolaor}, {Champion}, {Coster}, {D'Amico}, {Jameson},
  {Johnston}, {Keith}, {Kramer}, {Milia}, {Ng}, {Possenti}, {Stappers},
  {Thornton}, \& {van Straten}}]{2013MNRAS.434.1387L}
{Levin}, L., {Bailes}, M., {Barsdell}, B.~R., {Bates}, S.~D., {Bhat}, N.~D.~R.,
  {Burgay}, M., {Burke-Spolaor}, S., {Champion}, D.~J., {Coster}, P.,
  {D'Amico}, N., {Jameson}, A., {Johnston}, S., {Keith}, M.~J., {Kramer}, M.,
  {Milia}, S., {Ng}, C., {Possenti}, A., {Stappers}, B., {Thornton}, D., \&
  {van Straten}, W. 2013, \mnras, 434, 1387

\bibitem[{Lorimer {et~al.}(2015)}]{Lorimer:2015iga}
Lorimer, D.~R., {et~al.} 2015, Mon. Not. Roy. Astron. Soc., 450, 2185

\bibitem[{Macias \& Gordon(2014)}]{Macias:2013vya}
Macias, O., \& Gordon, C. 2014, Phys.Rev., D89, 063515

\bibitem[{{Macquart} \& {Kanekar}(2015)}]{2015ApJ...805..172M}
{Macquart}, J.-P., \& {Kanekar}, N. 2015, \apj, 805, 172

\bibitem[{Macquart \& Kanekar(2015)}]{Macquart:2015jfa}
Macquart, J.-P., \& Kanekar, N. 2015, Astrophys. J., 805, 172

\bibitem[{Manchester {et~al.}(2005)Manchester, Hobbs, Teoh, \&
  Hobbs}]{Manchester:2004bp}
Manchester, R.~N., Hobbs, G.~B., Teoh, A., \& Hobbs, M. 2005, Astron.J., 129,
  1993

\bibitem[{{Maron} {et~al.}(2000){Maron}, {Kijak}, {Kramer}, \&
  {Wielebinski}}]{2000A&AS..147..195M}
{Maron}, O., {Kijak}, J., {Kramer}, M., \& {Wielebinski}, R. 2000, \aaps, 147,
  195

\bibitem[{{Massaro} {et~al.}(2013){Massaro}, {D'Abrusco}, {Paggi}, {Masetti},
  {Giroletti}, {Tosti}, {Smith}, \& {Funk}}]{2013ApJS..206...13M}
{Massaro}, F., {D'Abrusco}, R., {Paggi}, A., {Masetti}, N., {Giroletti}, M.,
  {Tosti}, G., {Smith}, H.~A., \& {Funk}, S. 2013, \apjs, 206, 13

\bibitem[{{Massaro} {et~al.}(2014){Massaro}, {Masetti}, {D'Abrusco}, {Paggi},
  \& {Funk}}]{2014AJ....148...66M}
{Massaro}, F., {Masetti}, N., {D'Abrusco}, R., {Paggi}, A., \& {Funk}, S. 2014,
  \aj, 148, 66

\bibitem[{Massaro {et~al.}(2015)}]{2015ApJS..217....2M}
Massaro, F., {et~al.} 2015, Astrophys. J. Suppl., 217, 2

\bibitem[{{Mayer-Hasselwander} {et~al.}(1998){Mayer-Hasselwander}, {Bertsch},
  {Dingus}, {Eckart}, {Esposito}, {Genzel}, {Hartman}, {Hunter}, {Kanbach},
  {Kniffen}, {Lin}, {Michelson}, {Muecke}, {von Montigny}, {Mukherjee},
  {Nolan}, {Pohl}, {Reimer}, {Schneid}, {Sreekumar}, \&
  {Thompson}}]{1998A&A...335..161M}
{Mayer-Hasselwander}, H.~A., {Bertsch}, D.~L., {Dingus}, B.~L., {Eckart}, A.,
  {Esposito}, J.~A., {Genzel}, R., {Hartman}, R.~C., {Hunter}, S.~D.,
  {Kanbach}, G., {Kniffen}, D.~A., {Lin}, Y.~C., {Michelson}, P.~F., {Muecke},
  A., {von Montigny}, C., {Mukherjee}, R., {Nolan}, P.~L., {Pohl}, M.,
  {Reimer}, O., {Schneid}, E.~J., {Sreekumar}, P., \& {Thompson}, D.~J. 1998,
  \aap, 335, 161

\bibitem[{McCann(2015)}]{McCann:2014dea}
McCann, A. 2015, Astrophys. J., 804, 86

\bibitem[{{Muno} {et~al.}(2003){Muno}, {Baganoff}, {Bautz}, {Brandt}, {Broos},
  {Feigelson}, {Garmire}, {Morris}, {Ricker}, \&
  {Townsley}}]{2003ApJ...589..225M}
{Muno}, M.~P., {Baganoff}, F.~K., {Bautz}, M.~W., {Brandt}, W.~N., {Broos},
  P.~S., {Feigelson}, E.~D., {Garmire}, G.~P., {Morris}, M.~R., {Ricker},
  G.~R., \& {Townsley}, L.~K. 2003, \apj, 589, 225

\bibitem[{Ng {et~al.}(2014)}]{Ng:2014mca}
Ng, C., {et~al.} 2014, Mon. Not. Roy. Astron. Soc., 439, 1865

\bibitem[{O'Leary {et~al.}(2015)O'Leary, Kistler, Kerr, \&
  Dexter}]{O'Leary:2015gfa}
O'Leary, R.~M., Kistler, M.~D., Kerr, M., \& Dexter, J. 2015

\bibitem[{Pallanca {et~al.}(2012)Pallanca, Mignani, Dalessandro, Ferraro,
  Lanzoni, Possenti, Burgay, \& Sabbi}]{Pallanca:2012dc}
Pallanca, C., Mignani, R.~P., Dalessandro, E., Ferraro, F.~R., Lanzoni, B.,
  Possenti, A., Burgay, M., \& Sabbi, E. 2012, Astrophys. J., 755, 180

\bibitem[{{Perez} {et~al.}(2015){Perez}, {Hailey}, {Bauer}, {Krivonos}, {Mori},
  {Baganoff}, {Barri{\`e}re}, {Boggs}, {Christensen}, {Craig}, {Grefenstette},
  {Grindlay}, {Harrison}, {Hong}, {Madsen}, {Nynka}, {Stern}, {Tomsick}, {Wik},
  {Zhang}, {Zhang}, \& {Zoglauer}}]{2015Natur.520..646P}
{Perez}, K., {Hailey}, C.~J., {Bauer}, F.~E., {Krivonos}, R.~A., {Mori}, K.,
  {Baganoff}, F.~K., {Barri{\`e}re}, N.~M., {Boggs}, S.~E., {Christensen},
  F.~E., {Craig}, W.~W., {Grefenstette}, B.~W., {Grindlay}, J.~E., {Harrison},
  F.~A., {Hong}, J., {Madsen}, K.~K., {Nynka}, M., {Stern}, D., {Tomsick},
  J.~A., {Wik}, D.~R., {Zhang}, S., {Zhang}, W.~W., \& {Zoglauer}, A. 2015,
  \nat, 520, 646

\bibitem[{Petrovic {et~al.}(2014)Petrovic, Serpico, \&
  Zaharijas}]{Petrovic:2014uda}
Petrovic, J., Serpico, P.~D., \& Zaharijas, G. 2014, JCAP, 1410, 052

\bibitem[{{Petrovi{\'c}} {et~al.}(2015){Petrovi{\'c}}, {Serpico}, \&
  {Zaharijas}}]{Petrovic:2014xra}
{Petrovi{\'c}}, J., {Serpico}, P.~D., \& {Zaharijas}, G. 2015, \jcap, 2, 023

\bibitem[{{Prinz} \& {Becker}(2015)}]{2015arXiv151107713P}
{Prinz}, T., \& {Becker}, W. 2015, ArXiv e-prints

\bibitem[{{Ransom}(2001)}]{2001PhDT.......123R}
{Ransom}, S.~M. 2001, PhD thesis, Harvard University

\bibitem[{Ray {et~al.}(2012)Ray, Abdo, Parent, Bhattacharya, Bhattacharyya,
  {et~al.}}]{Ray:2012ue}
Ray, P., Abdo, A., Parent, D., Bhattacharya, D., Bhattacharyya, B., {et~al.}
  2012

\bibitem[{Ray {et~al.}(2013)}]{Ray:2012ms}
Ray, P.~S., {et~al.} 2013, Astrophys. J., 763, L13

\bibitem[{{Schinzel} {et~al.}(2015){Schinzel}, {Petrov}, {Taylor}, {Mahony},
  {Edwards}, \& {Kovalev}}]{2015ApJS..217....4S}
{Schinzel}, F.~K., {Petrov}, L., {Taylor}, G.~B., {Mahony}, E.~K., {Edwards},
  P.~G., \& {Kovalev}, Y.~Y. 2015, \apjs, 217, 4

\bibitem[{Stovall {et~al.}(2013)Stovall, Lorimer, \& Lynch}]{Stovall:2013gca}
Stovall, K., Lorimer, D., \& Lynch, R.~S. 2013, Class.Quant.Grav., 30, 224003

\bibitem[{Stovall {et~al.}(2014)}]{Stovall:2014gua}
Stovall, K., {et~al.} 2014, Astrophys. J., 791, 67

\bibitem[{Strong(2007)}]{Strong:2006hf}
Strong, A.~W. 2007, Astrophys. Space Sci., 309, 35

\bibitem[{{Tremaine} {et~al.}(1975){Tremaine}, {Ostriker}, \&
  {Spitzer}}]{1975ApJ...196..407T}
{Tremaine}, S.~D., {Ostriker}, J.~P., \& {Spitzer}, Jr., L. 1975, \apj, 196,
  407

\bibitem[{Venter {et~al.}(2014)Venter, Johnson, Harding, \&
  Grove}]{Venter:2014zea}
Venter, C., Johnson, T., Harding, A., \& Grove, J. 2014, in {Proceedings of
  SAIP2013, the 58th Annual Conference of the South African Institute of
  Physics, edited by Roelf Botha and Thulani Jili, ISBN: 978-0-620-62819-8}

\bibitem[{{Verbunt} \& {Hut}(1987)}]{1987IAUS..125..187V}
{Verbunt}, F., \& {Hut}, P. 1987, in IAU Symposium, Vol. 125, The Origin and
  Evolution of Neutron Stars, ed. D.~J. {Helfand} \& J.-H. {Huang}, 187

\bibitem[{Vitale \& Morselli(2009)}]{Vitale:2009hr}
Vitale, V., \& Morselli, A. 2009

\bibitem[{Wang(2005)}]{Wang:2005ti}
Wang, W. 2005, Chin.J.Astron.Astrophys.

\bibitem[{Yuan \& Ioka(2015)}]{Yuan:2014yda}
Yuan, Q., \& Ioka, K. 2015, Astrophys. J., 802, 124

\bibitem[{Yuan \& Zhang(2014)}]{Yuan:2014rca}
Yuan, Q., \& Zhang, B. 2014, JHEAp, 3-4, 1

\bibitem[{{Zhang} \& {Cheng}(2003)}]{2003A&A...398..639Z}
{Zhang}, L., \& {Cheng}, K.~S. 2003, \aap, 398, 639

\bibitem[{{Zhou} {et~al.}(2015){Zhou}, {Liang}, {Huang}, {Li}, {Fan}, {Feng},
  \& {Chang}}]{Zhou:2014lva}
{Zhou}, B., {Liang}, Y.-F., {Huang}, X., {Li}, X., {Fan}, Y.-Z., {Feng}, L., \&
  {Chang}, J. 2015, \prd, 91, 123010

\bibitem[{Zhou {et~al.}(2015)Zhou, Zhang, Huang, Li, Liang, Fu, Yan, \&
  Liu}]{Zhou:2015cta}
Zhou, J.~N., Zhang, P.~F., Huang, X.~Y., Li, X., Liang, Y.~F., Fu, L., Yan,
  J.~Z., \& Liu, Q.~Z. 2015, Mon. Not. Roy. Astron. Soc., 448, 3215

\end{thebibliography}
\bibliographystyle{apj}

\appendix

\section{Multi-wavelength study of MSP candidates in Fermi data} Based on a
spectral matching analysis, \cite{Bartels:2015aea} identified 13 sources in the
3FGL catalog \citep{TheFermi-LAT:2015hja} as candidates for MSPs in the inner
Galaxy ($|\ell|<12^\circ$ and $2^\circ<|b|<12^\circ$).  The criterion was that
the spectrum of the sources is roughly compatible with the spectrum of stacked
MSPs from \cite{Cholis:2014noa}, and they show no significant variability.  We
stress that the \emph{raison d'etre} for this source list is \emph{not} to find
the best MSP candidates for radio follow-up searches (this requires a more
detailed study that will be presented elsewhere), but simply to remove a bias
in the wavelet analysis from \cite{Bartels:2015aea} by unmasking some of the
3FGL sources that might be part of the bulge population.  However, we will here
analyze the properties of these 13 sources, as well as some of the other
wavelet peaks found in this analysis, to \emph{firstly} confirm that an MSP
interpretation of the 13 sources as well as the significant wavelet peaks is
compatible with multi-wavelength data, and \emph{secondly} demonstrate the
potential and limitations that such multi-wavelength studies of MSP candidates
in the inner Galaxy entail.

\subsection{Cross-correlation of gamma-ray MSP candidates and known radio pulsars}
\label{SedSec}

In the recent analysis of the inner Galaxy by \cite{Bartels:2015aea}, which
adopted a wavelet decomposition of the gamma-ray sky to search for
sub-threshold point sources, a significant clustering of photons compatible
with the unresolved gamma-ray emission from a bulge population of MSPs as
suggested by \Fermi-LAT data has been observed.  The region of interest (ROI)
of the analysis is defined by $|\ell|<12^{\circ}$ and
$2^{\circ}<|b|<12^{\circ}$.  The signal-to-noise ratio of the wavelet transform
at position $\Omega$, $\mathcal{S}(\Omega)$ \citep[Eq.~(2)
in][]{Bartels:2015aea}, is a rough measure for the local significance for
having a source at position $\Omega$, in units of standard deviations.  The
peaks in $\mathcal{S}(\Omega)$ considered in the wavelet search have
significances in the range $1\leq \mathcal{S}\leq10$.  In particular the ones
with $\mathcal{S}>3$ may be considered as
promising targets for radio follow-up searches for radio MSPs.

If the more significant gamma-ray wavelet peaks from \cite{Bartels:2015aea} are
indeed identified with a bulge MSP population, they should not be correlated
with foreground sources.  We explore this possibility by studying the
correlation between the radio pulsars in the ATNF catalog
\citep{Manchester:2004bp} and the wavelet peaks with $\mathcal{S}>2$ and
$\mathcal{S}>3$.  Within the main ROI, the pulsar ATNF catalog contains 331
pulsars with a measurement of the distance.  However, we will study potential
correlations not only in the inner Galaxy ROI, but also in the control regions
along the Galactic disk from \cite{Bartels:2015aea}, centered in $l = \pm k
\cdot 20^\circ$ and $b=0^{\circ}$, with $k$=1,2,3,4 and with the same extension
of the Galactic center region.  

\begin{figure*}
  \centering
  \includegraphics[width=0.45\textwidth]{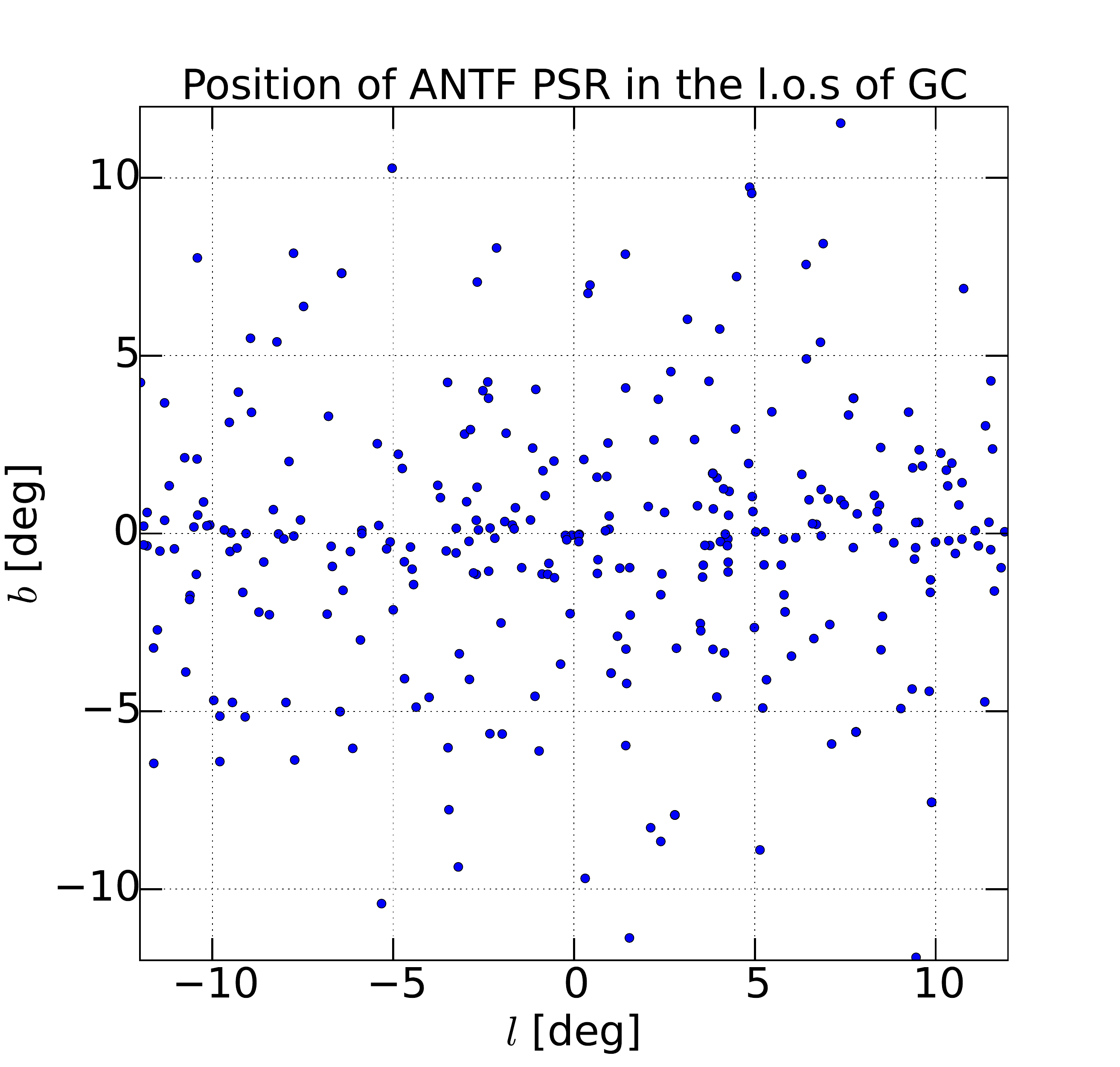}
  \includegraphics[width=0.45\textwidth]{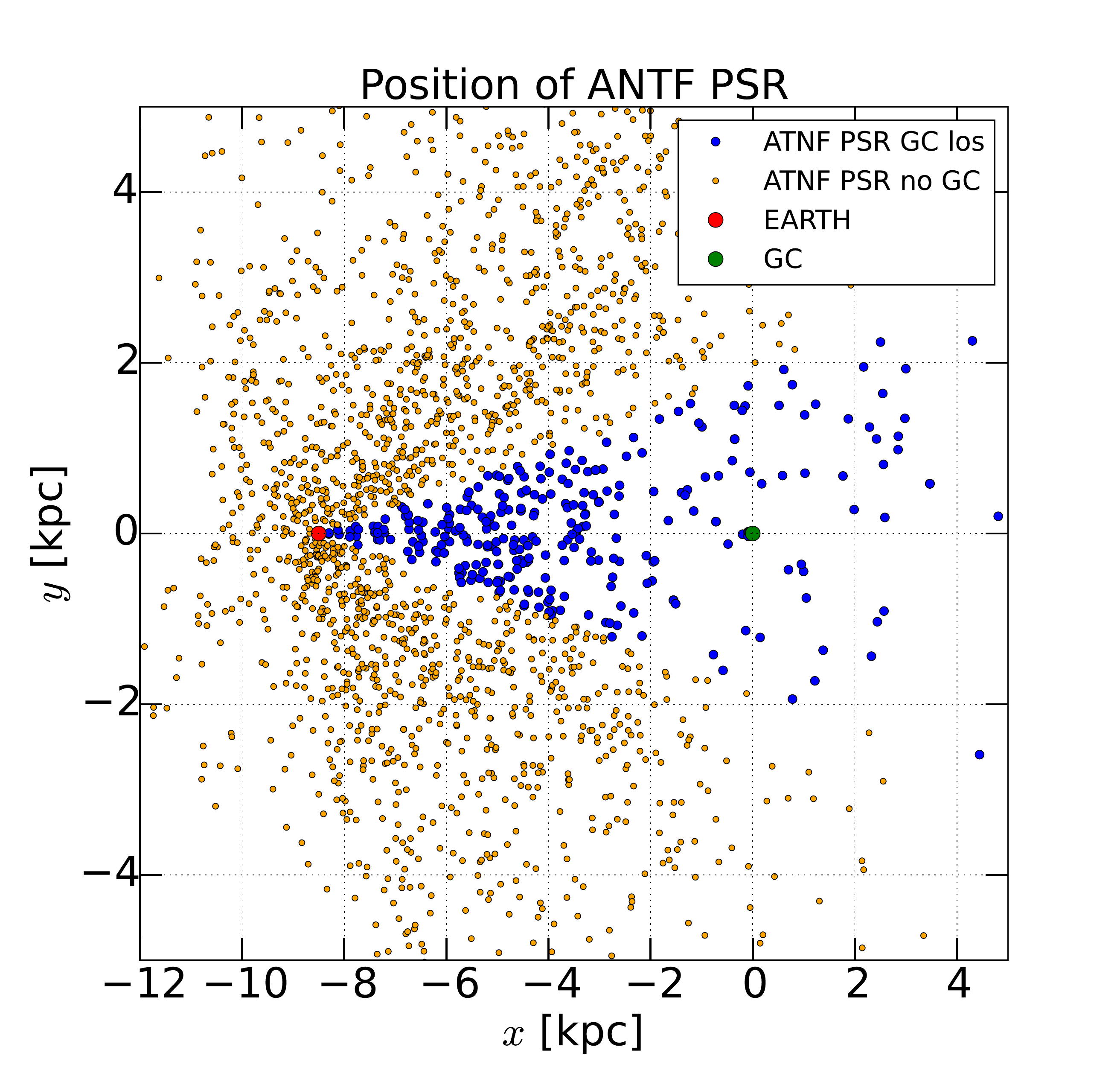}
  \caption{Positions of the ATNF catalog pulsars  in the $|l|<12^{\circ}$ and
    $2^{\circ}<|b|<12^{\circ}$ region.  Left panel: catalog sources are
    displayed as blue points in the longitude-latitude plane. Each position in
    the plane corresponds to an observation a line-of-sight.  Right panel: all catalog
    sources are displayed as gold points.  They are projected onto the Galactic
    plane, knowing their distance from the Earth (identified by a red point)
    and located by their Galactic coordinates (the Galactic center is
    identified by a green point).  The sources displayed in the left panel are
    depicted as blue points.}
  \label{fig:positionatnf}
  \medskip
\end{figure*}

\medskip

We consider here the same wavelet peaks as in \cite{Bartels:2015aea}.  That
means from the total number of identified wavelet peaks we subtract: (i) all
sources that spatially coincide with associated sources from the 3FGL catalog
\citep{TheFermi-LAT:2015hja}; (ii) all unassociated sources with a non-pulsar
spectrum, according to the same criterion as described in
\citep{Bartels:2015aea}.  

We derive for each ROI (main and control) the number of positional correlations
between the gamma-ray wavelet peaks and the ATNF sources.  As threshold
distance for the correlation, we tested two values, $0.1^\circ$ and
$0.2^\circ$. The first angle cut is equal to the largest value of the 95\%
containment angle (\texttt{Conf95\_SemiMajor} in the 3FGL catalog), which is an
indicator of the positional error of point sources.  The second value
$0.2^\circ$ has been considered because most of the gamma-ray peaks are just
below the detection threshold and so the 95\% containment angle parameter for
them is effectively larger.  However, we found similar results and will only
use $0.1^\circ$ in the following.

\begin{figure*}
  \centering
  \includegraphics[width=0.40\textwidth]{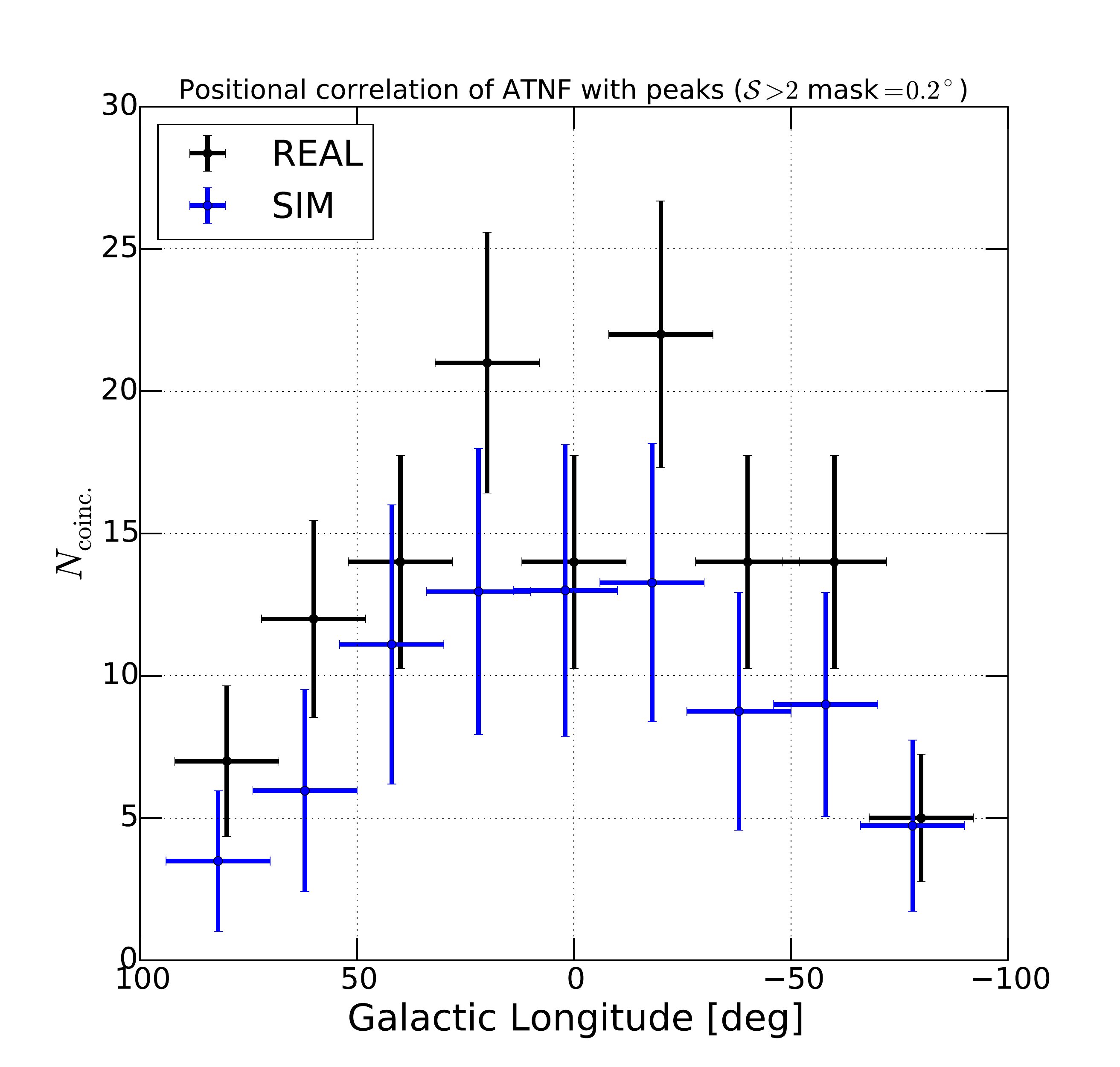}
  \includegraphics[width=0.40\textwidth]{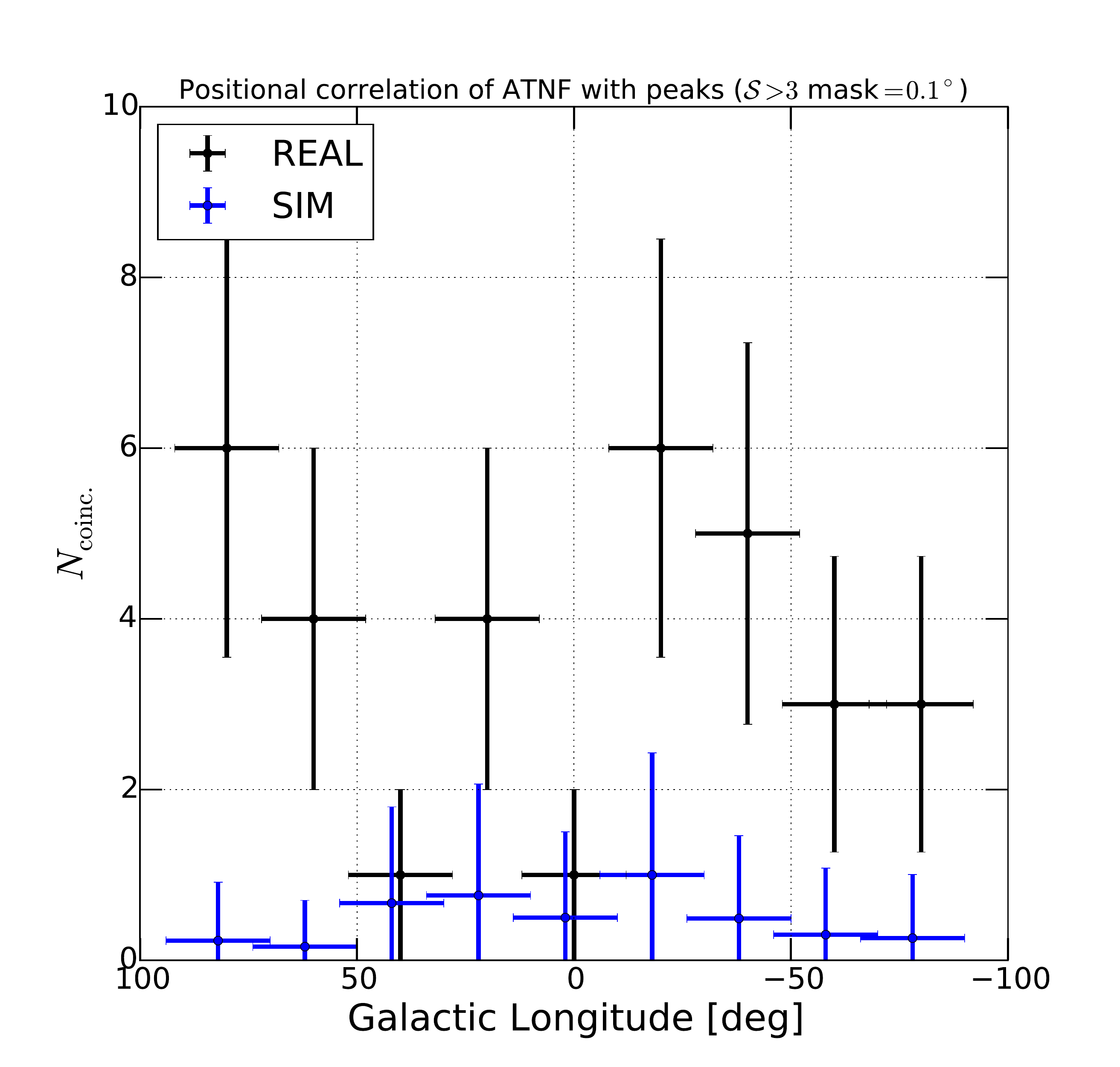}
    \caption{Number of positional correlations between the gamma-ray wavelet
      peaks and the sources in the ATNF catalog, as a function of the Galactic
      longitude, for latitudes $2^{\circ}<|b|<12^{\circ}$.  The left (right)
      panels correspond to peaks with significance $\mathcal{S}>2$
      ($\mathcal{S}>3$).  The black points represent the correlations found
      from the real gamma-ray wavelet peak catalog as discussed in the text,
      while the blue ones are derived from a reshuffling in latitude bins.  The
      analysis is performed for threshold angles $0.2^\circ$ (left) and
      $0.1^\circ$ (right).}
  \label{fig:corrpos}
  \medskip
\end{figure*}

In Fig.~\ref{fig:corrpos} we plot the number of positional correlations as a
function of the longitudinal ROI position.  For the gamma-ray wavelet peaks we
have chosen the significance $\mathcal{S}>2$ and $\mathcal{S}>3$.  The results
are plotted as black error bars, and actually fluctuate strongly from ROI to
ROI.  The error bars are defined as the Poissonian error on the number of
correlations.

We have also estimated the number of positional correlations that one would
expect from a random positioning of the wavelet peaks in each of the analyzed
sky regions.  In order to derive this test population, we used ``scrambled
data''  and changed the longitude and latitude of each wavelet peak randomly in
the interval $[l-2^\circ,l+2^\circ]$ and $[b-1^\circ,b+1^\circ]$.  In this way,
we largely preserve the observed spatial distribution of the peaks, which is
concentrated along the Galactic disk. 

The cross-correlation that we find between the ATNF sources and our scrambled
test wavelet sample are shown by the blue error bars in Fig.~\ref{fig:corrpos}.
Interestingly, for both $\mathcal{S}$ $> 2$ and even more $\mathcal{S}$ $> 3$,
we find in most ROIs an excess of correlations above what is randomly expected,
with the exception of the Galactic center and a region around $\ell\approx
40^\circ$.  This strongly suggests that some of the wavelet peaks are actually
caused by the emission of pulsars that are already part of the ATNF, but not
the 3FGL.  We note that the number of \emph{potential} correlations in each ROI
is much larger than what we find.  

The variations in the correlation between wavelet peaks and ATNF sources that
we find in most of the control regions away from the Galactic center suggest
that along the Galactic plane a number of radio pulsars remained below the
\Fermi\ detection threshold up to now, but showed up as wavelet peaks in our
analysis.  This effect depends on the general pulsar density in a certain
direction, and happens to be small towards the inner Galaxy.

\section{An analysis of 13 gamma-ray unassociated sources in the inner Galaxy}

We will in the following study in some detail the properties of the 13
unassociated 3FGL sources that were identified in \cite{Bartels:2015aea} as MSP
candidates (see their Table I).  We stress again that this does not imply that
these sources would be the best targets for radio follow-up searches.  Instead,
the discussion below will show what is in general possible with spectral and
multi-wavelength analyses.

\subsection{Gamma-ray spectral analysis}
\label{sec:spectraun}

\begin{table}
  \centering
  \begin{tabular}{ccccccc}
    \toprule
    3FGL Source & $\Gamma^{\rm{PSR}}$ & $E_{\rm{cut}}^{\rm{PSR}}$ (GeV) &
    $\tilde{\chi}^2_{\rm{PSR}}$ & $\Gamma^{\rm{AGN}}$ &
    $E_{\rm{cut}}^{\rm{AGN}}$ (GeV) & $\tilde{\chi}^2_{\rm{AGN}}$ \\
    \hline
    J1649.6-3007 &  $>$1.90 & $>$5.5  & 0.88  & 2.15  $\pm$  0.25 & 25 $\pm$ 5  & 0.15 \\
    J1703.6-2850 &  1.49  $\pm$  0.36 & $>$5.5  & 1.15  & 1.94  $\pm$  0.24 &  25 $\pm$ 4  & 0.32 \\
    J1740.5-2642 &  1.54  $\pm$  0.44 & 3.1 $\pm$ 1.6  & 0.08  & 1.94 $\pm$ 0.14 & $<$7  & 0.66 \\
    J1740.8-1933	&  $>$1.9 & $>$5.5  & 2.4  & 2.13  $\pm$  0.20 & $>$200 & 0.22 \\
    J1744.8-1557 & $>$1.9 & 4.7 $\pm$ 3.6 & 0.17  & 2.17 $\pm$ 0.58 & 10 $\pm$ 3 & 0.08 \\
    J1758.8-4108 &  $<$0.7  & 1.8 $\pm$ 0.3  & 1.91  & 1.85  $\pm$  0.35 & 21 $\pm$ 6  & 2.28 \\
    J1759.2-3848 &  1.52  $\pm$  0.22 & $>$5.5  & 0.18  & 1.96 $\pm$ 0.18 & $>270$  & 0.24 \\
    J1808.3-3357 &  1.37  $\pm$  0.32 & 2.5 $\pm$ 1.0  & 0.08  & 1.84$\pm$0.11 & $<$7  & 1.28 \\
    J1808.4-3519 &  $>$1.90 & $>$5.5  & 0.32  & 2.03  $\pm$  0.51 & 8.1 $\pm$ 3.0  & 0.27 \\
    J1808.4-3703 &  1.46  $\pm$  0.15 & 2.7 $\pm$  0.6 & 0.022  & 1.93$\pm$ 0.19& $<7$  & 0.64 \\
    J1820.4-3217  &  1.60  $\pm$  0.35 & 2.7 $\pm$ 1.0  & 0.41  & 2.05  $\pm$  0.13 & $<7$  & 0.21 \\
    J1830.8-3136 &  $<$0.70  & 1.8 $\pm$ 0.3  & 0.75  & $<1.75$ & 9.4 $\pm$ 3.0  & 1.80 \\
    J1837.3-2403  &  1.73  $\pm$  0.24 & $>$5.5  & 0.48  & 1.97  $\pm$  0.57 & 13 $\pm$ 5  & 0.50 \\
    \botrule
  \end{tabular}
  \caption{Results for the fits to the gamma-ray spectra of the 13 unassociated
  3FGL sources from \cite{Bartels:2015aea}, using 3FGL catalog spectral data
  and two different assumptions for the SED parameters (see text for details). }
  \label{Tab:gammafit}
\end{table}

We study here the gamma-ray spectral energy distribution (SED) of these MSP
candidates.  To this end, we perform a fit to their gamma-ray spectra as given
in the 3FGL catalog \citep{TheFermi-LAT:2015hja}, in the energy range $0.1-100$
GeV. We adopt a power-law with an exponential cutoff, which is the typical
gamma-ray SED of pulsars,
\begin{equation}
  \label{eq:plexp}
  \frac{dN}{dE} = K_0 \left( \frac{E}{E_0} \right)^{-\Gamma} \exp{\left(- \frac{E}{E_{\rm{cut}}} \right)},
\end{equation}
where $K_0$ is the normalization of the spectrum, $E_0$ is the pivot energy,
$\Gamma$ is the photon index and $E_{\rm{cut}}$ is the energy cutoff.  In order
to check if those sources could be spectrally associated with AGNs (although,
  as discussed in \cite{Bartels:2015aea}, this is \emph{a priori} not very
likely given the low average number density of AGNs in the Galactic disk), we
consider two different cases for the range of variability of the photon index
and the energy cut off.  We stress that for pulsars and AGNs, the model
parameters are usually strongly correlated, which we neglect here for
simplicity, however.

\begin{itemize}
  \item {\it Pulsar like}. The average value for $\Gamma$ and
    $E_{\rm{cut}}$ for pulsars in the {\it Fermi}-LAT catalogs (see e.g.
    \cite{TheFermi-LAT:2013ssa}) are $\Gamma=1.30\pm0.30$ and
    $\log_{10}(E_{\rm{cut}}/{\rm MeV}) = (3.38\pm0.18)$. We therefore restrict
    the photon index in range $\Gamma\in[0.70,1.90]$ and the energy cutoff
    $E_{\rm{cut}}\in[1.5,5.50]$ GeV, according to the $95\%$~CL limits of their
    observed distributions.  Note that this entails the spectra of both young
    and recycled pulsars.
  \item {\it Flat Spectrum Radio Quasar (FSRQ) like}. We have performed a fit to the FSRQ sources in the
    3FGL catalog \cite{TheFermi-LAT:2015hja} with a detection significance
    large than 6, with the SED assumed to be a power-law with an exponential
    cutoff (Eq.~\ref{eq:plexp}). The best fit parameters are $\Gamma =
    2.25\pm0.25$ and $E_{\rm{cut}}=30^{+120}_{-16}$ GeV, and the fit has a
    reduced chi-square $\tilde{\chi}^2=0.72$. We therefore restrict the photon
    index to the $95\%$~CL range $\Gamma\in[1.75,2.75]$ and
    $E_{\rm{cut}}\in[8.0,270]$ GeV.
\end{itemize}

The fit results are summarized in Tab.~\ref{Tab:gammafit} in terms of the
photon index $\Gamma$ and the exponential cutoff $E_{\rm{cut}}$ best fit values
for each of the 13 sources, both for the pulsar and the AGN priors on the free
parameters.  We also indicate the goodness-of-fit by the
$\tilde\chi^2=\chi^2/\text{dof}$, where the degrees of freedom are $\text{dof}=5-3$.  For most of the
sources, we find rather small values for $\tilde\chi^2$, which indicates that
the fluxes are over-fitted, likely related to the low number of energy bins or
the large statistical error bars of the fluxes, which precludes any statements
about what spectra are preferred.  In a few cases, the $\tilde\chi^2$ is
significantly above 1.0; values above around 2.3 would indicate a $90\%$~CL
tension between model and measured spectrum.  This is only the case for
J1740.8-1933, which is mildly inconsistent with a pulsar spectrum, and
J1758.8-4108, which is mildly inconsistent with a AGN spectrum.  We conclude
that spectral information alone, in the way we use it here, is not enough to
make strong statements about the nature of the source.  However, if we simply
interpret the results as indicative for a possible source type, 6 sources might
be more pulsar-like, and 6 source more AGN-like.  A more detailed study, taking
into account parameter correlations and a larger range of spectral bins, is
warranted but beyond the scope of the current work.

\bigskip

\subsection{Multi-wavelength properties from X-ray and radio}
\label{subsec:radiox}

Recent multi-frequency analyses \citep[see e.g.][]{2013ApJS..206...13M}
supported by optical follow up spectroscopic campaigns \citep[see
e.g.][]{2014AJ....148...66M} on different sample of unassociated gamma-ray
sources have been extremely successful to find new blazar-like counterparts as
well to exclude their presence \citep[see e.g.][and references
therein]{2015ApJS..217....2M}

For all the 13 unidentified gamma-ray sources in~\cite{Bartels:2015aea} we
investigated several catalogs and surveys, spanning the whole electromagnetic
spectrum, and searching for potential low-energy counterparts that could either
help to confirm or provide information on the pulsar-like nature/behavior of
these sources. We reduce the X-ray observations available in the SWIFT archive
and obtain with the follow up program on the unassociated \Fermi-LAT objects.

In particular, since each associated gamma-ray blazar has a radio counterpart
we first investigated the NRAO VLA Sky Survey that cover the footprint of these
13 objects \citep{1998AJ....115.1693C} to exclude or confirm the possible
presence of blazar-like potential counterparts within the Fermi positional
uncertainty. This has been also motivated by the success of the follow up radio
observations performed since the launch of Fermi
\citep[e.g.,][]{2015ApJS..217....4S}. We also searched in low frequency radio
observations (i.e., below $\sim$1~GHz) for blazar-like source.

\paragraph{3FGL J1703.6-2850} This {\it Fermi}-LAT source has a single
unidentified radio object (NVSS J170341-285343) lying within the positional
uncertainty region at 95\% level of confidence.  According to the NVSS radio
image NVSS J170341-285343 has compact radio structure also showing a jet-like
component that could resemble of a blazar-like nature.  This radio source has
also an optical counterpart in the USNO catalog. In the X-ray images obtained
by SWIFT there are no objects detected with a signal-to-noise ratio greater than
3.

\paragraph{3FGL J1740.5-2642} There are two radio sources lying within the
positional uncertainty region of this unassociated {\it Fermi}-LAT object.
However the first source: NVSS J174012-264422 is a planetary nebula (aka ESO
520 PN-015) and thus is unlikely to be the low-energy counterpart of 3FGL
J1740.5-2642.  The other one, NVSS J174039-264541 is a simple, bright (flux
density at 1.4 GHz of 14.7 mJy), radio source with a compact structure having
also an optical correspondence in the USNO catalog.

\paragraph{3FGL J1740.8-1933} For 3FGL J1740.8-1933 as in the previous case
there are two compact radio sources lying within the positional uncertainty
region at 95\% level of confidence: NVSS J174051-193011 and NVSS
J174105-193006. None of them has an optical counterpart but the latter is also
detected in the WISE all-sky survey, even if its IR colors are not consistent
with those of the {\it Fermi}-LAT detected blazars. No sources are detected in
the X-rays as paper in the SWIFT observations.

\paragraph{3FGL J1744.8-1557} There are 5 radio sources in the NVSS catalog
that lie within the positional uncertainty region of 3FGL J1744.8-1557.  Two of
them are also detected in the WISE all-sky survey: NVSS J174509-155000 and NVSS
J174443-160531 but they do not have IR colors similar to the {\it Fermi}-LAT
blazars.  In addition, NVSS J174437-160253 shows an extended structure while
all the others appear to be compact in the NVSS radio images. None of them is
detected in the X-rays.

\paragraph{3FGL J1759.2-3848} 3 radio sources reported in the NVSS catalog, all
compact, are present in the line-of-sight of this source. The most interesting
one is probably NVSS J175926-384753 that lies only 136 arcsec from the
gamma-ray position of 3FGL J1759.2-3848 and has both an IR and an optical
counterpart.  None of them is indeed detected in the X-rays. There is only one
source in the SWIFT-XRT image but it corresponds to a bright star in the
field of view clearly detected in the optical and ultraviolet images of  the UVOT
instrument on board of SWIFT.

\paragraph{3FGL J1808.4-3703} This source is remarkably interesting because
within its positional uncertainty region at 95\% level of confidence there
is a known X-ray transient: SAX J1808.4-3658.  This is an accreting
MSPS, in which the neutron star is orbiting around a brown dwarf
companion. A recent and detailed X-ray analysis of all the archival SWIFT-XRT
observations is presented in~\cite{Campana:2008vf}.

\paragraph{3FGL J1820.4-3217} This is the unique source of our sample for which
the gamma-ray spectral properties have been investigated with a statistical
approach. The results provided by a classification tree method support the idea
that the gamma-ray behavior of this source resembles that of an active galaxy
rather than a pulsar.  There is a radio source (i.e., NVSS J182045-321621)
lying within its positional uncertainty region that presents a faint extended
structure and has an infrared and an optical potential counterpart at $\sim$12
arcsec distance from the radio core position. This NVSS object is not detected
in the X-rays.

\paragraph{3FGL J1830.8-3136} Four radio sources are detected within the region
of interest for 3FGL J1830.8-3136, in particular NVSS J183027-313738 shows a
compact structure but the other two radio objects: NVSS J183038-313506 and NVSS
J183033-313608 appear to be knots of a jet-like extended structure of 0.06
degrees length.  NVSS JNVSS J183027-313738 is also detected in the optical but
does not have an IR counterpart in the WISE all sky survey.

\paragraph{3FGL J1837.3-2403} Approximately 0.2 degrees from the position of
the {\it Fermi}-LAT source, and less than 0.1 degree distance from the border
of its elliptical positional uncertainty region having a major axis of 0.2
degrees there is a well known globular cluster: M22.  Unfortunately the SWIFT
XRT image is centered on the globular cluster and thus it is covering
completely the {\it Fermi}-LAT region of interest, so it is not possible to
know if there are X-ray sources detected that could be potential counterpart of
the gamma-ray object.

\paragraph{3FGL J1649.6-3007, 3FGL J1758.8-4108 and 3FGL J1808.4-3519} No X-ray
sources are detected within the positional uncertainty region of this {\it
Fermi}-LAT source in the SWIFT image.  In addition there are no radio sources
within the same region of interest and no WISE sources with IR colors similar
to gamma-ray blazars.  

\paragraph{3FGL J1808.3-3357} There are 3 X-ray sources and among them one is
NOVASGR20093.

\end{document}